\documentclass[]{aastex631}

\usepackage{amsmath}
\usepackage{bm}
\usepackage[shortlabels]{enumitem}

\shorttitle{Polarization properties of energetic pulsars}
\shortauthors{Mitra, Basu \& Melikidze}

\begin{document}

\title{Polarization Properties of Energetic Pulsars at Meterwavelengths}

\author[0000-0002-9142-9835]{Dipanjan Mitra}
\affiliation{National Centre for Radio Astrophysics, Tata Institute of Fundamental Research, Pune 411007, India.}
\affiliation{Janusz Gil Institute of Astronomy, University of Zielona G\'ora, ul. Szafrana 2, 65-516 Zielona G\'ora, Poland.}

\author[0000-0003-1824-4487]{Rahul Basu}
\affiliation{Janusz Gil Institute of Astronomy, University of Zielona G\'ora, ul. Szafrana 2, 65-516 Zielona G\'ora, Poland.}

\author[0000-0003-1879-1659]{George I. Melikidze}
\affiliation{Janusz Gil Institute of Astronomy, University of Zielona G\'ora, ul. Szafrana 2, 65-516 Zielona G\'ora, Poland.}
\affiliation{Evgeni Kharadze Georgian National Astrophysical Observatory, 0301 Abastumani, Georgia.}

\begin{abstract}
Polarization behaviour shows a transition in the pulsar population, where 
energetic sources with higher spin-down energy loss, $\dot{E} > 10^{34}$ 
erg~s$^{-1}$, often have fractional linear polarisation ($L/I$) close to 
100\%, while below this range $L/I$ is usually lower than 50\%. The 
polarisation behaviour has been primarily studied at higher frequencies above 1 
GHz, and in this work we explore the single pulse polarisation behaviour in 
pulsars with $\dot{E} > 5\times10^{33}$ erg~s$^{-1}$ at a lower frequency range
of 300-750 MHz. The polarisation behaviour can be divided into two categories, 
the first with $L/I>$ 70\% where the polarisation position angle (PPA) follows 
a single track, and a second group with $L/I <$ 70\% and scattered PPA 
behaviour with or without orthogonal modes. However, there are some single 
pulses in the first category that also have lower $L/I$ and exhibit the 
presence of two polarisation modes along orthogonal tracks. The radio emission
in pulsars arises due to coherent curvature radiation (CCR) from charge 
bunches, which develops due to non-linear instabilities in the pulsar plasma 
forming charge separated  envelope solitons. The CCR excites orthogonally 
polarised X and O modes oriented perpendicular and parallel to the magnetic 
field line planes, that detach in the plasma and propagate independently. The 
O-mode is seven times stronger than the X-mode but gets damped in the medium. 
We show that incoherent mixing of the X and O modes with different levels of 
damping can reproduce the observed polarisation features in the energetic 
pulsar population.
\end{abstract}


\section{Introduction} \label{sec:intro}
The coherent radio emission in normal pulsars, with periods ($P$) $>$ 20 
milliseconds, arises due to plasma instabilities excited in the inner 
magnetosphere close to the neutron star surface. The emission mechanism excites
plasma modes which propagates in the magnetospheric medium and detaches from 
the plasma as electromagnetic waves before entering the interstellar medium on 
their way to the observer \citep{1975ARA&A..13..511G,1995JApA...16..137M,
2024Univ...10..248M}. Several aspects of the emission process are still 
ambiguous, like the modification and eventual escape of the emission modes from
the magnetospheric plasma, and the polarisation features are expected to 
provide important clues in their understanding. The rotating vector model 
\cite[RVM,][]{1969ApL.....3..225R} predicts the electric vector of the emergent
radiation from pulsars to trace the variation of the magnetic field line 
planes, and as the star rotates the observer's line of sight cuts different 
magnetic planes such that the polarisation position angle (PPA) has a S-shaped 
curve. Although the PPA in the average profile often show wide deviation from 
the RVM, careful measurements using time samples with high levels of linear 
polarisation, i.e. polarisation fraction in excess of 80\%, show that the PPA 
follow S-shaped curve that closely match the RVM \citep{2023MNRAS.521L..34M,
2024MNRAS.530.4839J}. In some pulsars two parallel tracks are seen, separated 
by phase difference of 90$\degr$, and in certain instances the PPA jumps 
between these two tracks. These represent the orthogonal polarisation modes 
(OPM) and can be identified with the extraordinary (X) and ordinary (O) modes 
of the strongly magnetized pair plasma, where the electric vector corresponding
to the X-mode is perpendicular to the plane containing the propagation vector, 
$\bm{k}$, and the local magnetic field, $\bm{B}$, while electric vector of the 
O-mode lies in this plane \citep[see e.g.][]{1976ApJ...204L..13C,
1977PASA....3..120M}. The observation of OPM showing alignment with RVM 
provides evidence for radio emission mechanism in pulsars to be coherent 
curvature radiation (CCR) excited by charge bunches, with CCR capable of 
exciting X and O modes, respectively \citep{2023MNRAS.521L..34M}. In the CCR 
mechanism the two modes are excited simultaneously at the wave generation 
region with the O-mode being seven times stronger than the X-mode, and they 
split up in the medium due to different refractive indices and propagate as 
100\% linearly polarised waves \citep{1986ApJ...302..120A}. A distinguishing 
feature of the CCR mechanism is that the O and X modes after disentangling are 
polarised with orientation parallel and perpendicular to a specific 
$\bm{k}$-$\bm{B}$ plane that coincide with the plane of the curved magnetic 
field line.

The high linearly polarised time samples are seen less often in lower energetic
pulsars with spindown energy loss $\dot{E} < 5\times10^{33}$ ergs~s$^{-1}$, 
where the typical linear polarisation fraction is around 25\% 
\citep{2024MNRAS.530.4839J}. In these low polarised signals the PPAs are 
usually scattered, although in certain instances they follow the RVM tracks, 
which points towards incoherent mixing of X and O modes from a large number of 
sources before the emission detaches from the pulsar medium. Even in the pulsar
profiles formed after averaging several thousand pulses the polarisation 
fraction rarely exceeds 50\% in the low energetic population, suggesting the 
dominant role of mode mixing in the magnetosphere that contribute to 
de-polarisation in these sources. 

The polarisation behaviour shows a clear transition in more energetic pulsars, 
$\dot{E} > 5\times10^{33}$ ergs~s$^{-1}$, where the polarisation fraction 
increases significantly. The change in polarisation properties with pulsar 
energetics has been primarily reported at higher frequencies above 1 GHz 
\citep[see e.g.][]{1998A&A...336..209V,2001AJ....122.2001C,2018MNRAS.474.4629J,
2019MNRAS.489.1543O,2023MNRAS.520.4582P}. A sharp change in the linear 
polarisation level was reported by \cite{2008MNRAS.391.1210W} at $\dot{E} = 
10^{34.5\pm0.08}$ erg~s$^{-1}$, with less than 50\% fractional polarisation 
in the average profiles below this value and more than 50\% above it. 
Subsequent studies have found this transition to be more gradual, for e.g. 
\citet{2023RAA....23j4002W} reports a large scatter in the fractional 
polarisation above $\dot{E} > 5\times10^{33}$ the fractional polarisation. At 
lower frequencies below 1 GHz the above trend has been less clear, with some 
studies showing no clear evolution of polarisation behaviour 
\citep{1998A&A...336..209V,2009ApJS..181..557H}, while others have found 
evidence of higher polarisation fraction in more energetic pulsars 
\citep{2016ApJ...833...28M,2019MNRAS.489.1543O}. The high level of fractional 
linear polarisation suggests that in the emergent emission the 100\% linearly 
polarised X or O modes are able to escapes the pulsar magnetosphere without 
significant mode mixing. Although other pulsars in this regime showing low 
polarisation levels are also expected to have OPM as well as internal mode 
mixing.

In addition to the polarisation behaviour there are other physical properties
that also seem to evolve with pulsar energetics. It has been observed that 
pulsars with higher $\dot{E}$ primarily have relatively simpler profile shapes
with either a single component or merged features \citep{2006MNRAS.368.1856J}. 
In contrast the pulsars with several distinct components in their profiles tend
to have lower $\dot{E}$ values \citep{2023MNRAS.520.4961O}. The nature of
the PPA variation across the average profile has also been linked to the 
complexity of the profile shapes, with simpler single component profiles 
generally having PPA along well defined RVM tracks \citep{2006MNRAS.368.1856J}.
Recent studies have also found clear dependence between single pulse features 
like subpulse drifting and periodic amplitude modulation with $\dot{E}$ 
\citep{2016ApJ...833...29B,2019MNRAS.482.3757B,2020ApJ...889..133B}.
The drifting behaviour is limited to pulsars below $\dot{E} < 10^{33}$ 
erg~s$^{-1}$, while the periodic amplitude modulation is seen over a much wider
range. Several highly energetic pulsars are also prominent sources of 
$\gamma$-rays \citep{Smith_2023}. 

 The pulsar radio emission appear in bursts that usually occupy less than 
10\% of the rotational period. The individual pulses are made up of distinct 
structures known as subpulses and the average profile typically comprises of 
more than one component. To explain the observed emission properties 
\citet{1975ApJ...196...51R} suggested the the presence of isolated sparking 
discharges in a inner vacuum gap above the polar caps. The sparks are expected 
to undergo variable $\bm{E}$x$\bm{B}$ drift in the gap region and this gives 
rise to the phenomenon of subpulse drifting that has been extensively observed 
in the single pulse sequences of many pulsars \citep{2006A&A...445..243W,
2016ApJ...833...29B}. However, later studies have revealed several limitations 
of spark formation in a vacuum gap, particularly highlighting that isolated 
sparks with discharge free zones in-between are highly unstable \citep{CR80,
2024ApJ...974L..32C}. This issue has been addressed in the partially screened 
gap (PSG) model that considers a non-dipolar polar cap where the back-flowing 
particles produced during sparking heats the surface to critical temperatures 
($T_i \sim 10^6$ K) for free ionic discharge \citep{2003A&A...407..315G}. A 
steady outflow of positively charged ions from the heated surface is setup to 
screen the potential difference in the gap region. When the surface temperature
goes below $T_i$ the sparking discharge is ignited to thermostatically regulate
the actual surface temperature and the ionic free flow. In the PSG the sparks 
are formed in tightly packed configurations without any discharge free zone 
between them. The sparks are arranged along concentric rings, with the 
outermost ring closely bordering the polar cap boundary, along with a central 
spark \citep{2020MNRAS.496..465B,2022ApJ...936...35B}. The PSG model has been 
used to explain a number of observational features like subpulse drifting 
\citep{GMZ06,BMM23}, the unique emission from the long period pulsar 
J2144$-$3933 located at the so called `death line' of pulsars 
\citep{2020MNRAS.492.2468M}, the X-ray emission from the surface 
\citep{2019MNRAS.489.4589A,SG20,2020MNRAS.491...80P}, etc. The PSG model 
predicts a variation in the different plasma parameters as a function of 
$\dot{E}$, and has potential to further study the polarisation variations with
pulsar energetics.

Investigating the physical origin of polarisation behaviour in pulsars and 
their variation within the population, particularly when exploring the role of
the PSG model, requires measurements of single pulse emission from a 
statistically significant sample over multiple frequencies. Due to the 
sensitivity limitations of most radio telescopes in the past, the single pulse 
polarisation studies have only been possible in a limited number of pulsars, 
primarily in the low $\dot{E}$ range, and there exists a marked shortage of 
such studies from energetic pulsars with $\dot{E} > 10^{33}$ erg~s$^{-1}$. The 
advent of modern telescopes with wide-band receiver systems like the upgraded 
GMRT, MeerKAT, FAST, etc., have made it possible at present to conduct single
pulse studies in more pulsars. In this work we aim to address the shortage of
polarisation measurements, particularly at the sub-GHz frequency range, by 
carrying out detailed observations of energetic pulsars using GMRT at the two 
relatively wide frequency ranges between 300-500 MHz and 550-750 MHz. In the 
subset of pulsars where time samples could be measured with sufficient 
sensitivity, we further explore the single pulse polarisation behaviour to 
identify a mechanism that leads to de-polarisation in the radio emission from 
energetic pulsars.

\section{Observation and Data Analysis}

The GMRT is an interferometric array consisting of 30 antennas, each of 45 
meters diameter, arranged in a Y-shaped configuration with 14 antennas spread 
around a central square kilometer area and the remaining 16 antennas along 
three arms within a circle of 25 km diameter \citep{1991CuSc...60...95S}. The 
telescope receiver system \citep{2017CSci..113..707G} have undergone an upgrade
in recent years and currently operates in four nearly contiguous wide-frequency
bands between 120-250 MHz (Band2), 250-500 MHz (Band-3), 550-850 MHz (Band-4), 
and 1050-1450 MHz (Band-5). The emission from pulsars are usually measured in 
the phased-array mode where the signals from a subset of the antennas in the 
array are co-added, after correcting for phase differences between them, to 
gather high time resolution recordings of the emitted signal.

We selected a sample of pulsars with $\dot{E} > 5\times10^{33}$ erg~s$^{-1}$ 
and $P > 0.05$ s in order to measure their polarisation properties (see Table 
\ref{tab1}). The GMRT can effectively measure sources above declination range
of -45$\degr$ and we also limited the sample to dispersion measures (DM) less
than 250~pc~cm$^{-3}$ primarily to avoid effects of strong scattering at 
meterwavelengths. All pulsars were observed on two separate occasions, the 
first between 13 and 14 of September, 2021 at the lower frequency range of 
300-500 MHz (Band-3), and the second on 21 September, 2021 at the higher 
frequency range of 550-750 MHz (Band-4). On the first session we were able to 
use 18 antennas in the phased-array, comprising of 12 available central square 
antennas and 6 nearby antennas from the three arms. On the second day 13 
central square antennas were operational resulting in 19 antennas being used 
for the observations. The extreme arm antennas are generally not used in 
phased-array setups as their phase response show fluctuations over short 
timescales due to ionospheric changes and thereby require more frequent phase 
corrections, which significantly reduces the available observing time on the 
target sources. The data were recorded in full stokes polarisation filterbank 
mode with the 200 MHz frequency band divided into 2048 spectral channels. Each 
pulsar was observed for approximately 2000 single pulses at each frequency band
with a time resolution of $t_{res} = 327.68\mu$s. 

The Band-3 and Band-4 receiver systems of GMRT are equipped with dual-linear 
polarisation feeds that are converted into left and right hand circular 
polarisation signals using a quadrature hybrid. Data from PSR J0953+0755 and 
PSR J1932+1059 were also recorded during these observations and they served as 
polarisation calibrators. The auto and cross-polarised signals of each time 
sample were suitably calibrated to produce the four Stokes parameters: I, Q, U
and V for each frequency channel \citep[see][for details]{2016ApJ...833...28M}.
Finally, using the known values of the dispersion measure and rotation measure
from the catalogue\footnote{ATNF Pulsar Catalogue \citep{2005AJ....129.1993M} 
available at http://www.atnf.csiro.au/research/pulsar/psrcat.} the systematic 
changes in the polarisation behavior across frequencies were corrected and 
subsequently averaged to produce the polarised single-pulse time series. The
effect of cross-coupling in the antenna feeds has not been addressed during 
this analysis and can result in systematic errors of up to 10\% of the measured
Stokes parameters, which has been included in the error estimates. The final 
data products were folded with topocentric corrected pulsar period and average 
profiles were produced. We were able to measure 16 pulsars from Band-3 and 32 
pulsars from Band-4 with sufficient detection sensitivity, for a total of 35 
unique sources as listed in Table~\ref{tab1}.

\begin{deluxetable}{cccccccccc}
\tablecaption{Physical properties and average Polarization features of observed pulsars}
\label{tab1}
\tabletypesize{\footnotesize}
\tablehead{\colhead{Name} & \colhead{$P$} & \colhead{$\dot{P}$} & \colhead{DM} & \colhead{$\dot{E}$} & \colhead{Frequency} & \colhead{$\%L$} & \colhead{$\%V$} & \colhead{\%$\mid$$V$$\mid$} & \colhead{W$_{5\sigma}$} \\ 
\colhead{} & \colhead{(sec)} & \colhead{(s~s$^{-1}$)} & \colhead{(pc~cm$^{-3}$)} & \colhead{(ergs~s$^{-1}$)} & \colhead{(MHz)} & \colhead{} & \colhead{} & \colhead{} & \colhead{(deg)}}
\startdata
 J0117+5914 &  0.1014 & 5.85$\times$10$^{-15}$ & ~49.4 & 2.2$\times$10$^{35}$ & 550-750 & 14.4$\pm$1.7 & -3.9$\pm$1.7 &  5.1$\pm$1.3 & 25.6$\pm$1.7 \\
          &         &              &         &         & 300-500 & 20.0$\pm$1.5 & -5.0$\pm$1.6 &  6.4$\pm$1.3 & 29.1$\pm$1.7 \\
 J0139+5814 &  0.2724 & 1.07$\times$10$^{-14}$ & ~73.8 & 2.1$\times$10$^{34}$ & 550-750 & 90.2$\pm$0.1 & 13.0$\pm$0.1 & 13.1$\pm$0.1 & 22.5$\pm$0.6 \\
          &         &              &         &         & 300-500 & 84.4$\pm$0.1 & 11.8$\pm$0.2 & 11.9$\pm$0.1 & 21.2$\pm$0.6 \\
 J1637$-$4553\tablenotemark{a} & 0.1187 & 3.19$\times$10$^{-15}$ & 193.2 & 7.5$\times$10$^{34}$ & 550-750 & 76.0$\pm$2.0 & -0.1$\pm$1.1 & 4.8$\pm$0.8 & 24.9$\pm$1.4 \\
          &         &             &        &      & 300-500 & 62.4$\pm$5.1 &  6.2$\pm$4.7 & 11.8$\pm$2.9 & 40.8$\pm$1.5 \\
 J1705$-$3950 &  0.3189 & 6.06$\times$10$^{-14}$ & 207.3 & 7.4$\times$10$^{34}$ & 550-750 & 52.6$\pm$3.9 & -26.0$\pm$4.0 & 27.2$\pm$3.7 & 24.8$\pm$0.5 \\
 J1709$-$4429 &   0.1024  & 9.30$\times$10$^{-14}$ & ~75.6 & 3.4$\times$10$^{36}$ & 550-750 & 85.6$\pm$1.0 & -25.4$\pm$1.1 & 25.6$\pm$1.0 & 78.5$\pm$1.6 \\
          &           &              &        &        & 300-500 & 76.0$\pm$1.4 & -14.8$\pm$1.8 & 17.1$\pm$1.3 & 83.1$\pm$1.7 \\
 J1718$-$3825 &  0.0746 & 1.32$\times$10$^{-14}$ & 247.5 & 1.2$\times$10$^{36}$ & 550-750 & 67.9$\pm$6.0 & -16.9$\pm$9.5 & 26.3$\pm$6.7 & 68.2$\pm$2.3 \\
 J1733$-$3716 & 0.3375  & 1.50$\times$10$^{-14}$ & 153.2 & 1.5$\times$10$^{34}$ & 550-750 & 79.6$\pm$0.7 & -12.4$\pm$0.8 & 13.5$\pm$0.7 & 77.9$\pm$0.5 \\
          &         &             &        &         & 300-500 & 56.5$\pm$2.5 & -14.2$\pm$3.3 & 19.0$\pm$2.4 & 75.1$\pm$0.5 \\
 J1739$-$2903\tablenotemark{a} & 0.3228 & 7.87$\times$10$^{-15}$ & 138.6 & 9.2$\times$10$^{33}$ & 550-750 & 5.0$\pm$0.8 &  5.8$\pm$1.0 &  5.9$\pm$0.9 & 27.8$\pm$0.5 \\
 J1739$-$3023 & 0.1143 & 1.14$\times$10$^{-14}$ & 170.5 & 3.0$\times$10$^{35}$ & 550-750 & 76.5$\pm$3.2 & -5.5$\pm$2.6 &  8.7$\pm$1.9 & 27.9$\pm$1.5 \\
 J1740+1000 & 0.1540 & 2.13$\times$10$^{-14}$ & ~23.9 & 2.3$\times$10$^{35}$ & 550-750 & 93.2$\pm$1.5 & -6.4$\pm$1.7 &  9.1$\pm$1.2 & 53.6$\pm$1.1   \\
 J1740$-$3015 & 0.6068 & 4.66$\times$10$^{-13}$ & 152.0 & 8.2$\times$10$^{34}$ & 550-750 & 61.3$\pm$0.2 & -19.2$\pm$0.2 & 19.7$\pm$0.2 & 32.5$\pm$0.3   \\
 J1757$-$2421 & 0.2341 & 1.30$\times$10$^{-14}$ & 179.5 & 4.0$\times$10$^{34}$ & 550-750 & 19.8$\pm$0.6 &  5.1$\pm$0.6 &  5.5$\pm$0.5 & 57.0$\pm$0.7  \\
          &        &             &         &          & 300-500 & 22.2$\pm$2.2 &  7.3$\pm$2.2 &  9.8$\pm$1.5 & 84.7$\pm$0.7       \\
 J1803$-$2137 & 0.1336 & 1.34$\times$10$^{-13}$ & 234.0 & 2.2$\times$10$^{36}$ & 550-750\tablenotemark{b} & 51.4$\pm$1.4 & 33.1$\pm$1.7 & 33.9$\pm$1.4 & 149.1$\pm$1.3    \\
          &         &             &         &          & 300-500\tablenotemark{b} & 37.8$\pm$1.6 & 29.2$\pm$1.9 & 30.2$\pm$1.5 & 181.8$\pm$1.3         \\
 J1809$-$1917 & 0.0827 & 2.55$\times$10$^{-14}$ & 197.1 & 1.8$\times$10$^{36}$ & 550-750 & 63.5$\pm$6.3 & 40.5$\pm$8.6 & 44.9$\pm$6.6 & 102.9$\pm$2.1   \\
 J1824$-$1945 & 0.1893 & 5.24$\times$10$^{-15}$ & 224.4 & 3.0$\times$10$^{34}$ & 550-750 & 12.4$\pm$0.2 &  1.0$\pm$0.2 &  1.4$\pm$0.1 & 56.2$\pm$0.9  \\
          &        &             &         &          & 300-500 & 21.8$\pm$1.1 &  5.5$\pm$1.2 &  6.4$\pm$0.9 & 94.2$\pm$0.9  \\       
 J1826$-$1334 & 0.1014 & 7.53$\times$10$^{-14}$ & 231.0 & 2.8$\times$10$^{36}$ & 550-750 & 67.7$\pm$6.1 & 36.4$\pm$7.8 & 39.4$\pm$6.3 & 59.4$\pm$1.7  \\
 J1835$-$1020 & 0.3024 & 5.92$\times$10$^{-15}$ & 115.9 & 8.4$\times$10$^{33}$ & 550-750 & 20.9$\pm$0.6 & -6.0$\pm$0.5 &  6.2$\pm$0.5 & 17.9$\pm$0.6  \\
 J1841$-$0345 & 0.2040 & 5.79$\times$10$^{-14}$ & 194.3 & 2.7$\times$10$^{35}$ & 550-750\tablenotemark{b} & 51.6$\pm$3.0 & -10.4$\pm$5.0 & 17.8$\pm$3.1 & 52.7$\pm$0.8  \\
 J1847$-$0402 & 0.5978 & 5.17$\times$10$^{-14}$ & 142.0 & 9.6$\times$10$^{33}$ & 550-750 &  9.8$\pm$0.2 &  4.9$\pm$0.1 &  5.4$\pm$0.1 & 27.4$\pm$0.3\\
 J1856+0113 & 0.2674 & 2.08$\times$10$^{-13}$ & ~96.1 & 4.3$\times$10$^{35}$ & 550-750 & 32.9$\pm$1.2 &  0.3$\pm$1.2 &  4.7$\pm$1.1 &  9.3$\pm$0.6 \\
 J1913+0904 & 0.1632 & 1.76$\times$10$^{-14}$ & ~96.9 & 1.6$\times$10$^{35}$ & 550-750 & 43.0$\pm$3.3 &  7.0$\pm$2.9 &  9.8$\pm$2.1 & 24.6$\pm$1.0  \\
 J1915+1009 & 0.4045 & 1.53$\times$10$^{-14}$ & 241.7 & 9.1$\times$10$^{33}$ & 300-500 & 40.4$\pm$1.1 & 23.6$\pm$0.9 & 23.6$\pm$0.9 & 38.2$\pm$0.4   \\
 J1921+0812 & 0.2106 & 5.36$\times$10$^{-15}$ & ~82.8 & 2.3$\times$10$^{34}$ & 550-750 & 82.4$\pm$4.9 & 10.2$\pm$4.5 & 15.0$\pm$3.5 & 12.3$\pm$0.8   \\
 J1922+1733 & 0.2361 & 1.34$\times$10$^{-14}$ & 234.0 & 4.0$\times$10$^{34}$ & 550-750 & 48.0$\pm$2.6 & -14.6$\pm$2.5 & 16.1$\pm$2.1 & 40.0$\pm$0.7  \\
 J1932+2020 & 0.2682 & 4.22$\times$10$^{-15}$ & 211.1 & 8.6$\times$10$^{33}$ & 550-750 & 10.3$\pm$0.6 &  3.9$\pm$0.3 &  3.9$\pm$0.3 & 30.8$\pm$0.6  \\
          &         &             &        &           & 300-500 &  8.9$\pm$1.5 & -3.6$\pm$2.2 &  4.9$\pm$1.6 & 40.0$\pm$0.6  \\
 J1932+2220 & 0.1444 & 5.76$\times$10$^{-14}$ & 219.2 & 7.5$\times$10$^{35}$ & 300-500 & 76.3$\pm$1.8 & 13.2$\pm$1.4 & 13.2$\pm$1.4 & 18.8$\pm$1.2  \\
 J1935+2025\tablenotemark{a} & 0.0801 & 6.08$\times$10$^{-14}$ & 181.4 & 4.7$\times$10$^{36}$ & 550-750 & 81.4$\pm$5.7 & -3.6$\pm$4.5 &  9.9$\pm$2.7 & 29.5$\pm$2.1 \\
 J1938+2213 & 0.1661 & 4.24$\times$10$^{-14}$ & ~92.8 & 3.7$\times$10$^{35}$ & 550-750 & 39.9$\pm$3.3 & 13.9$\pm$2.1 & 14.2$\pm$1.9 & 24.1$\pm$1.0 \\
 J2002+3217 & 0.6967 & 1.05$\times$10$^{-13}$ & 142.2 & 1.2$\times$10$^{34}$ & 550-750 & 30.0$\pm$0.6 & -3.4$\pm$0.3 &  3.8$\pm$0.2 & 15.2$\pm$0.2  \\
          &         &             &        &           & 300-500 & 25.7$\pm$0.9 & -4.4$\pm$0.9 &  5.5$\pm$0.7 & 20.2$\pm$0.2  \\
 J2006+3102 & 0.1636 & 2.49$\times$10$^{-14}$ & 106.8 & 2.2$\times$10$^{35}$ & 550-750 & 72.7$\pm$3.5 & 12.6$\pm$4.5 & 31.7$\pm$3.4 & 30.3$\pm$1.0  \\
          &         &             &        &           & 300-500 & 88.6$\pm$18.3 &  2.3$\pm$18.3 & 28.8$\pm$12.5 & 24.5$\pm$1.0    \\
 J2013+3845 & 0.2301 & 8.85$\times$10$^{-15}$ & 238.2 & 2.9$\times$10$^{34}$ & 550-750 & 56.0$\pm$0.3 &  1.1$\pm$0.3 &  2.7$\pm$0.2 & 66.2$\pm$0.7  \\
          &         &             &        &           & 300-500 & 55.1$\pm$0.8 &  3.0$\pm$0.8 &  5.0$\pm$0.4 & 80.0$\pm$0.7   \\
 J2043+2740 & 0.0961 & 1.27$\times$10$^{-15}$ & ~21.0 & 5.6$\times$10$^{34}$ & 550-750 & 74.1$\pm$0.6 & -1.3$\pm$0.4 &  1.5$\pm$0.3 & 33.2$\pm$1.8    \\
          &         &             &        &           & 300-500 & 73.9$\pm$0.8 &  4.9$\pm$0.8 &  5.4$\pm$0.6 & 33.2$\pm$1.8   \\
 J2150+5247 & 0.3322 & 1.01$\times$10$^{-14}$ & 148.9 & 1.1$\times$10$^{34}$ & 550-750 & 19.0$\pm$0.4 & -0.6$\pm$0.3 &  5.4$\pm$0.2 & 27.4$\pm$0.5   \\
          &         &             &        &           & 300-500 & 20.0$\pm$0.9 &  4.8$\pm$0.5 &  6.0$\pm$0.3 & 32.3$\pm$0.5    \\
 J2229+6114 & 0.0516 & 7.74$\times$10$^{-14}$ & 205.1 & 2.2$\times$10$^{37}$ & 550-750 & 14.3$\pm$2.7 & -6.3$\pm$3.7 &  7.2$\pm$3.4 & 59.6$\pm$3.5  \\
 J2337+6151 & 0.4953 & 1.93$\times$10$^{-13}$ & ~58.4 & 6.3$\times$10$^{34}$ & 550-750 & 40.8$\pm$1.6 & -0.7$\pm$0.4 &  3.0$\pm$0.3 & 29.1$\pm$0.3   \\
          &        &             &        &           & 300-500 & 31.0$\pm$4.1 & -6.3$\pm$2.3 & 9.9$\pm$1.6 & 24.8$\pm$0.3 \\
\enddata
\tablenotetext{a}{Reporting the main pulse behaviour in a pulsar with interpulse emission.}
\tablenotetext{b}{Three consecutive time samples along the pulse longitude have been averaged to improve detection sensitivity.}
\end{deluxetable}

\section{Results}
\subsection{Average polarisation behaviour of energetic pulsars}\label{avgprop}
The band averaged time series measurements of each pulsar consists of $n_s$
single pulses, identified by the coefficient $i = 1, 2, ... n_s$. Each single 
pulse is divided into $n_l = \textrm{Int}(P/t_{res})$ phase bins across 
the pulsar period such that the longitudinal phase is $\phi_j = 2\pi j/n_l$, 
where $j = 1, 2, ... n_l$. The unique time sample identified by the 
coefficients $(i, j)$ has four distinct measurements of the Stokes parameters 
$I_i(\phi_j)$, $Q_i(\phi_j)$, $U_i(\phi_j)$, $V_i(\phi_j)$ and at each 
longitude, $\phi_j$, the average intensity, $I(\phi_j)$, the average linear 
polarisation, $L(\phi_j)$, the average circular polarisation, $V(\phi_j)$, and 
the mean PPA, $\psi(\phi_j)$, can be obtained as :
\begin{eqnarray}
I(\phi_j) & ~ = & ~~ \cfrac{\sum_{i=1}^{n_s} I_i(\phi_j)}{n_s}, \label{eq:Iavg} \\
L(\phi_j) & ~ = & ~~ \cfrac{\left(\left\{\sum_{i=1}^{n_s} Q_i(\phi_j)\right\}^2 + \left\{\sum_{i=1}^{n_s} U_i(\phi_j)\right\}^2\right)^{1/2}-~\mu_L}{n_s}, \label{eq:Lavg}\\
V(\phi_j) & ~ = & ~~ \cfrac{\sum_{i=1}^{n_s} V_i(\phi_j)}{n_s}, \label{eq:Vavg} \\
\psi(\phi_j) ~ & = & ~~ \cfrac{1}{2}~\tan^{-1}\left(\cfrac{\sum_{i=1}^{n_s} U_i(\phi_j)}{\sum_{i=1}^{n_s} Q_i(\phi_j)}\right), \label{eq:PPAavg}
\end{eqnarray}
where $\mu_L$ is the mean baseline level of the uncorrected average linear 
polarisation profile. The emission window of the pulsed emission is determined 
by considering the baseline noise levels in the average profile and identifying
the region which has significant emission above the baseline, i.e. more than 5 
times the standard deviation of noise. The pulse phase corresponding to leading
edge of the emission window is termed as $\phi_l$ and trailing side as 
$\phi_t$, and the width is estimated as $W_{5\sigma} = \phi_t - \phi_l$. Adding
the emission features within this window gives estimates of the average 
polarisation features like fractional linear polarisation $L/I = \sum_j 
L(\phi_j)/I(\phi_j)$, fractional circular polarisation $V/I = \sum_j 
V(\phi_j)/I(\phi_j)$ and fractional absolute circular polarisation 
$\mid$$V$$\mid/I = \sum_j \mid$$V(\phi_j)$$\mid/I(\phi_j)$.

\begin{figure}
\begin{center}
\includegraphics[scale=0.4]{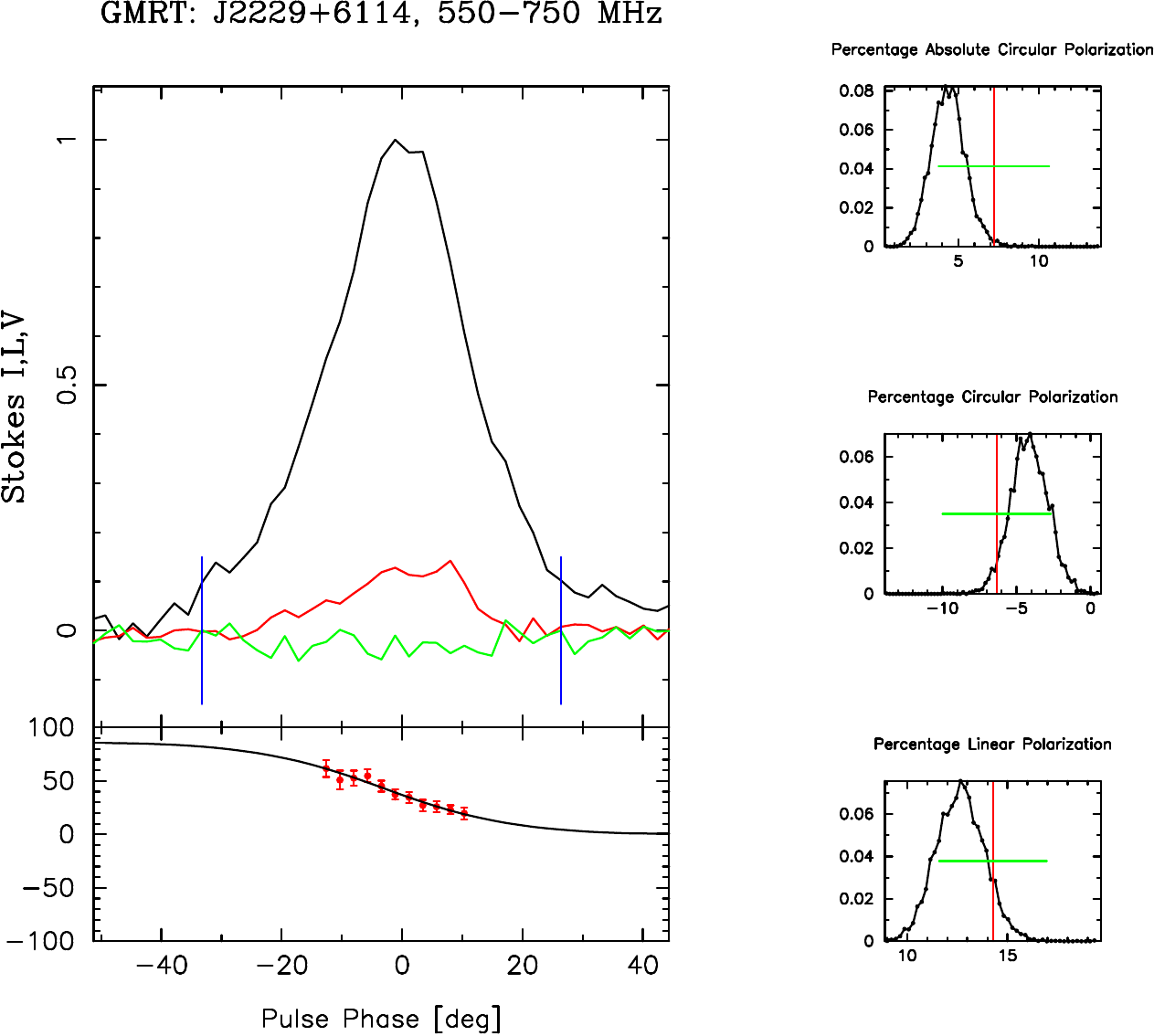}\\
\caption{An example of the average polarisation features in pulsars, from PSR 
J2229+6114 observed at the frequency range 550-750 MHz. The time-averaged 
polarisation profile corresponding to $I$, $L$ and $V$, and polarisation 
position angle variation is shown in the left window. The right windows show 
the distribution of \%$L$, \%$V$ and \% $|V|$ due to profile baseline noise, 
along with their median (red vertical line) and standard deviation (green 
horizontal line). 
}
\label{fig:avgpol}
\end{center}
\end{figure}

Fig.~\ref{fig:avgpol} shows an example of the average profile and polarisation
features estimated in eq.(\ref{eq:Iavg}) -- (\ref{eq:PPAavg}), across the 
longitude range within the pulsed window. Although the measured four Stokes 
parameters have Gaussian distributed errors, the derived quantities have 
distinctly non-Gaussian error characteristics. A scheme to estimate their 
errors has been devised by \cite{2016ApJ...833...28M} using the baseline noise 
variations in the four Stokes parameters. The right panels in the figure shows 
the estimated distributions of the percentage linear, circular and absolute 
circular polarisations using this method, where the median values (vertical red 
line) and standard deviations (horizontal green line) are reported as the 
corresponding measurements and errors, respectively, of each quantity in 
Table~\ref{tab1}. The PPA variations in the figure has been further selected to
represent longitudes with significant emission intensity, i.e. exceeding three 
times the standard deviation of baseline noise distribution. An RVM fit to to 
the PPA is also shown in the plot, which is estimated from the inclination 
angle between the rotation axis and magnetic axis, $\alpha$, and the impact 
angle specifying the angular separation between the magnetic axis and the LOS 
during its closest approach, $\beta$ :
\begin{equation}
\psi_{\rm RVM}(\phi_j) = \psi_{\circ} + \tan^{-1} \left(\cfrac{\sin{\alpha}\sin{(\phi_j-\phi_{\circ})}}{\sin{(\alpha + \beta)}\cos{\alpha} - \sin{\alpha}\cos{(\alpha+\beta)}\cos{(\phi_j-\phi_{\circ})}}\right).
\label{req1}
\end{equation}
Here $\psi_{\circ}$ and $\phi_{\circ}$ are the phase offsets of the PPA and
the pulse phase, respectively. Although in principle it is possible to obtain 
estimates of $\alpha, \beta, \phi_{\circ}$ and $\psi_{\circ}$ from the RVM 
fits, in practise the estimated $\alpha$ and $\beta$ are highly correlated and 
cannot be constrained from this exercise \citep{2001ApJ...553..341E,
2004A&A...421..215M}. When the PPA traverse exhibit prominent S-shape curve the
phase offsets, $\phi_{\circ}$ and $\psi_{\circ}$, can be estimated with 
relative accuracy, but these estimates become more uncertain when the shape 
becomes flatter. 

When the effect of pulsar rotation is considered the propagation direction of 
the emergent radio emission from pulsars has an additional shift in the 
perpendicular direction with respect magnetic field line planes due to 
abberation-retardation (A/R) effect \citep{1991ApJ...370..643B}. This implies 
that a phase shift is introduced between the PPA curve and the profile with the
shift being directly proportional to the radio emission height ($h_E$). This 
allows a completely independent estimate of the emission height in pulsars, 
provided suitable reference phases can be attributed to the PPA traverse and 
the emission window to measure the phase shift. The reference point for the 
profile center is obtained from the leading and trailing edges of the profile, 
$\phi_c = (\phi_l + \phi_t)/2$, while the steepest gradient point at 
$\phi_{\circ}$ is the reference point for the PPA \citep{2017JApA...38...52M}. 
The phase shift is obtained as $\Delta\phi = \phi_{\circ} - \phi_c$, which 
gives estimate of the emission height \citep{2008MNRAS.391..859D} : 
\begin{equation}
h_E = \frac{c P \Delta\phi}{8\pi} \approx ~208.2~\left(\frac{P}{\text{s}}\right) \left(\frac{\Delta\phi}{\text{deg}}\right)~\text{km}.
\label{eq:h_AR}
\end{equation}
A positive value of $\Delta \phi$ is necessary to obtain sensible estimates
of the emission height, however there are pulsars where it may not be possible
to determine the phase shifts accurately making height estimations untenable
from these sources \citep[see discussions in][for likely 
reasons]{2004A&A...421..215M,2023ApJ...952..151M}. We were able to find 
suitable RVM fits in 27 average profiles corresponding to 21 pulsars as 
reported in Table~\ref{tab2}. The table shows the different phase estimates as 
well as the measured emission height, wherever relevant. A brief description of
the average profile features of each pulsar is also provided in appendix 
\ref{sec:app}.

\begin{figure}
\begin{center}
\includegraphics[scale=0.4]{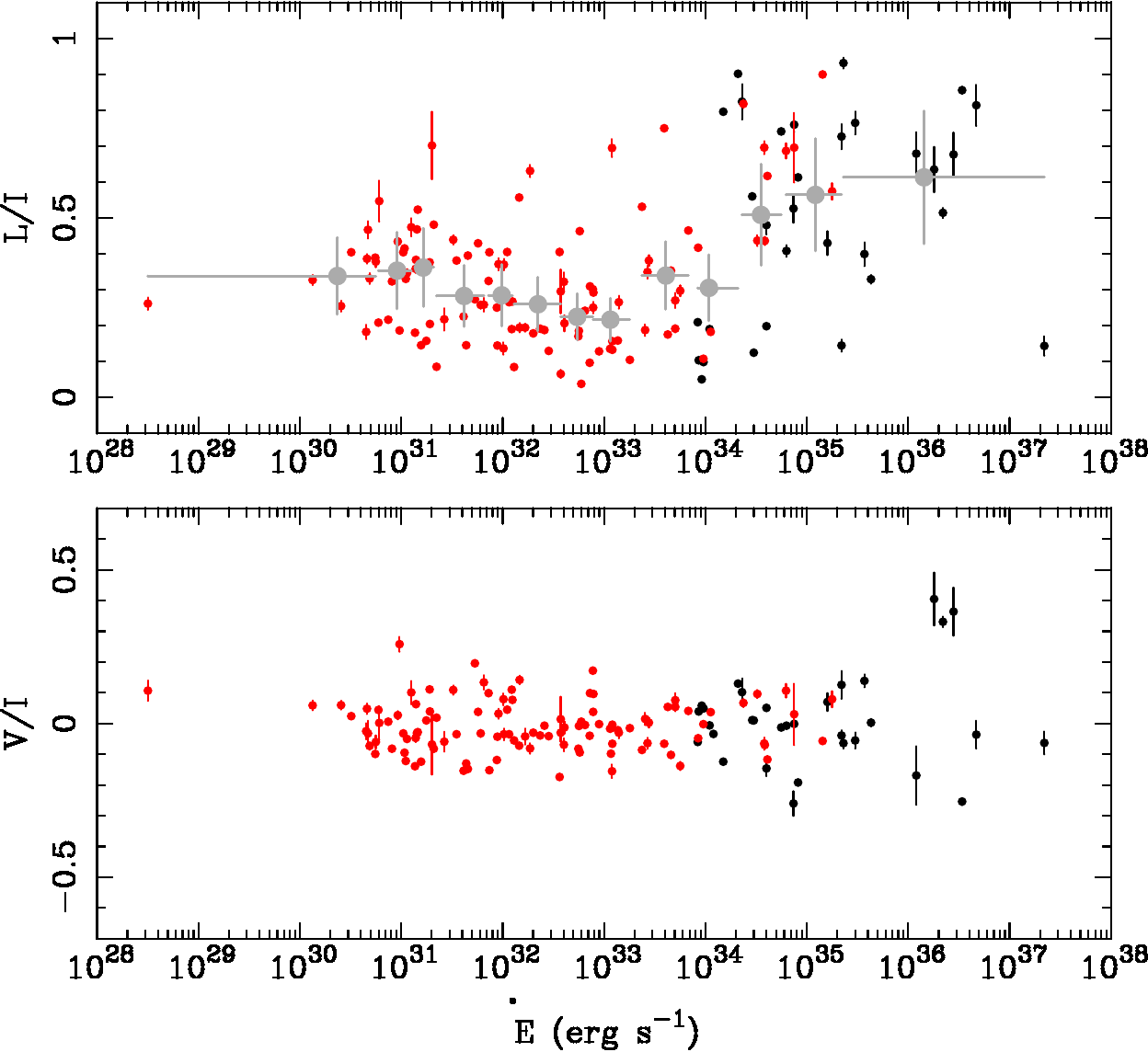}
\caption{The top panel shows the variation of the fractional linear 
polarisation in pulsars as a function of $\dot{E}$. The plot includes archival
measurements at 610 MHz (red points) as well as the more energetic pulsars 
measured at 550-750 MHz (black points). The grey error bars in the top panel
represent the median linear polarisation levels within specific intervals of
$\dot{E}$. The bottom panel shows the equivalent measurements for the 
fractional circular polarisation from the two observations.}
\label{fig:fracdist}
\end{center}
\end{figure}

The measurements of the average polarisation in the more energetic pulsars
have made it possible to determine the evolution of the polarisation fraction 
with $\dot{E}$ at the sub-GHz frequency range. Previously the polarisation 
behaviour in pulsars has also been studied in the Meterwavelength Single-pulse 
Polarimetric Emission Survey \citep[MSPES,][]{2016ApJ...833...28M} at 325 MHz 
and 610 MHz, primarily for lower energetic sources. We combine the previous 
measurements of 102 pulsars at 610 MHz with the 32 pulsars measured between 
550-750 MHz in this work to find the evolution of the polarisation fraction 
with pulsar energetics. Fig.~\ref{fig:fracdist} shows the distribution of linear 
(top panel) and circular (bottom panel) polarisation fraction with $\dot{E}$, 
where it is evident that above $\dot{E} > 10^{34}$ erg~s$^{-1}$ the fractional
linear polarisation exceeds 50\%, similar to the high frequency behaviour. 
However, there is significant spread in the polarisation levels with many 
energetic pulsars having lower polarisation fractions as well, for e.g. PSR 
J2229+6114 despite being the most energetic pulsar in the list has a relatively 
low fractional linear polarisation of $\sim$14\%. To better highlight the 
change in the polarisation fraction with pulsar energetics, we have divided the
$\dot{E}$ range into regular intervals and estimated the median level of 
polarisation in each interval, to reduce the fluctuation. The median 
polarisation levels are within 20\% -- 40\% below $10^{34}$ erg~s$^{-1}$ and 
are between 50\% -- 60\% levels above this range. Although a few pulsars in the
high $\dot{E}$ range show significant levels of circular polarisation, unlike 
the linear polarisation behaviour no clear trend is visible for the circular
polarisation. 

\renewcommand{\arraystretch}{1.4}
\begin{deluxetable}{ccccccccc}
\tablecaption{Average Profile Properties and Radio Emission Height}
\label{tab2}
\tablehead{\colhead{Jname} & \colhead{$P$} & \colhead{Frequency} & \colhead{$\phi_{l}$} & \colhead{$\phi_{t}$} & \colhead{$\phi_c$} & \colhead{$\phi_{\circ}$} & \colhead{$\Delta\phi$} & \colhead{$h_E$} \\
\colhead{} & \colhead{(s)}  & \colhead{(MHz)} & \colhead{(deg)} & \colhead{(deg)} & \colhead{(deg)} & \colhead{(deg)} & \colhead{(deg)} & \colhead{(km)} }
\startdata
  J0117+5914  & 0.1014 & 550-750 & -15.7$\pm$1.2 & 9.9$\pm$1.1 & -2.9$\pm$1.1 & -9.0$^{\ 7.4}_{\ 2.0}$   & ---  & --- \\
   &   & 300-500 & -15.7$\pm$1.2 & 13.3$\pm$1.1 & -1.3$\pm$1.1 & 4.5$^{\ 19.5}_{\ 15.5}$ & 5.8$^{\ 20.6}_{\ 16.6}$ & 122$^{\ 435}_{\ 350}$ \\
  J0139+5814  & 0.2724 & 550-750 & -13.2$\pm$0.4 & 9.3$\pm$0.4 & -1.95$\pm$0.4 & 3.5$^{\ 0.6}_{\ 0.7}$ & 5.5$^{\ 1.0}_{\ 1.1}$ & 309$^{\ 57}_{\ 62}$ \\
   &   & 300-500 & -9.3$\pm$0.4 & 11.9$\pm$0.4 & 1.3$\pm$0.4 & 3.3$^{\ 0.8}_{\ 2.0}$ & 2.0$^{\ 1.2}_{\ 2.4}$ & 113$^{\ ~68}_{\ 136}$ \\
 J1709$-$4428 & 0.1024 & 550-750 & -43.8$\pm$1.1 & 34.6$\pm$1.1 & -4.6$\pm$1.1 & 4.6$^{\ 2.8}_{\ 1.4}$ & 9.2$^{\ 3.9}_{\ 2.5}$ & 196$^{\ 83}_{\ 53}$ \\
   &   & 300-500 & -38.0$\pm$1.1 & 45.0$\pm$1.1 & 3.5$\pm$1.1 & -5.3$^{\ 0.1}_{\ 1.1}$ & --- & --- \\
 J1718$-$3825 & 0.0746 & 550-750 & -18.2$\pm$1.6 & 49.9$\pm$1.6 & 15.9$\pm$1.6 & 19.5$^{\ 3.7}_{\ 0.8}$ & 3.6$^{\ 5.3}_{\ 2.4}$  & 57$^{\ 82}_{\ 37}$ \\
 J1733$-$3716 & 0.3375 & 550-750 & -18.1$\pm$0.3 & 59.7$\pm$0.3 & 20.8$\pm$0.3 & 28.7$^{\ 1.6}_{\ 2.0}$ & 7.9$^{\ 1.9}_{\ 2.3}$ & 555$^{\ 134}_{\ 162}$ \\
 J1739$-$3023 & 0.1143 & 550-750 & -11.8$\pm$1.0 & 15.9$\pm$1.0 & 2.1$\pm$1.0 & -16.8$^{\ 93}_{\ 26}$ & ---  & --- \\
  J1740+1000  & 0.1540 & 550-750 & -26.8$\pm$0.8 & 26.8$\pm$0.7 & 0.0$\pm$0.8 & 10.7$^{\ 25}_{\ ~5}$ & 10.7$^{\ 25.8}_{\ ~5.8}$ & 343$^{\ 827}_{\ 186}$ \\
 J1803$-$2137 & 0.1336 & 550-750 & -112.9$\pm$0.9 & 36.1$\pm$0.8 & -38.4$\pm$0.9 & -33$^{\ 3.2}_{\ 4.0}$ & 5.4$^{\ 4.1}_{\ 4.9}$ & 150$^{\ 114}_{\ 136}$ \\
 J1809$-$1917 & 0.0827 & 550-750 & -17.1$\pm$1.4 & 30.0$\pm$1.4 & 6.5$\pm$1.4 & -29.1$^{\ 7.2}_{\ 7.7}$ & ---  & --- \\
 J1835$-$1020 & 0.3024 & 550-750 & -7.2$\pm$0.3 & 10.7$\pm$0.3 & 1.8$\pm$0.3 & 0.4$^{\ 0.6}_{\ 1.1}$ & --- & --- \\
  J1913+0904  & 0.1632 & 550-750 & -8.6$\pm$0.7 & 15.9$\pm$0.7 & 3.7$\pm$0.7 & 7.5$^{\ 16}_{\ ~5}$ & 3.8$^{\ 16.7}_{\ ~5.7}$ & 129$^{\ 567}_{\ 194}$ \\
  J1921+0812  & 0.2106 & 550-750 & -10.6$\pm$0.5 & 1.6$\pm$0.5 & -4.5$\pm$0.5 & -8.2$^{\ 7}_{\ 5}$ & --- & --- \\
  J1922+1733  & 0.2361 & 550-750 & -21.0$\pm$0.5 & 19.0$\pm$0.5 & -1.0$\pm$0.5 & 16.1$^{\ 14}_{\ 1.3}$ & 17.1$^{\ 14.5}_{\ ~1.8}$ & 841$^{\ 713}_{\ ~88}$ \\
  J1932+2220  & 0.1444 & 300-500 & -5.3$\pm$0.8 & 13.4$\pm$0.8 & 4.1$\pm$0.8 & -15$^{\ 78}_{\ 35}$ & ---  & --- \\
  J1935+2025  & 0.0801 & 550-750 & -17.7$\pm$1.4 & 11.8$\pm$1.4 & -3.0$\pm$1.4 & -5.0$^{\ 19.8}_{\ 18.0}$ & ---  & --- \\
  J2006+3102  & 0.1636 & 550-750 & -7.5$\pm$0.7 & 22.7$\pm$0.7 & 7.6$\pm$0.7 & 15.9$^{\ 2.5}_{\ 5.0}$ & 8.3$^{\ 3.2}_{\ 5.7}$ & 283$^{\ 109}_{\ 194}$ \\
  J2013+3845  & 0.2301 & 550-750 & -33.3$\pm$0.5 & 32.8$\pm$0.5 & -0.3$\pm$0.5 & 33.8$^{\ 15.8}_{\ ~8.4}$ & 34.1$^{\ 16.3}_{\ ~8.9}$ & 1633$^{\ 781}_{\ 426}$ \\
   &   & 300-500 & -30.2$\pm$0.5 & 49.7$\pm$0.5 & 9.8$\pm$0.5 & 70.9$^{\ 69}_{\ 90}$ & 61.1$^{\ 69.5}_{\ 90.5}$  & 2927$^{\ 3330}_{\ 4336}$ \\
  J2043+2740  & 0.0961 & 550-750 & -15.3$\pm$1.2 & 17.8$\pm$1.2 & 1.3$\pm$1.2 & 11.3$^{\ 1.2}_{\ 3.2}$ & 10.0$^{\ 2.4}_{\ 4.4}$ & 200$^{\ 48}_{\ 88}$ \\
   &   & 300-500 & -16.5$\pm$1.2 & 16.5$\pm$1.2 & 0.0$\pm$1.2 & 10.0$^{\ 0.0}_{\ 4.5}$ & 10.0$^{\ 1.2}_{\ 5.7}$ & 200$^{\ ~24}_{\ 114}$ \\
  J2150+5247  & 0.3322 & 550-750 & -9.0$\pm$0.3 & 18.3$\pm$0.3 & 4.7$\pm$0.3 & 8.8$^{\ 1.0}_{\ 0.5}$ & 4.1$^{\ 1.3}_{\ 0.8}$ & 284$^{\ 90}_{\ 55}$ \\
  J2229+6114  & 0.0516 & 550-750 & -33.2$\pm$2.6 & 26.3$\pm$2.3 & -3.5$\pm$2.5 & -3.0$^{\ ~4.2}_{\ 18.9}$ & 0.5$^{\ ~6.7}_{\ 21.4}$ & 5$^{\ ~72}_{\ 230}$ \\
  J2337+6151  & 0.4953 & 550-750 & -17.9$\pm$0.2 & 11.0$\pm$0.2 & -3.5$\pm$0.2 & 0.3$^{\ 0.8}_{\ 0.7}$ & 3.8$^{\ 1.0}_{\ 0.9}$ & 391$^{\ 103}_{\ ~93}$ \\
   &   & 300-500 & -15.1$\pm$0.2 & 9.6$\pm$0.2 & -2.8$\pm$0.2 & 0.4$^{\ 1.8}_{\ 1.5}$ & 3.2$^{\ 2.0}_{\ 1.7}$ & 330$^{\ 206}_{\ 175}$ \\
\enddata
\end{deluxetable}

\subsection{Single pulse polarisation behaviour of energetic pulsars}
\label{splin}

The single pulse studies are possible if the individual time samples, with 
temporal resolution $t_{res}$, are detected with sufficient sensitivity, i.e. 
above three times the standard deviation of baseline noise levels. In addition,
to ensure proper statistical analysis we have restricted our studies to sources
with at least 100 significantly polarised time samples during the observing 
duration. This left us with 13 pulsars that are further divided into two 
groups: class $C_1$ comprising of 5 pulsars where the average fractional linear
polarisation is more that 70\%, and class $C_2$ with 8 pulsars having average 
linear polarisation fraction below this level.\\

\begin{figure}
\gridline{\fig{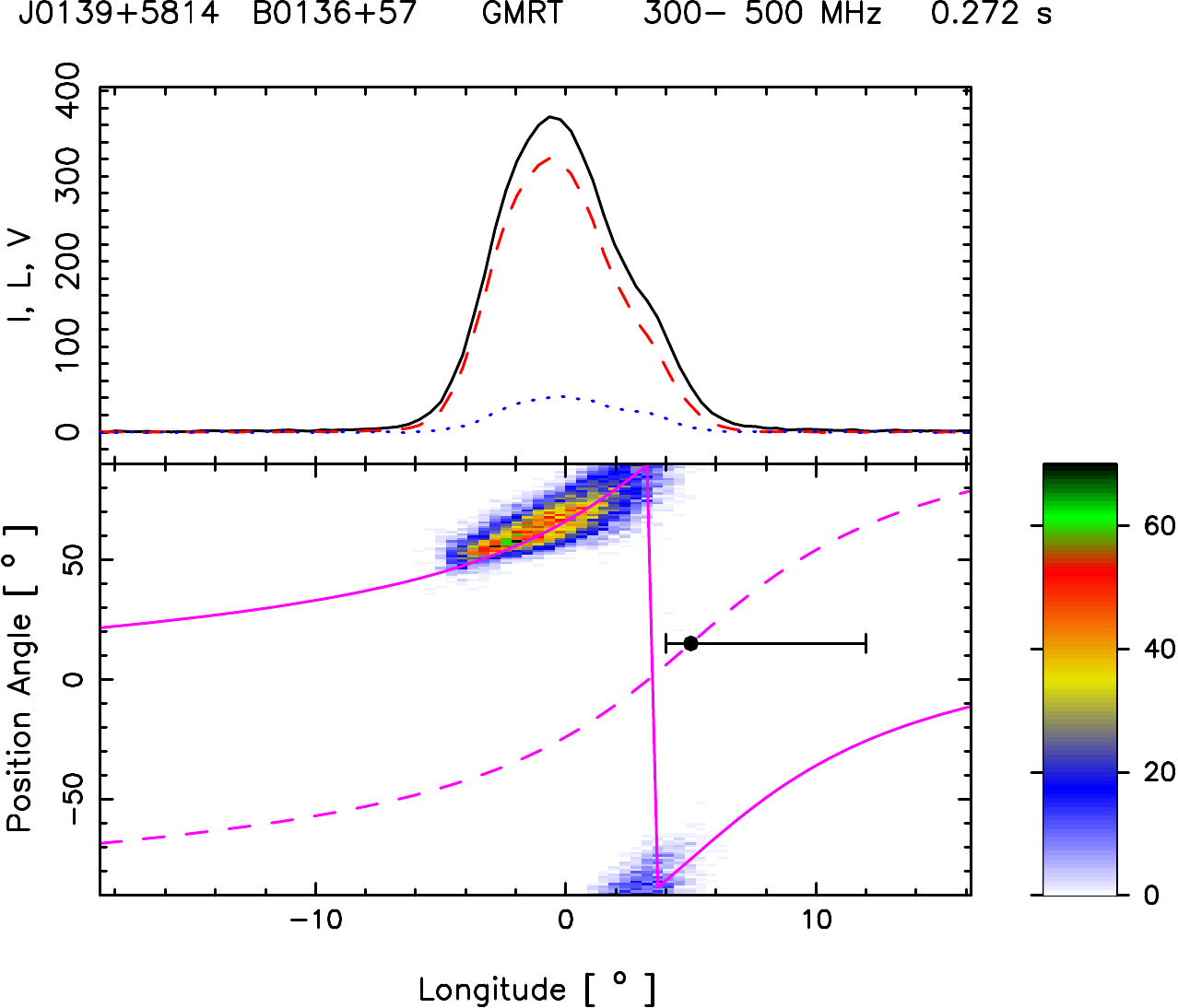}{0.3\textwidth}{(a)}
          \fig{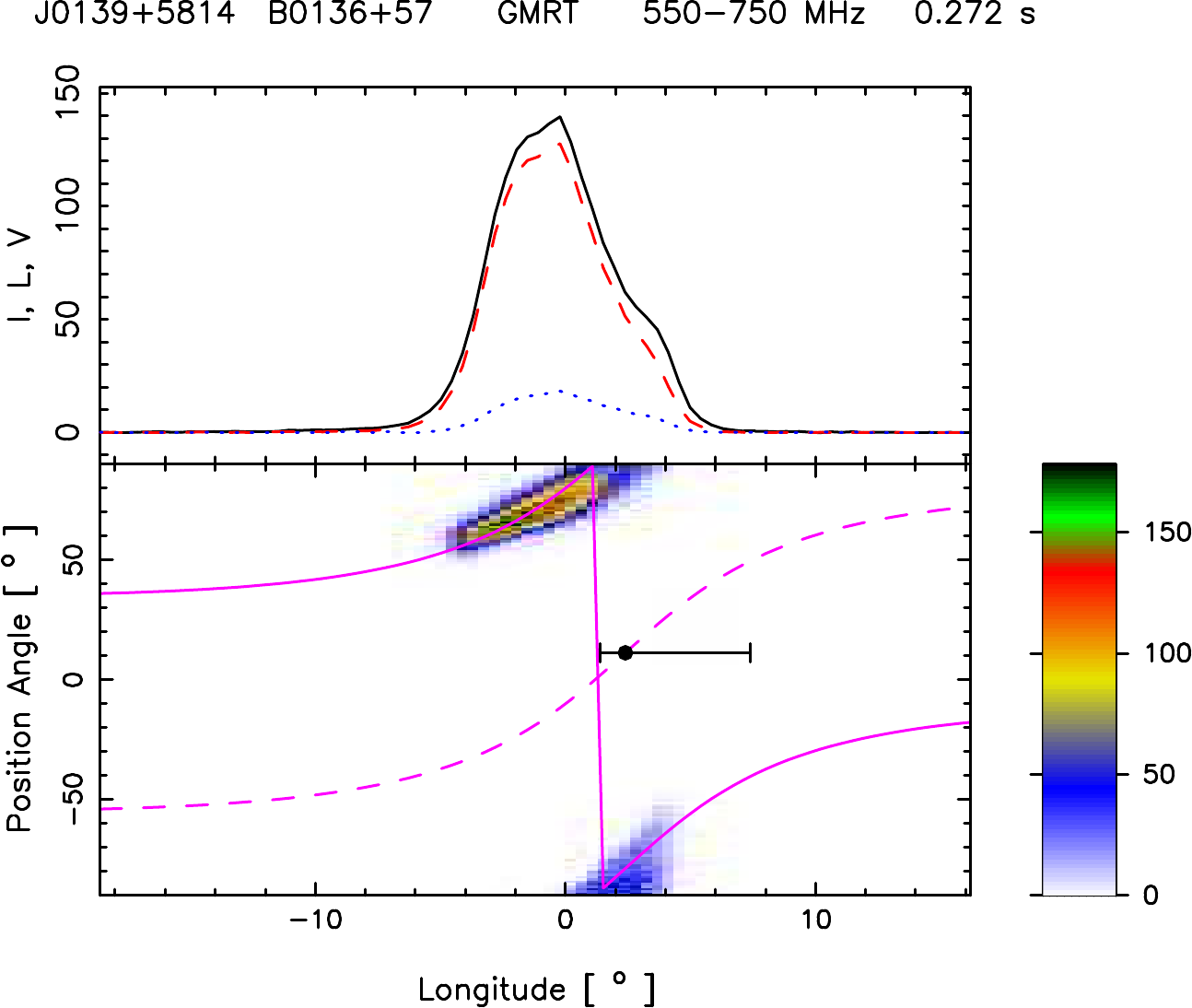}{0.3\textwidth}{(b)}
          \fig{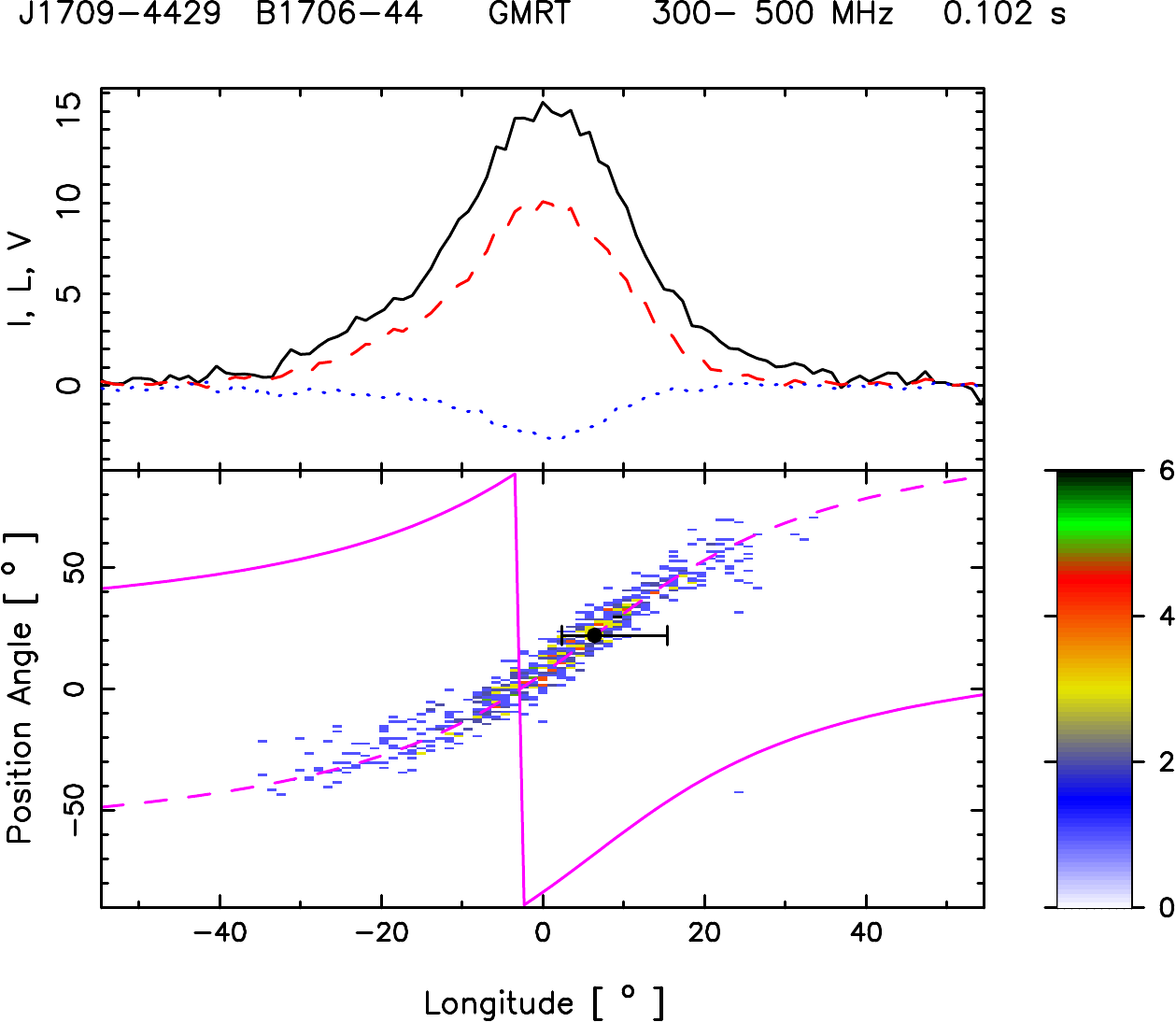}{0.3\textwidth}{(c)}}
\gridline{\fig{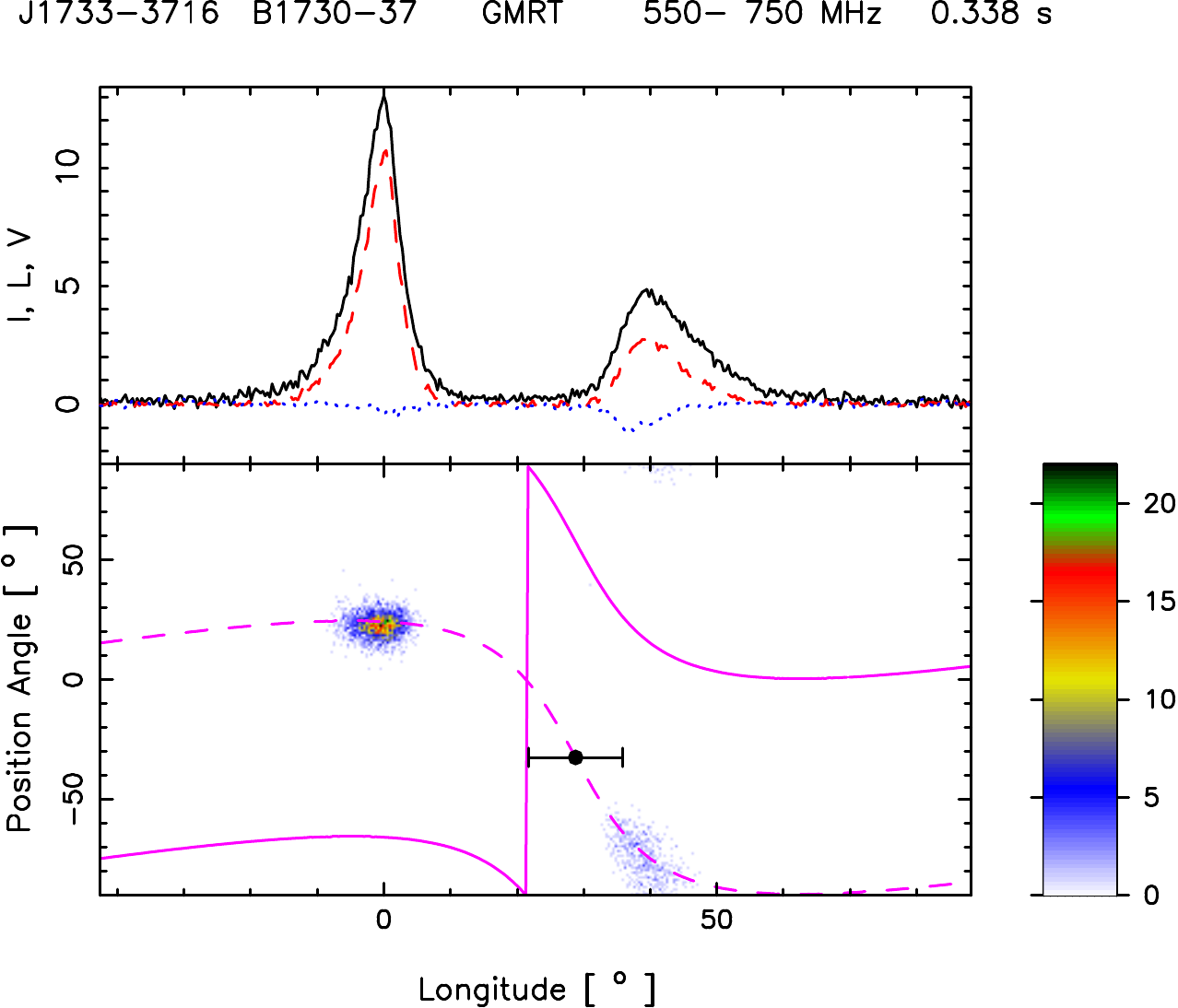}{0.3\textwidth}{(d)}
          \fig{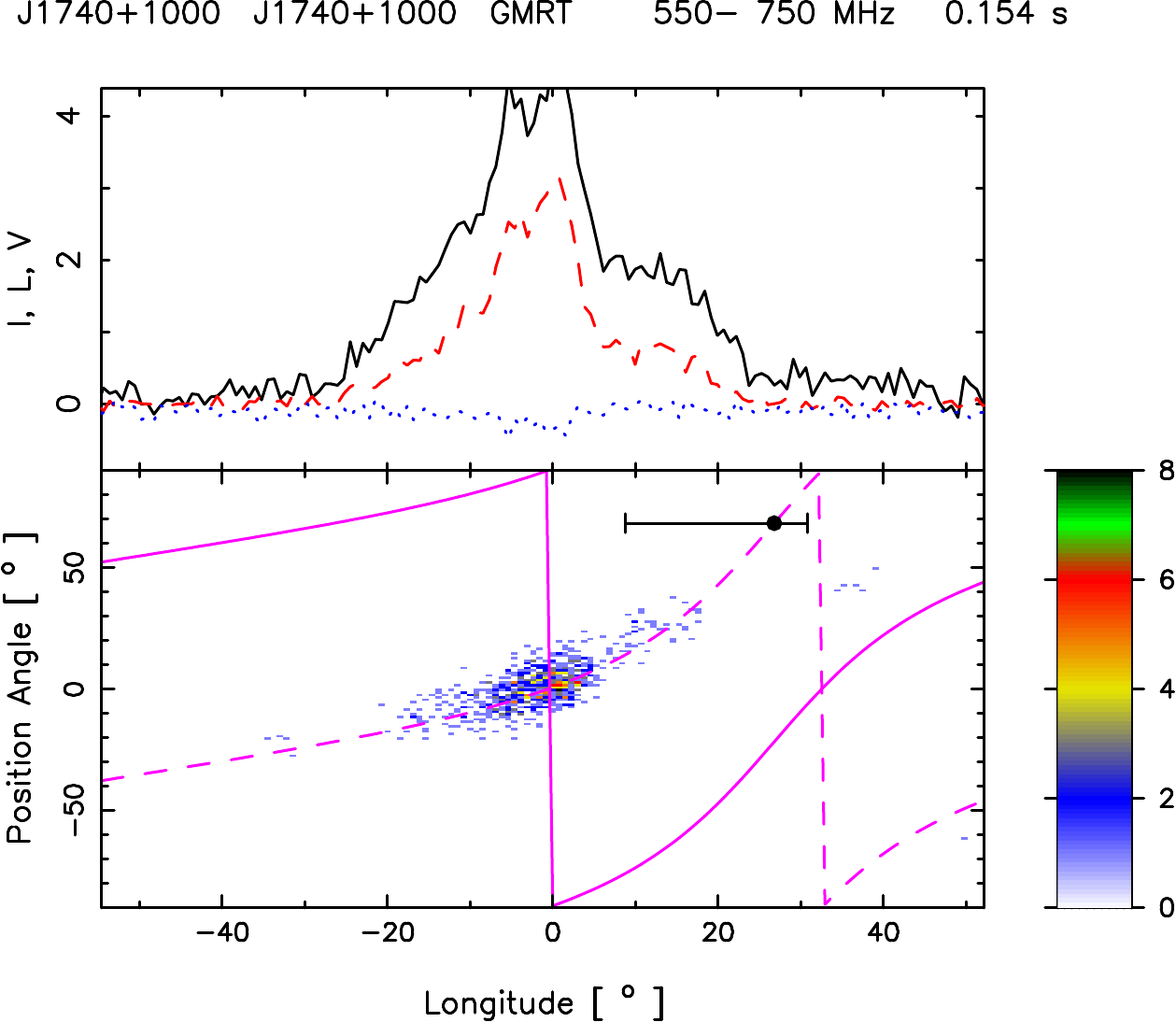}{0.3\textwidth}{(e)}
          \fig{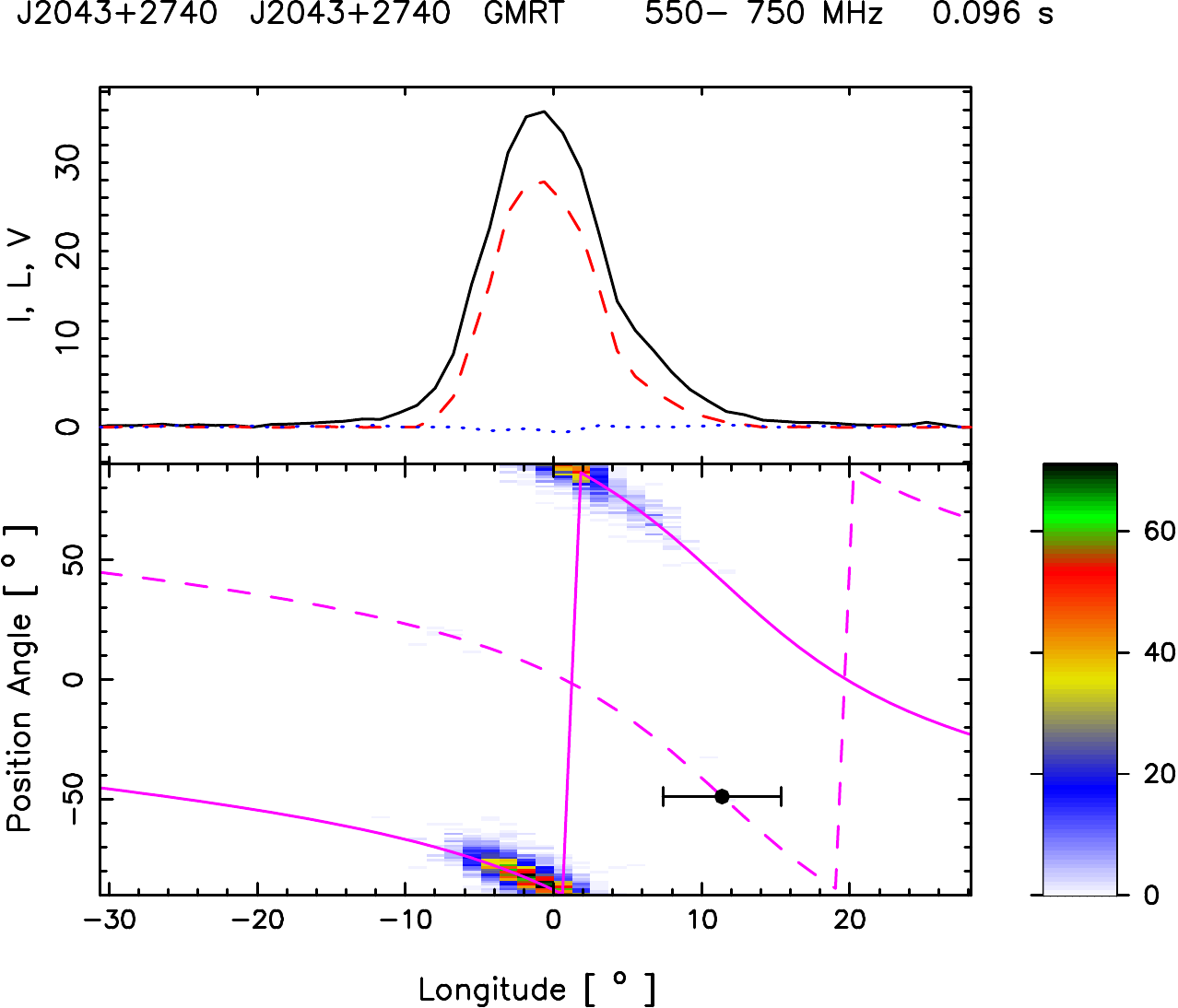}{0.3\textwidth}{(f)}}
\caption{The single pulse PPA distribution of pulsars belonging to the $C_1$ 
class with average fractional linear polarisation larger than 70\%. In each 
plot the top panel shows the average pulse profile where black, red and blue 
curves correspond to total intensity, linear and circular polarisation 
respectively. The bottom panel shows the PPA distribution obtained from the
individual time samples in the single pulses, where the number of points at any
location within the window is represented by the colour bar next to the 
profile. The RVM fits to the orthogonal PPA modes are also shown in the plot as
solid and dashed magenta lines.}
\label{fig:PPAhistC1}
\end{figure}

\begin{figure}
\gridline{\fig{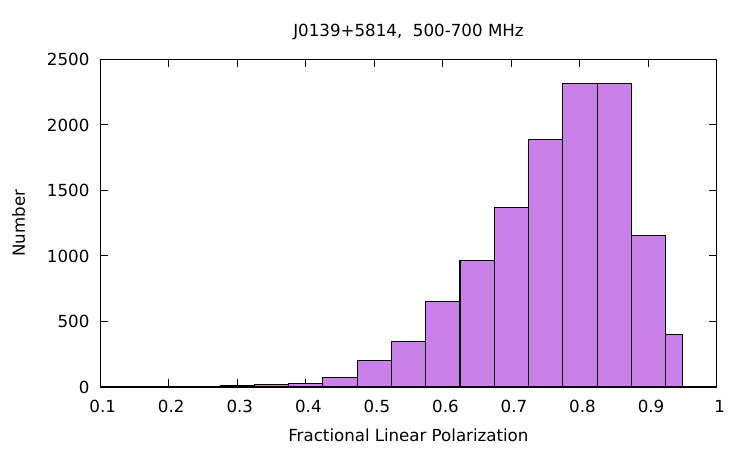}{0.33\textwidth}{(a)}
          \fig{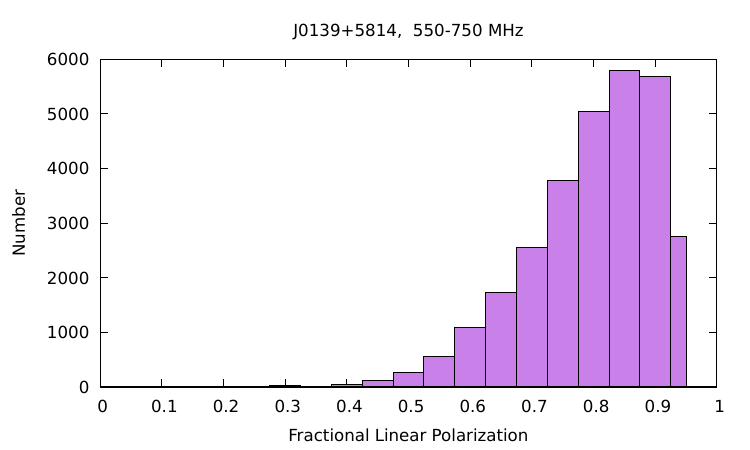}{0.33\textwidth}{(b)}
          \fig{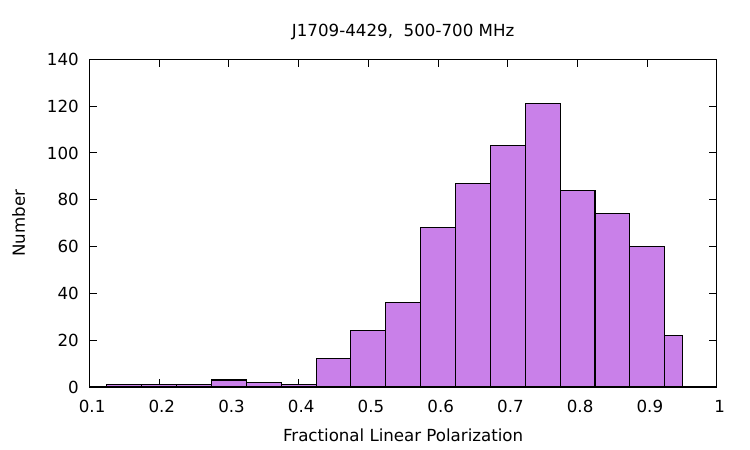}{0.33\textwidth}{(c)}}
\gridline{\fig{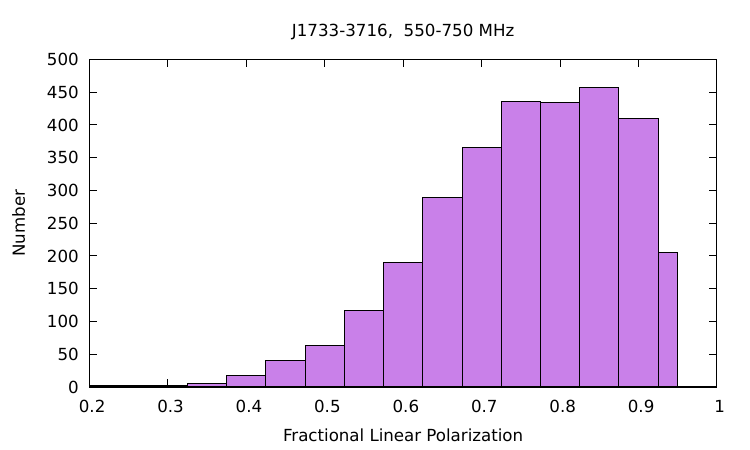}{0.33\textwidth}{(d)}
          \fig{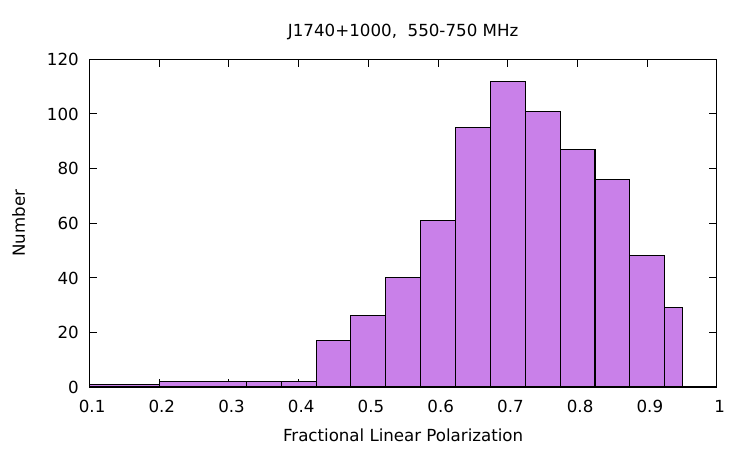}{0.33\textwidth}{(e)}
          \fig{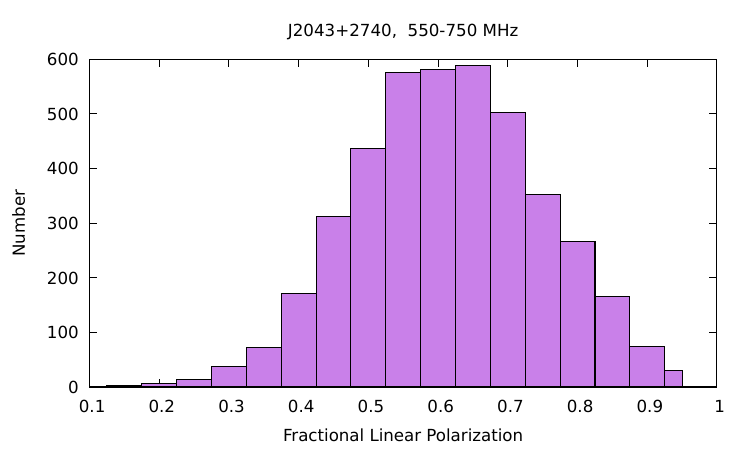}{0.33\textwidth}{(f)}}
\caption{The distribution of the linear polarisation fraction for time samples 
in the $C_1$ class of pulsars with average fractional linear polarisation 
larger than 70\%.}
\label{fig:fracpolC1}
\end{figure}

\begin{figure}
\gridline{\fig{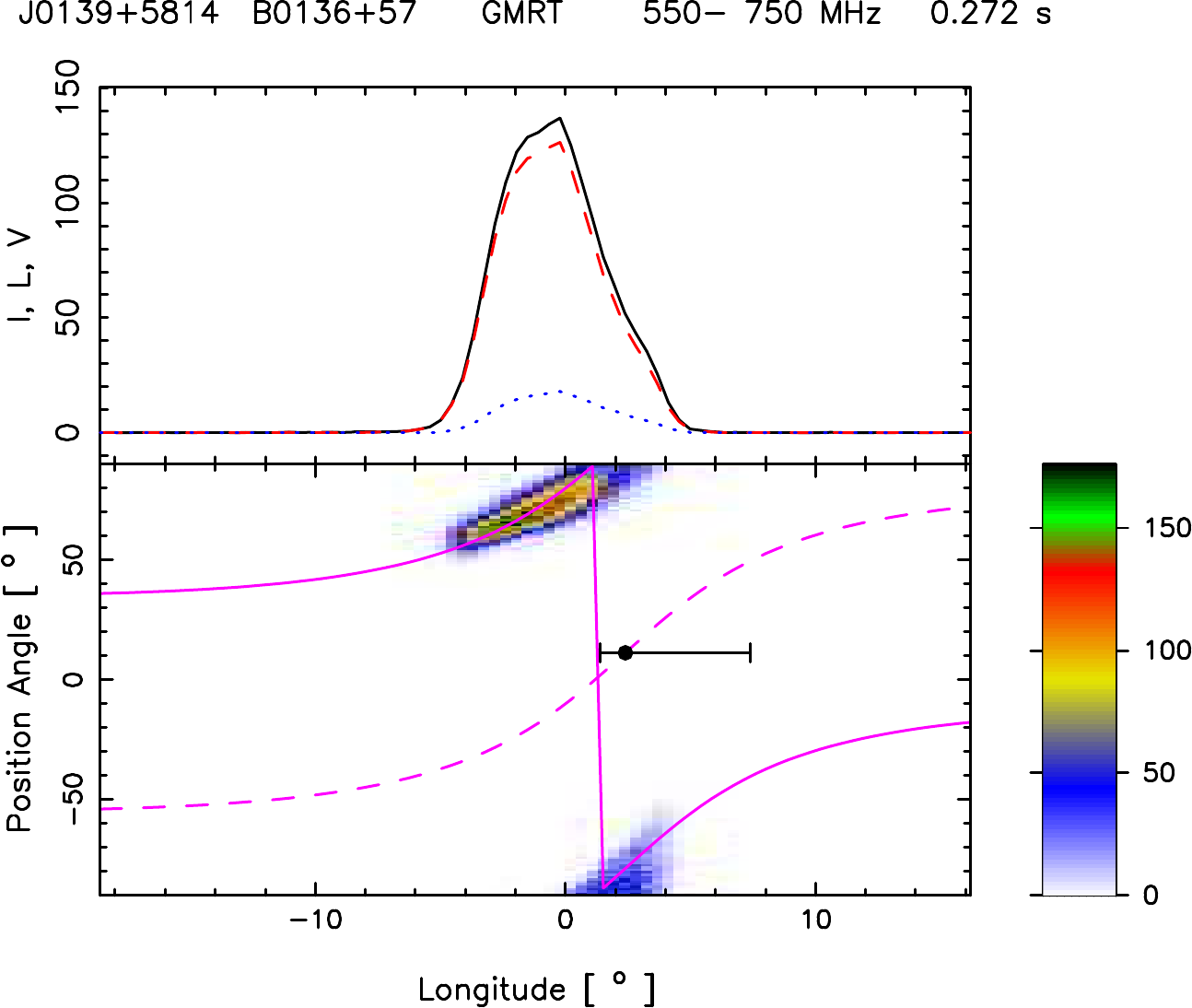}{0.28\textwidth}{(a)}
          \fig{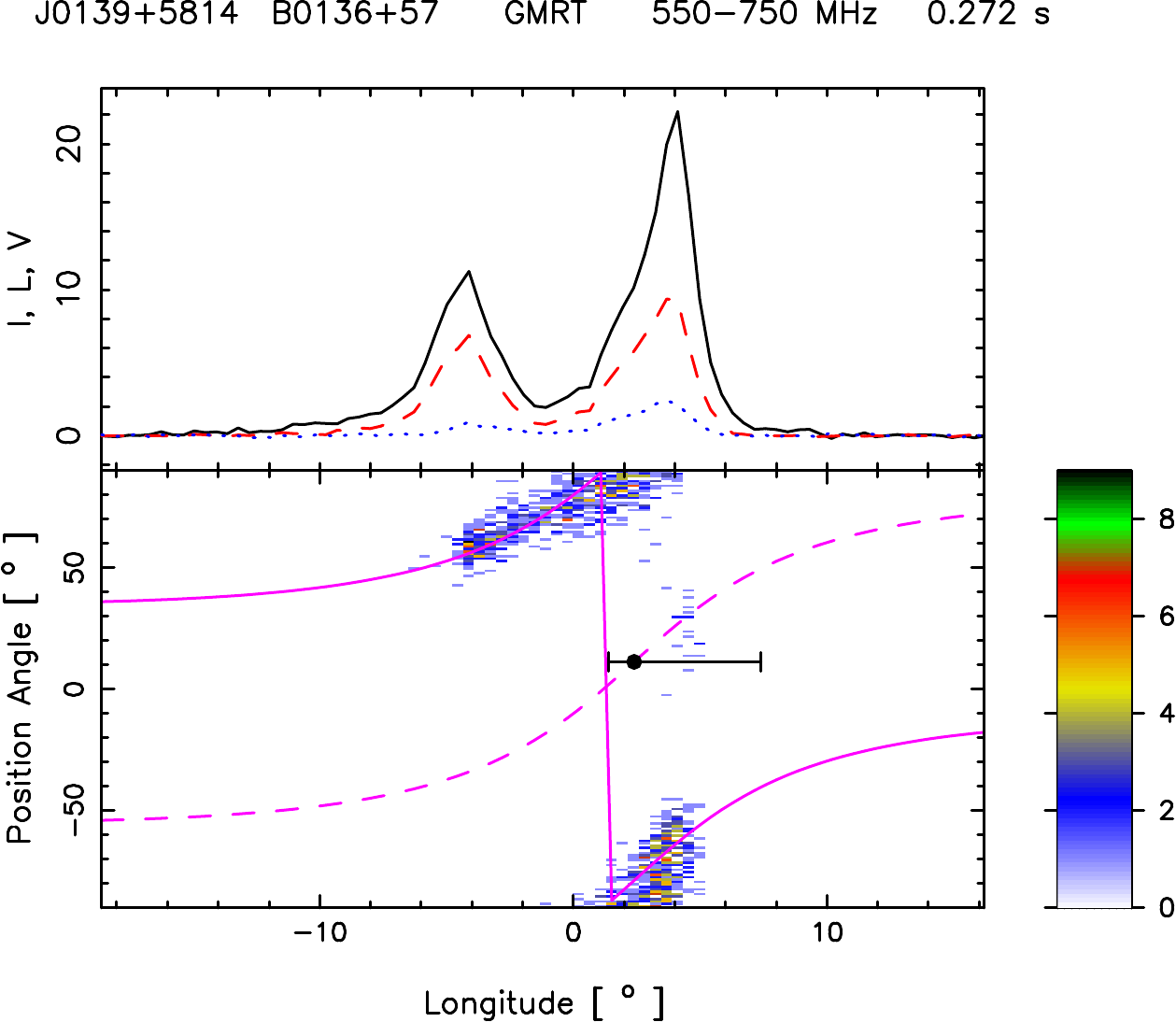}{0.28\textwidth}{(b)}
          \fig{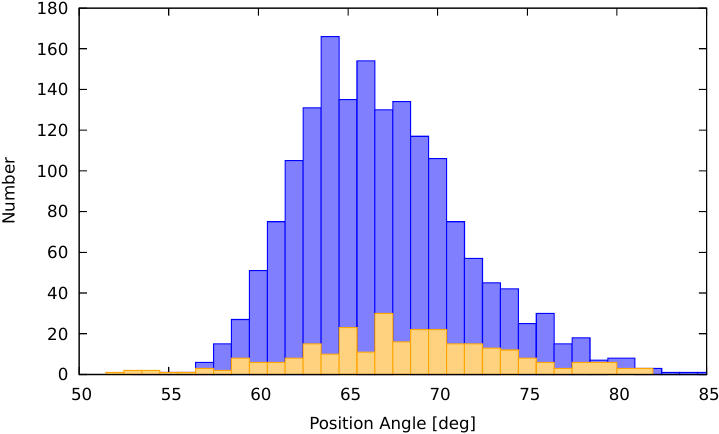}{0.38\textwidth}{(c)}}
\caption{The left panel shows the polarisation properties of PSR B0136+57 at 
the frequency range 550-750 MHz for the highly polarised time samples with 
linear polarisation fraction between 70--100\%, while the middle panel 
represents time samples with low levels of linear polarisation less than 70\%.
See caption in Fig.~\ref{fig:PPAhistC1} for additional description. The right 
panel shows the histograms representing the PPA distribution at the specific 
longitudinal phase of -1.52$\degr$, where the blue and orange colours correspond
to the highly polarised time samples and the less polarised cases, 
respectively.}	
\label{fig:B0136_band4}
\end{figure}

\subsubsection{ $C_1$ class of pulsars }
Fig.~\ref{fig:PPAhistC1} shows the distribution of the PPAs from the individual
time samples of the five pulsars in this group with six separate measurements. 
PSR J0139+5814 has sufficient detection sensitivity at both frequency bands, PSR 
J1709$-$4428 is reported at the lower frequency range between 300--500 MHz, 
while measurements from PSRs J1733$-$3718, J1740+1000 and J2043+2740 is carried
out at the 550--750 MHz frequency band. The RVM fits to the OPM are shown along
with the PPA distributions of the time samples. In all cases the PPA traverses
are confined to a single polarisation mode, along with an intrinsic spread 
around the RVM fits. The distribution of the linear polarisation fraction of
the time samples is shown in Fig.~\ref{fig:fracpolC1}. In all six cases the 
distributions show asymmetrical structures with single peaks, spread out in the
leading sides and relatively sharp declines toward the trailing ends. The peak 
polarisation levels are above 70\% as expected based on the classification 
criterion, while the leading edge extends to 10--20\% levels, suggesting that 
at certain instances emission with low polarisation fraction also emerge from 
these sources. 

The PPA behaviour is best explored by considering specific example of PSR 
J0139+5814 at the 550--750 MHz frequency range, that has 29504 time samples
with significant polarised emission across the observing duration. In around 
87\% of these time samples the emission is highly polarised with polarisation 
fraction exceeding 70\%. Separate average profiles are constructed using the 
the high and low polarisation emission, as shown in 
Fig.~\ref{fig:B0136_band4}(a) and (b), respectively, and show very different 
shapes. The highly polarised profile has a single peaked structure, similar to 
the average profile, but the lower polarised emission show two distinct peaks 
in the average profile, with the outer parts of the emission window being more
prominent than the center. There also appears to be presence of some OPM in
the PPA distribution of the time samples with lower linear polarisation. In 
both cases the PPAs are distributed in narrow clusters with small spread, 
without any significant spread within the window. Fig.~\ref{fig:B0136_band4}(c)
shows the histograms representing the spread of PPA at the longitude phase 
-1.52$\degr$ for both the high and low polarisation time samples. The 
distributions peak between polarisation angles of 60--70$\degr$ with a spread 
of around 15$\degr$. 

The $C_1$ class is unique in the pulsar population as such high levels of 
average polarisation are only observed in sources with $\dot{E} > 5 \times 
10^{33}$ ergs/s. The single pulse results presented here suggest that while 
majority of the time samples have high levels of fractional linear 
polarisation, there also exist samples with lower polarisation levels that do 
not contribute towards any significant de-polarisation of the average emission. 
Our studies of PSR J0139+5814 also show the presence of a few time samples 
with low polarisation levels having OPM, suggesting that even in the $C_1$ 
category of pulsars the waves escaping the pulsar magnetosphere is a mixture of
orthogonal modes, where one mode usually dominates. In other pulsars in this 
class we were not able to detect clear OPM, which is primarily a result of the 
weaker detection sensitivity of single pulses along with lower prevalence of 
time samples with low polarisation levels. We believe that longer observations 
and higher detection sensitivity will also reveal the presence of OPM in most 
pulsars belonging to the $C_1$ class.\\

\begin{figure}
\gridline{\fig{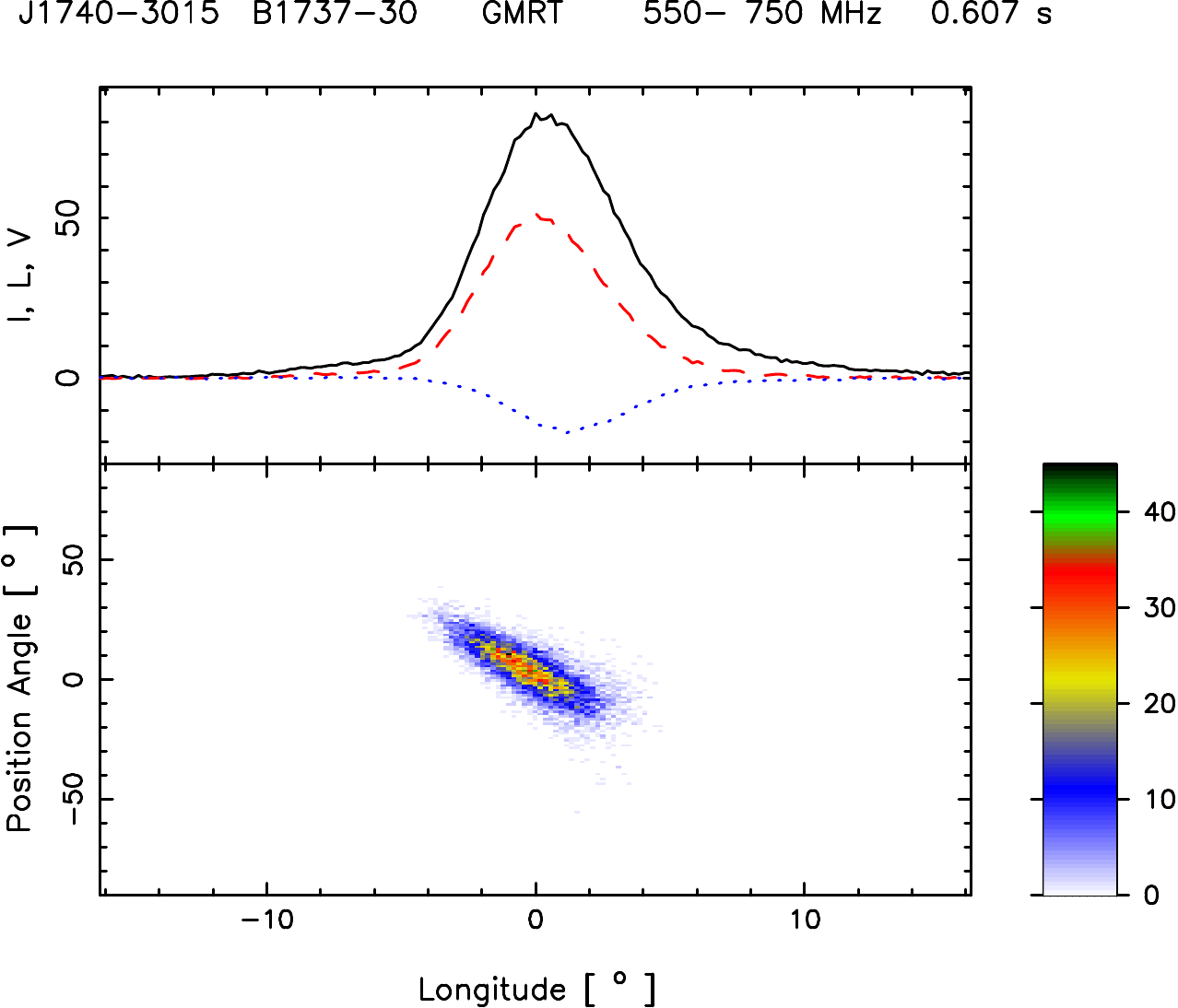}{0.32\textwidth}{(a)}
          \fig{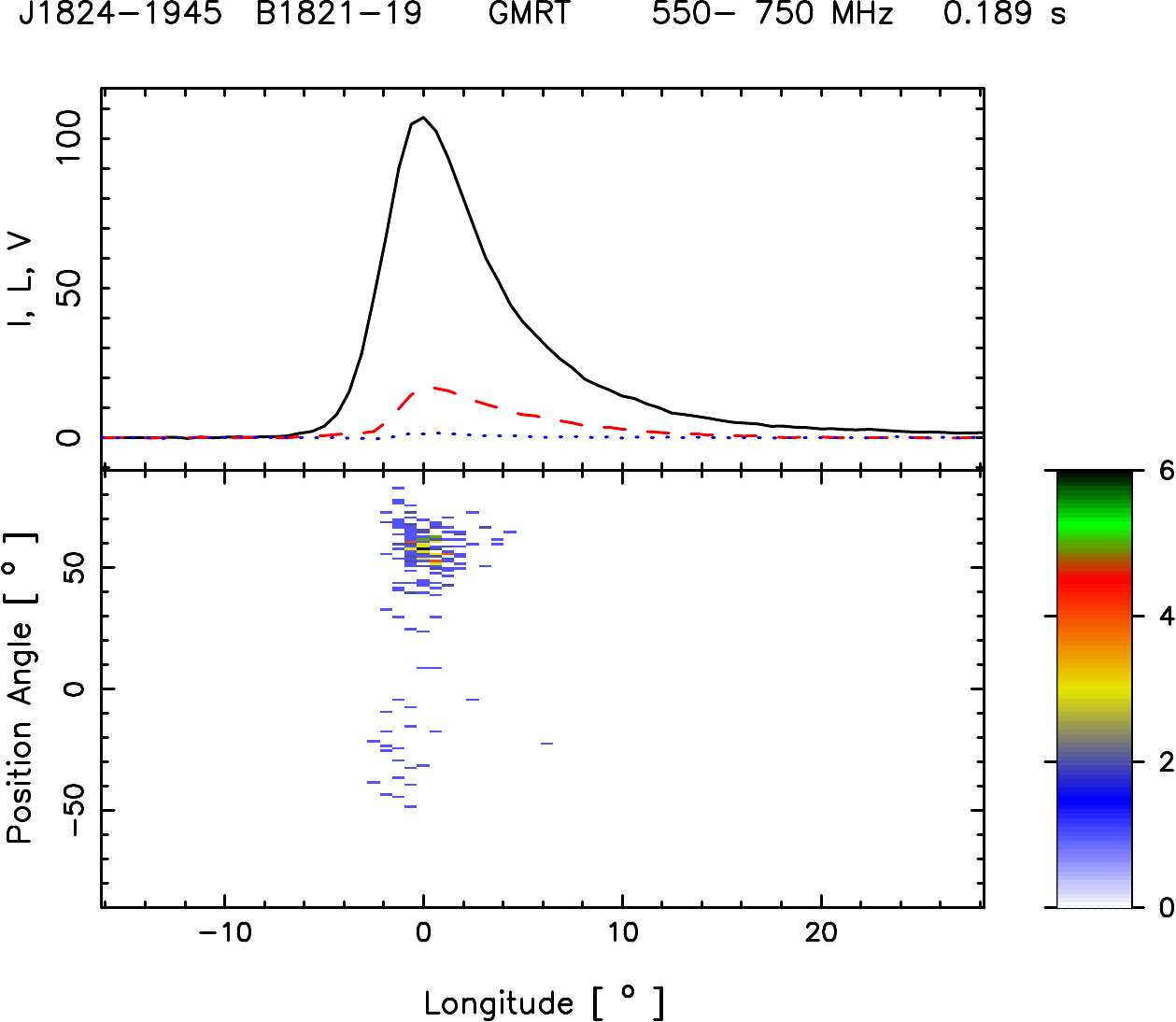}{0.32\textwidth}{(b)}
          \fig{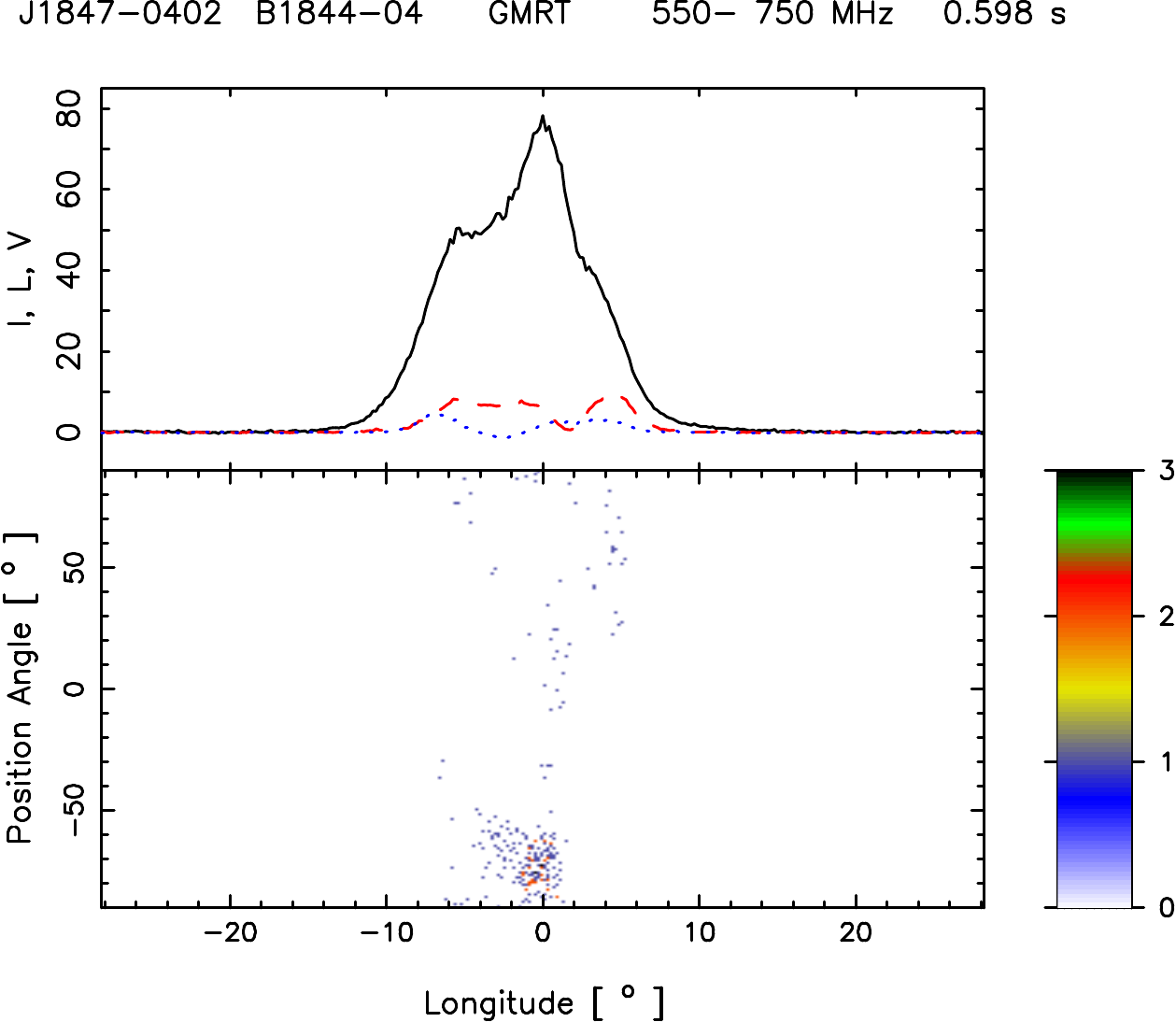}{0.32\textwidth}{(c)}}
\gridline{\fig{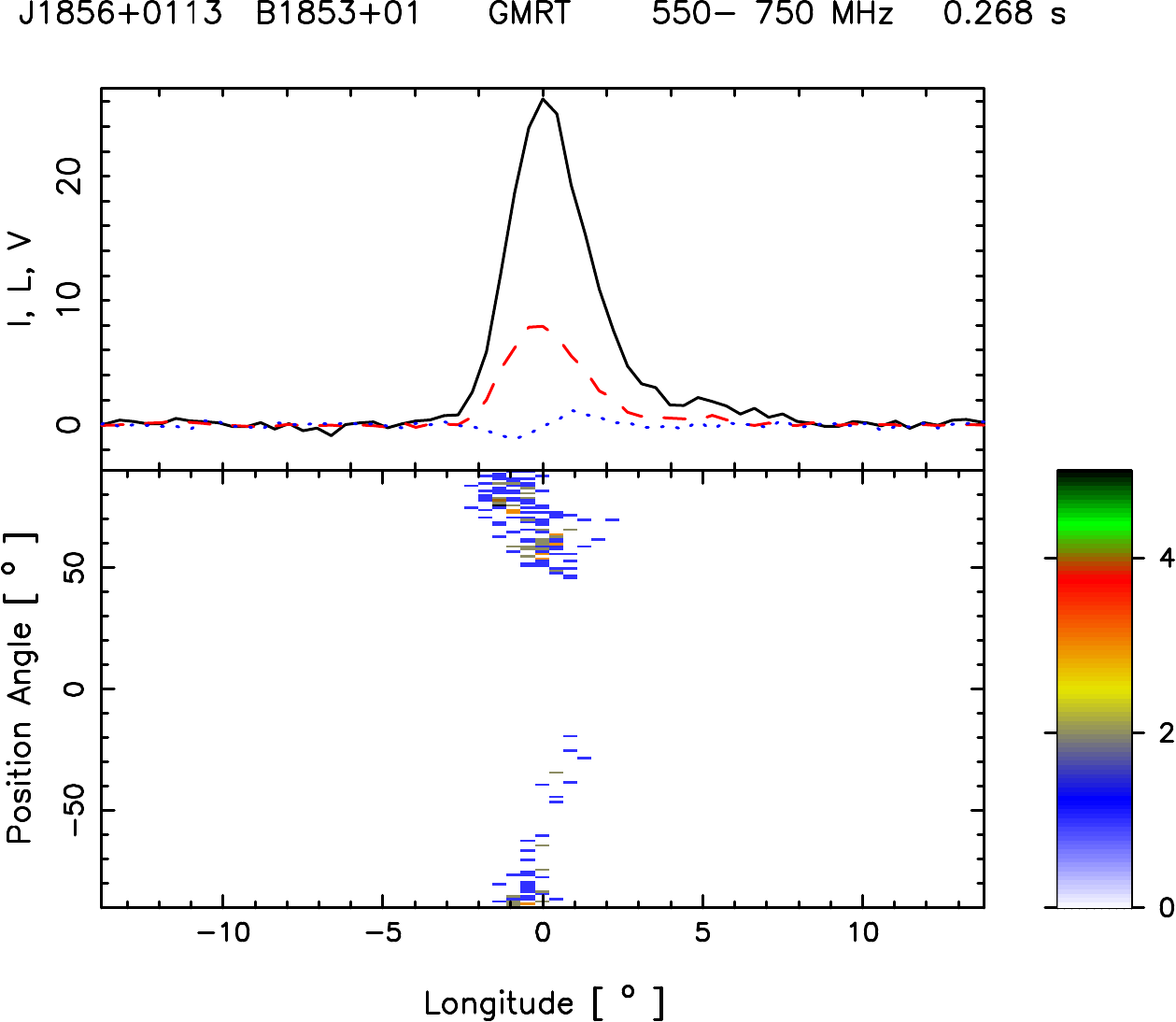}{0.32\textwidth}{(d)}
          \fig{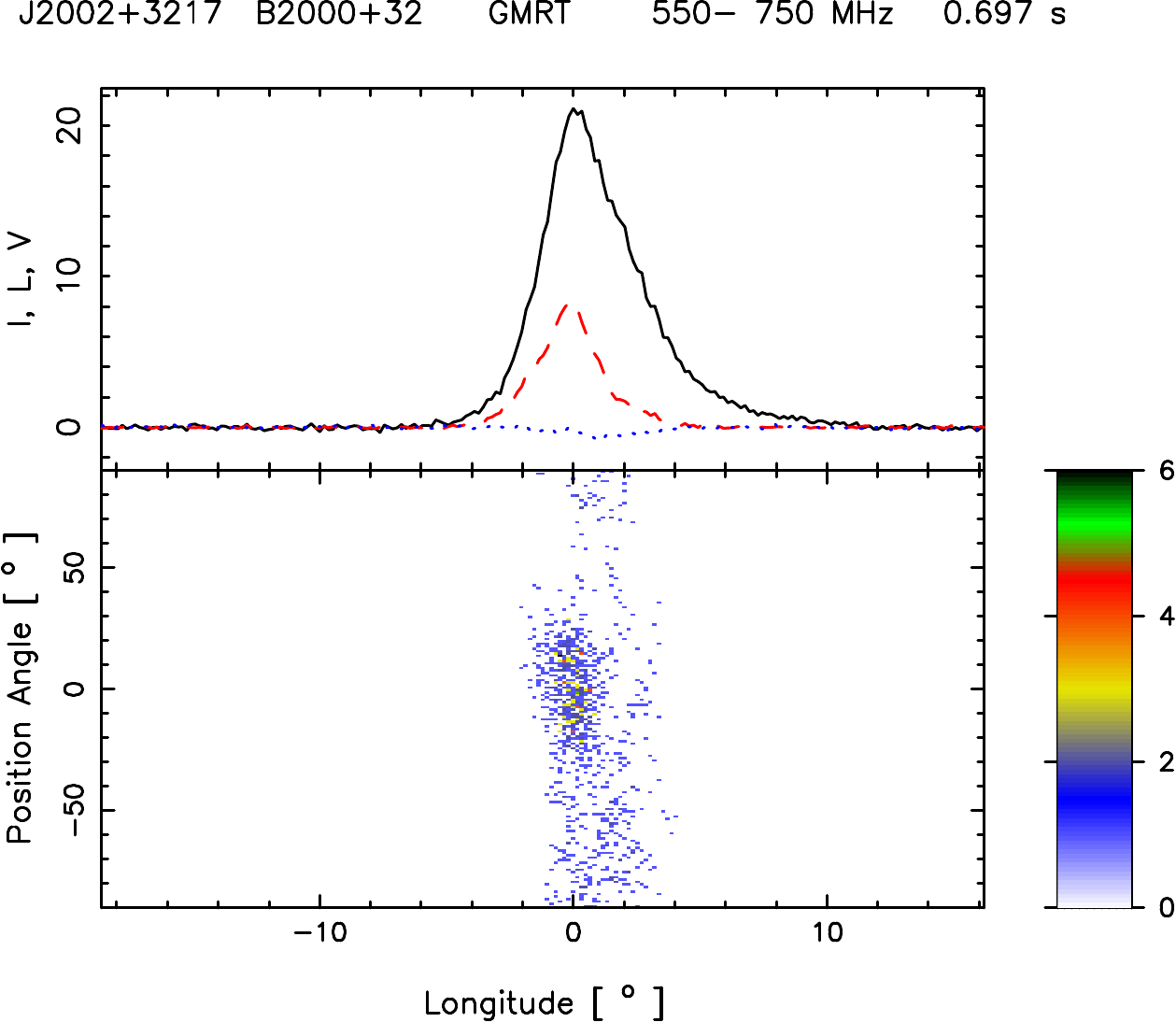}{0.32\textwidth}{(e)}
          \fig{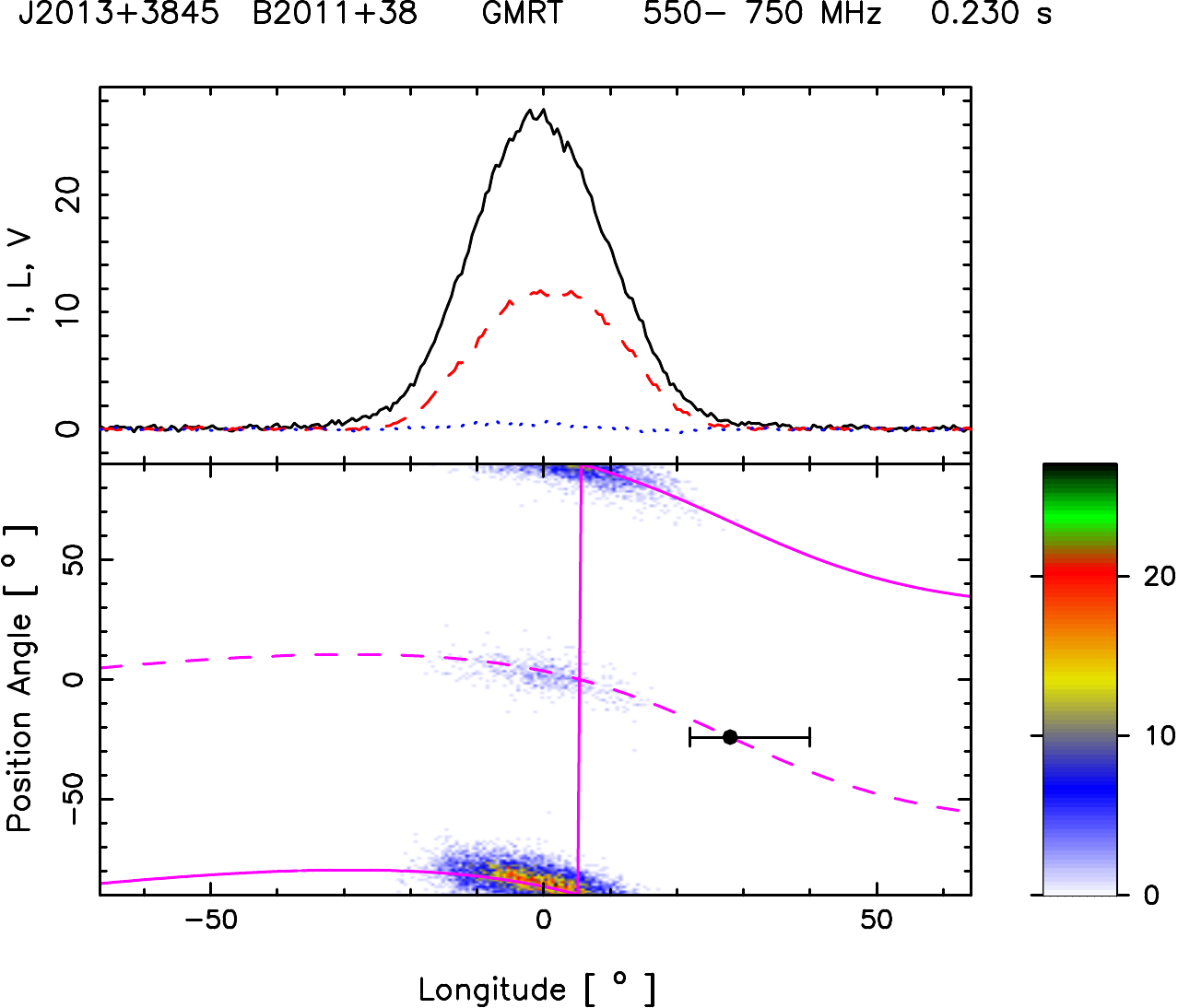}{0.32\textwidth}{(f)}}
\gridline{\fig{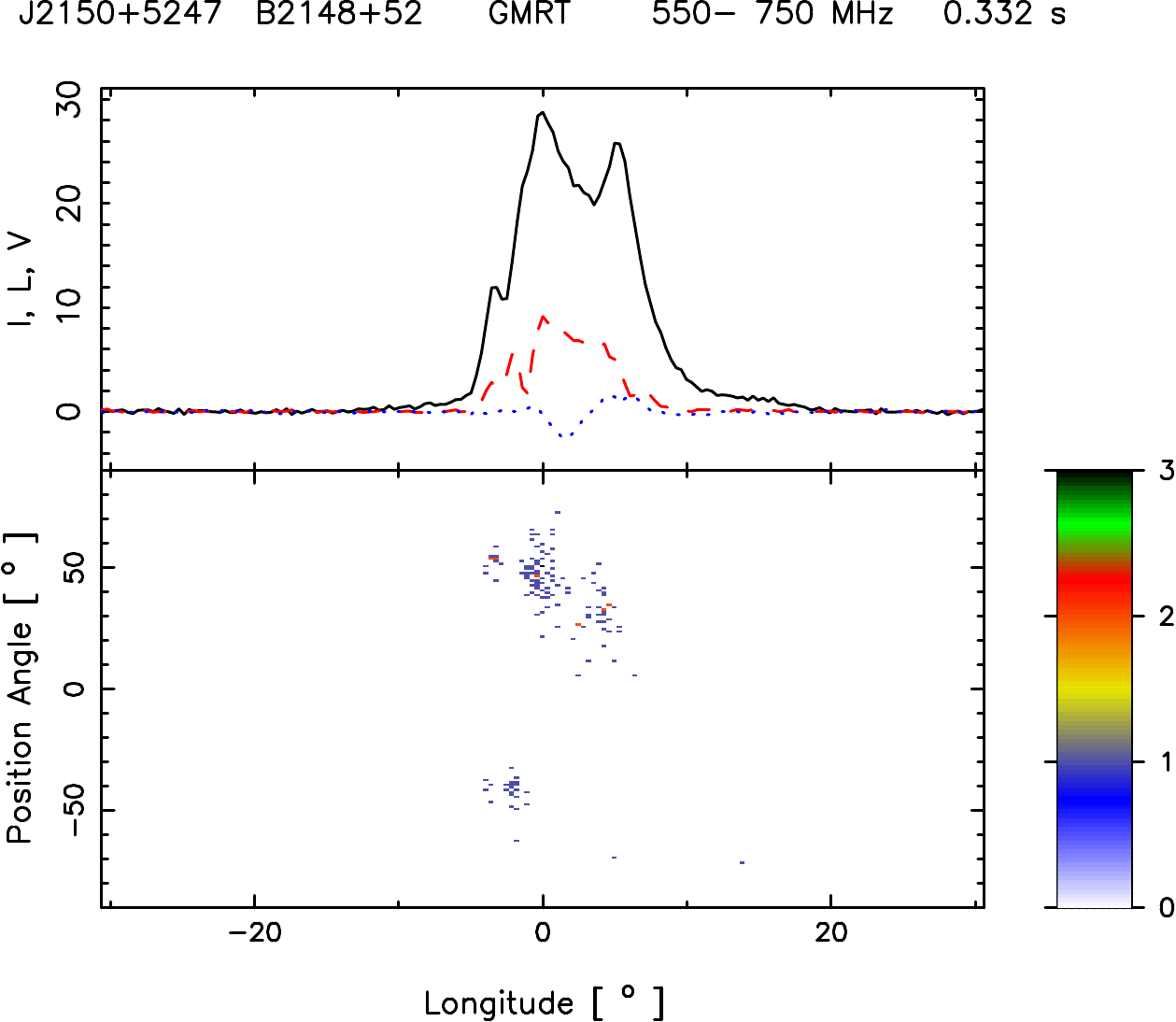}{0.32\textwidth}{(g)}
          \fig{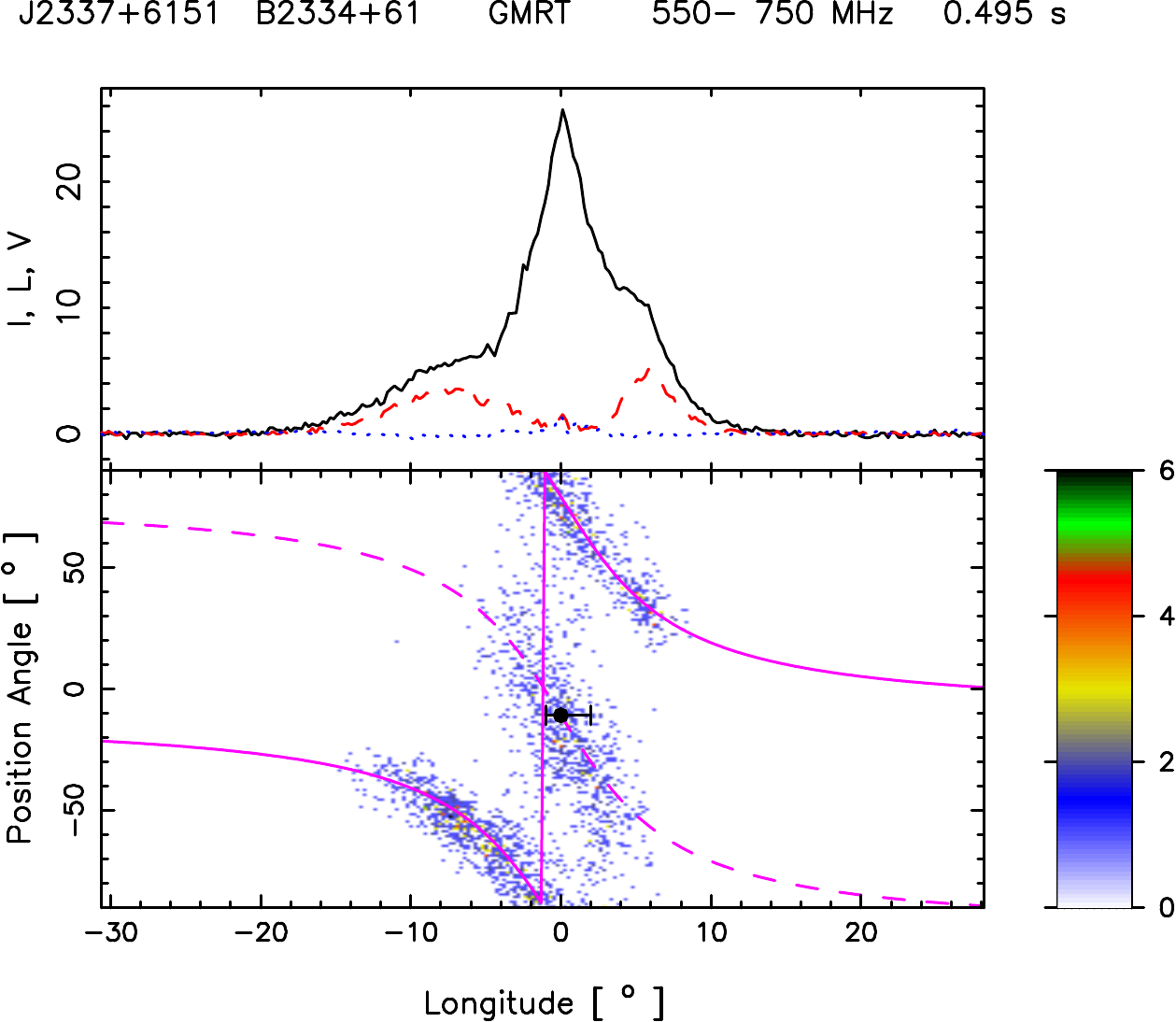}{0.32\textwidth}{(h)}}
\caption{The single pulse PPA distribution of pulsars belonging to the $C_2$
class with average fractional linear polarisation less than 70\%. See caption 
in Fig.\ref{fig:PPAhistC1} for additional description.}
\label{fig:PPAhistC2}
\end{figure}

\begin{figure}
\gridline{\fig{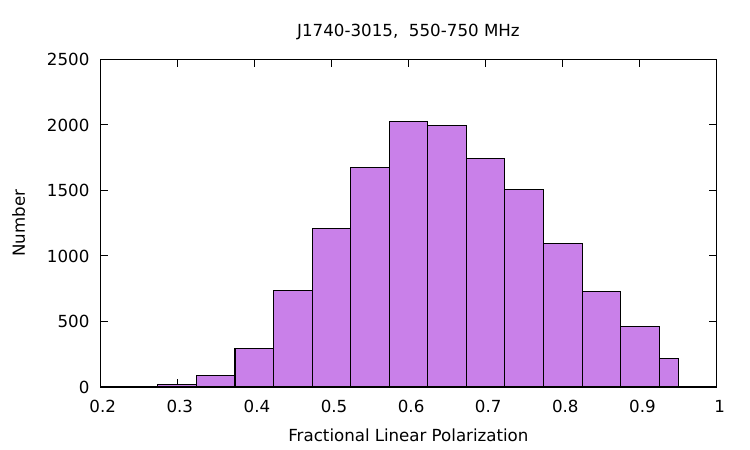}{0.33\textwidth}{(a)}
          \fig{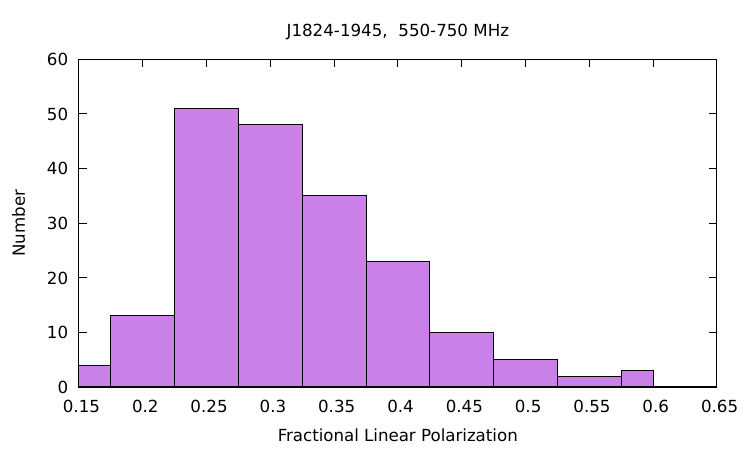}{0.33\textwidth}{(b)}
          \fig{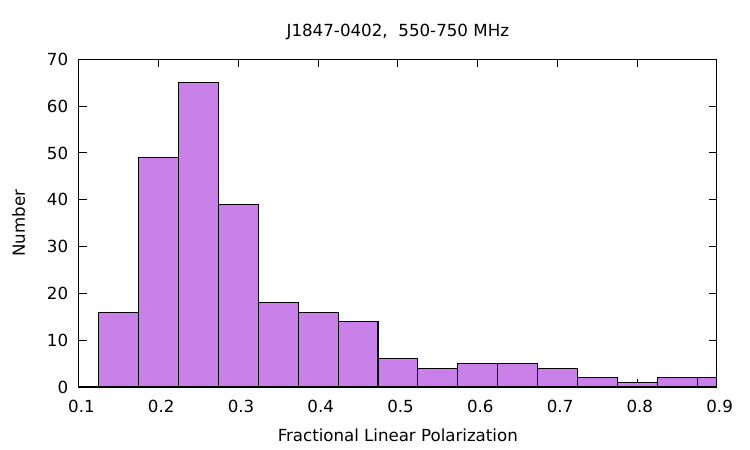}{0.33\textwidth}{(c)}}
\gridline{\fig{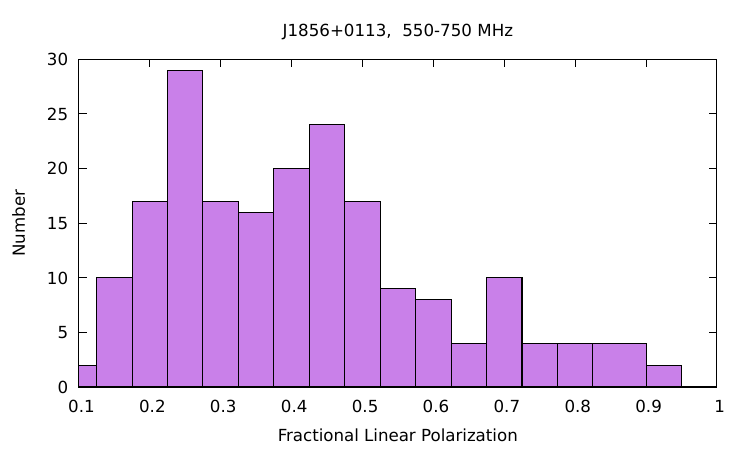}{0.33\textwidth}{(d)}
          \fig{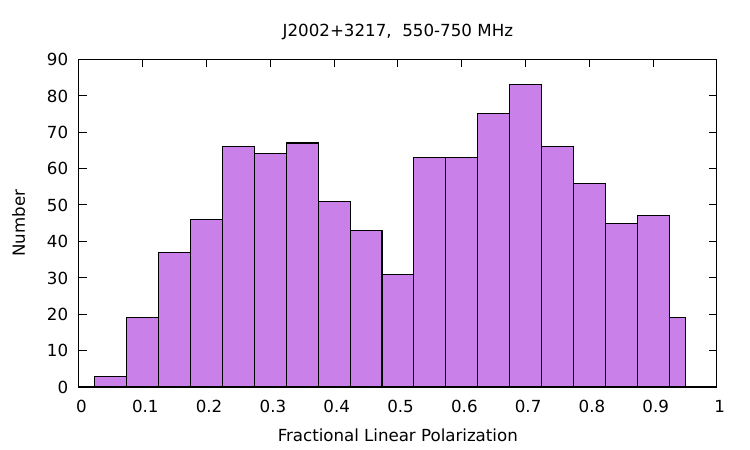}{0.33\textwidth}{(e)}
          \fig{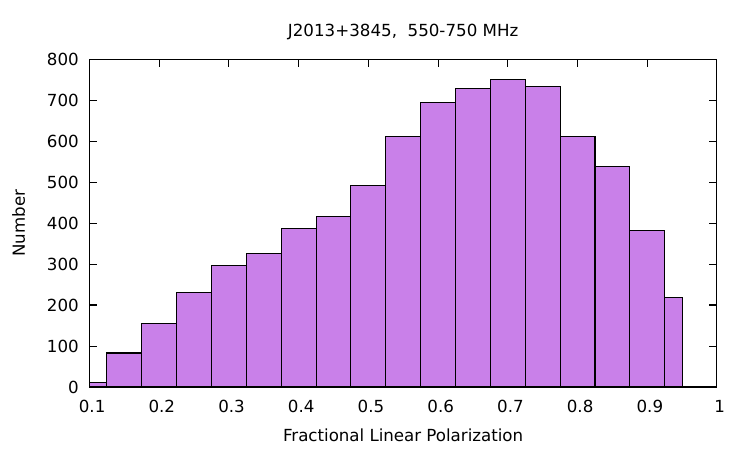}{0.33\textwidth}{(f)}}
\gridline{\fig{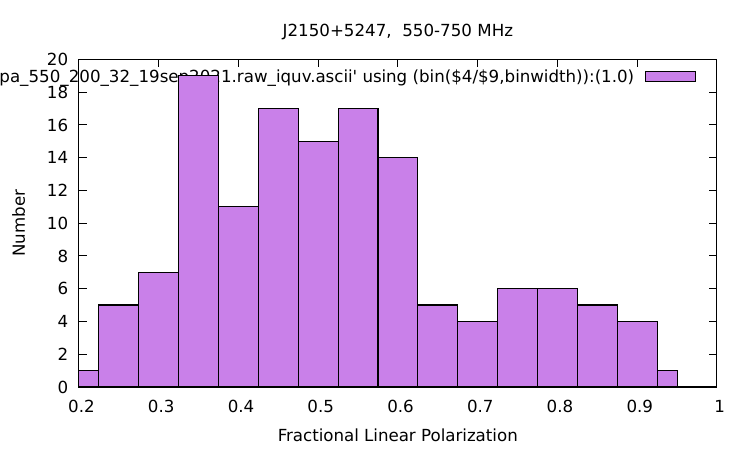}{0.33\textwidth}{(g)}
          \fig{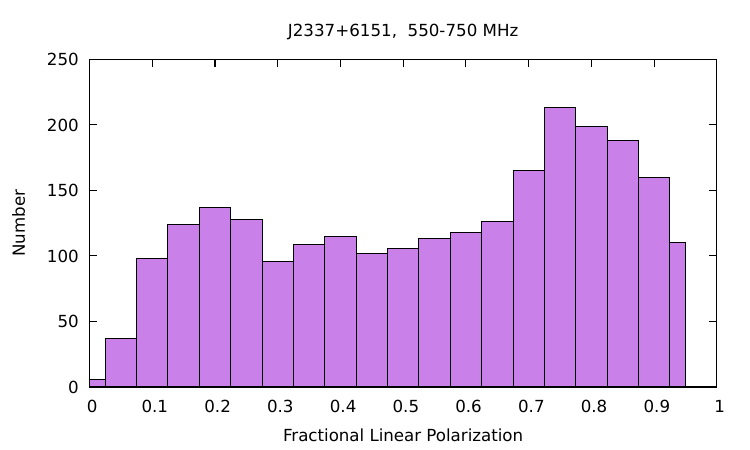}{0.33\textwidth}{(h)}}
\caption{The distribution of the linear polarisation fraction for time samples 
in the $C_2$ class of pulsars with average fractional linear polarisation
less than 70\%.}
\label{fig:fracpolC2}
\end{figure}

\begin{figure}
\gridline{\fig{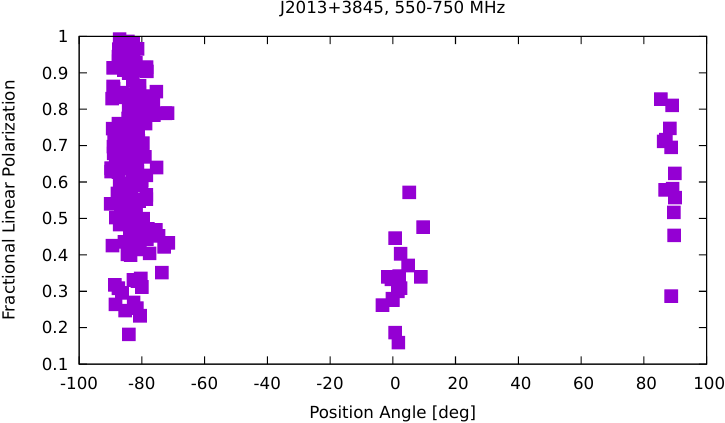}{0.45\textwidth}{(a)}
          \fig{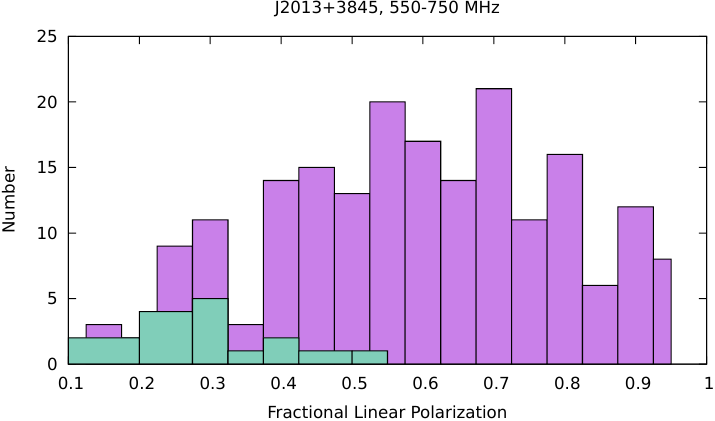}{0.45\textwidth}{(b)}}
\caption{The left panel shows the fractional polarisation level of each time 
sample of PSR J2013+3845 at longitude phase of -0.5$\degr$ as a function of the
PPA value, showing the presence of the two OPMs. The right panel shows the 
histograms representing the distribution of fractional linear polarisation 
distribution corresponding to the dominant (magenta) and less frequent (green) 
OPMs.} 
\label{fig:J2013_0.5deg}
\end{figure}

\subsubsection{ $C_2$ class of pulsars }
Fig.~\ref{fig:PPAhistC2} shows the PPA distributions of the time samples in 
the eight pulsars belonging to this category, all measured at the 550-750 MHz
frequency band, and unlike the previous case there is a wide variety in the 
nature of the distributions. Similarly, in Fig.~\ref{fig:fracpolC2} the 
fractional polarisation distributions also have different characteristics 
within this group. In PSR J1740$-$3015 the polarisation behaviour is most like 
the $C_1$ class, with majority of the PPA time samples clustered around a 
single polarisation mode. The fractional linear polarisation has a more 
symmetrical distribution peaking around the 60\% level. At higher frequencies 
this pulsar exhibits three components and the PPAs follow the RVM track 
\citep{2021MNRAS.504..228S}. At the 550--750 MHz frequency range the average
profile has a single component and the PPA distribution shows deviation from 
RVM nature near the trailing side of the profile. There is hint of interstellar
scattering tail in the profile that may explain the deviation from the RVM 
nature \citep{2009MNRAS.392L..60K}, while the presence of OPM jumps in the trailing side of the emission 
window is also a possibility. 

PSR J2013+3845 also shows clustered PPA distributions with small spread, but 
unlike the previous case both OPM tracks are visible. The fractional 
polarisation distribution also resembles the $C_1$ class of pulsars, with a
single peaked, asymmetric structure stretched along the leading side and having 
steeper descent at the trailing edge. But the peak of the distribution is lower
at around 70\% level, while the minimum polarisation fraction reaches 
10\% level. To further highlight the nature of OPMs in this pulsar we have 
selected a specific longitudinal phase of -0.5$\degr$ and plotted the PPA of 
the time samples in this longitude as a function of their linear polarisation 
fraction, as shown in Fig.~\ref{fig:J2013_0.5deg}(a). The distribution of the 
fractional polarisation levels is shown in Fig.~\ref{fig:J2013_0.5deg}(b), 
where the dominant mode (magenta) has higher levels of polarisation extending 
up to 95\% and a wider peak between 50--70\%. The less dominant mode has much 
lower polarisation levels, spread between 10--50\%, with the distribution 
peaking around the 30\% level.

In the remaining six pulsars the PPA distributions are more scattered and have 
wider spread, although there seems to be two OPMs in PSR J2337+6151 despite the 
noisy nature of the distribution. The fractional polarisation distribution of 
the individual time samples in PSR J1824$-$1945 and J1847$-$0402 have similar 
structures, with peaks between 20--30\% levels and longer tails. In PSRs 
J2002+3217 and J2337+6151 the polarisation fractions have wide distributions 
between 10--90\% with double peaked structure, where the lower peak is between
20-30\% level while a second grouping is seen at higher polarisation fraction
above 70\%. In the other two cases, PSRs J1856+0113 and J2150+5247, the 
distributions are irregular but mostly comprise of time samples with less than 
50\% polarisation level. The polarisation behaviour in these pulsars seem to
be comparable with the lower energetic population that also exhibit low levels
of polarisation in the radio emission.

\subsection{Statistical model of polarisation using orthogonal modes} 
\label{sec:stat}

\begin{figure}
\gridline{\fig{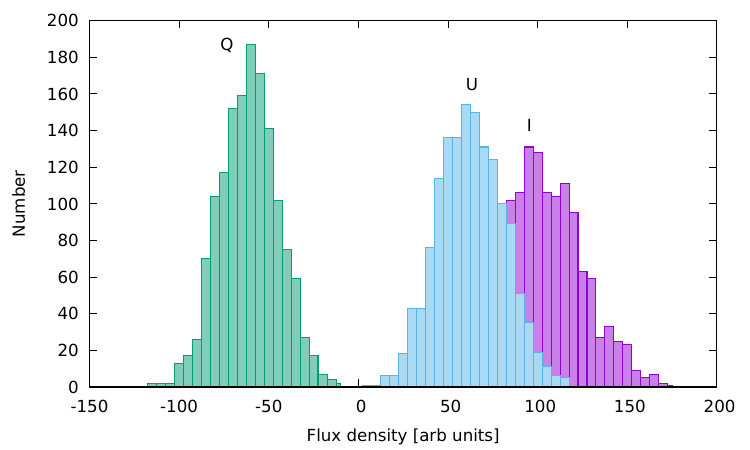}{0.42\textwidth}{(a)}
          \fig{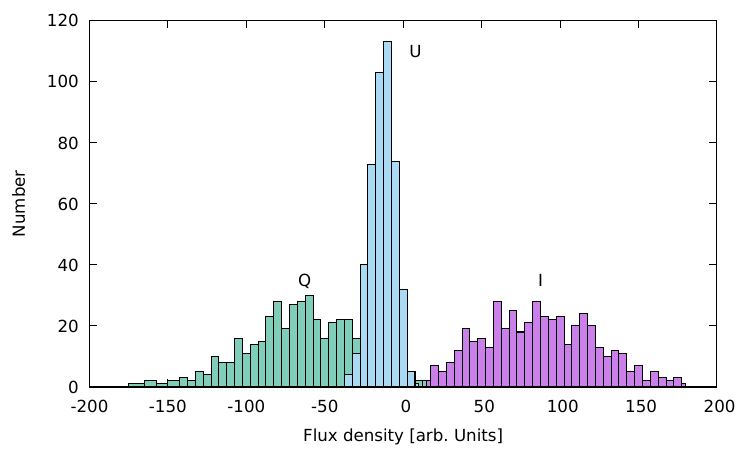}{0.42\textwidth}{(b)}}
\gridline{\fig{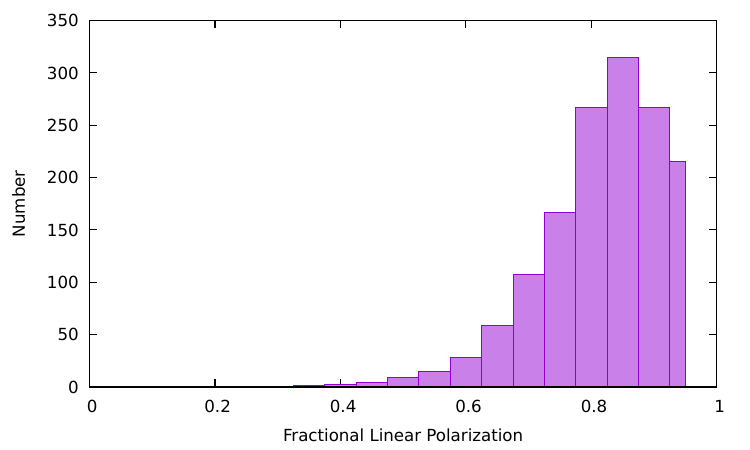}{0.42\textwidth}{(c)}
          \fig{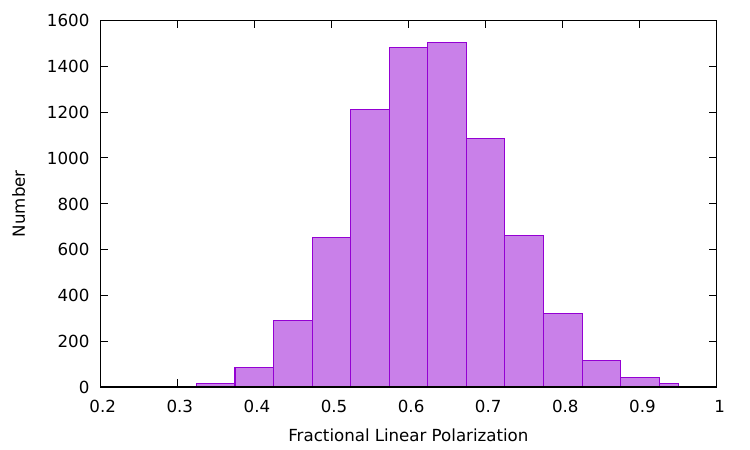}{0.42\textwidth}{(d)}}
\gridline{\fig{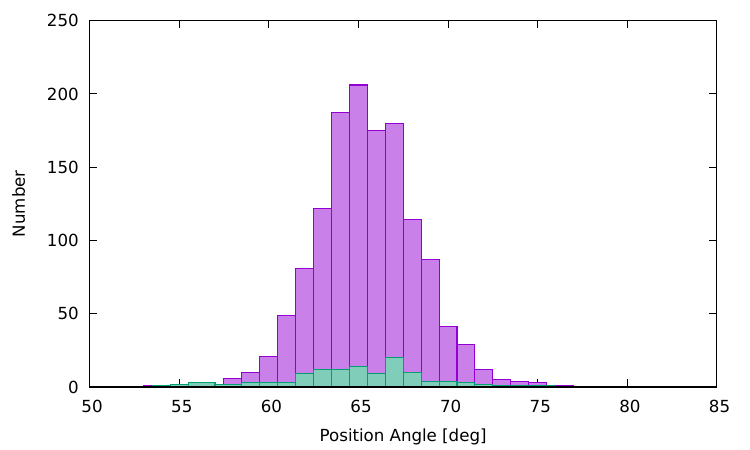}{0.42\textwidth}{(e)}
          \fig{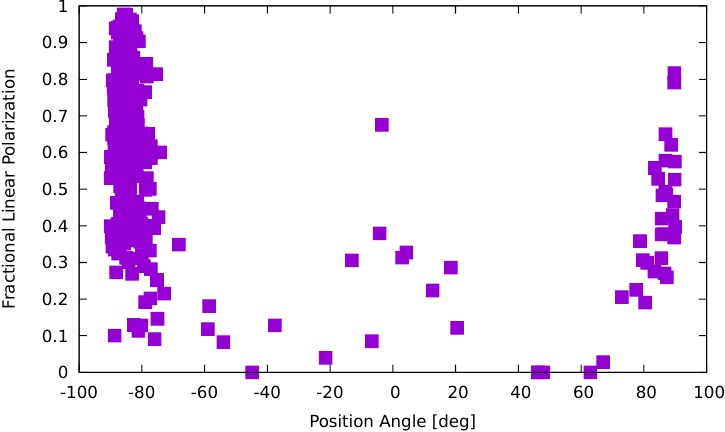}{0.42\textwidth}{(f)}}
\caption{The simulated polarisation behaviour in pulsars where the individual 
time samples have been estimated by combining two normally distributed 
orthogonal modes. Three distinct configurations : I, II and III have been setup
with different means and standard deviations of the two orthogonal modes 
(see text for details). The top panels show the distribution of the Stokes 
parameters I, Q and U, corresponding to configurations I (left panel) and III 
(right panel). The middle panels show the fractional polarisation distributions 
in the single pulses for configuration I (left panel) and II (right panel). The
bottom panel in the left shows the distribution of the PPA corresponding to the
high levels of fractional linear polarisation (magenta) and low polarisation
level (green) from configuration I, while the right panel shows the fractional 
polarisation spread as a function of the PPA from configuration III, showing 
the two OPMs in the PPA.}
\label{fig:siml_pol}
\end{figure}

A statistical approach towards understanding the nature of polarisation 
behaviour seen in pulsars has been proposed by \cite{1998ApJ...502..883M}, 
where the resultant polarisation behaviour is modelled by combining two fully 
polarised modes, orthogonal to each other and having stochastic distributions.
If we consider the flux densities of the two modes to be represented by the 
variables $X_1$ and $X_2$, then according to this scheme the four Stokes 
parameters can be estimated as : I = $X_1 + X_2$, Q = $X_1 - X2$, and U = V = 
0. $X_1$ and $X_2$ are drawn from random variables that have mean levels 
$\mu_1$ and $\mu_2$ with standard deviations $\sigma_1$ and $\sigma_2$, 
respectively. The statistical relationship between the mode distributions is 
described by the parameters $M = \mu_1/\mu_2$ and $R=|\mu_1-\mu_2|/3(\sigma_1 +
\sigma_2)$. The observed polarisation properties like PPA distribution, 
presence of OPMs, non-orthogonal PPA behaviour, etc., can be replicated in 
several ways. This involve selecting suitable distributions of $X_1$ and $X_2$,
adding random baseline noise to the Stokes parameters, introducing rotations in
the PPA by redistributing the intensities in Q and U, etc. We have used this 
method to study the polarisation behaviour of the energetic pulsars reported in
this work. Three specific configurations has been simulated in order to 
reproduce the polarisation properties seen in the $C_1$ and $C_2$ classes of 
pulsars, as described below.

\begin{enumerate}[I.]
\item In the first configuration the $X_1$ distribution is characterised 
by $\mu_1 = 95$ and $\sigma_1 = 22$, while $X_2$ is determined by $\mu_2 = 5$ 
and $\sigma_2 = 3.5$, such that $M = 19$ and $R = 1.2$. The baseline noise 
distribution specified by $\mu_N = 0$ and $\sigma_N = 8$ is added to the Stokes
parameters and additional rotation of linear polarisation by $\psi_{\circ} = 
66\degr$ is carried out to obtain the distributions of time samples I, Q and U 
as as shown in Fig.~\ref{fig:siml_pol}(a). The fractional polarisation 
distribution from this setup, shown in Fig.~\ref{fig:siml_pol}(c), peaks around
80\% level and has a relatively stretched leading side that resembles the $C_1$
class of pulsars. Fig.~\ref{fig:siml_pol}(e) shows the distribution of the PPA 
corresponding to time samples with high polarisation fraction above 70\% level 
(magenta) and low polarised signals below this level (green). The 66$\degr$ 
phase rotations have been introduced to resemble the PPA distributions of PSR 
B0136+57~in Fig.~\ref{fig:B0136_band4}(c). 
\item In the second case the distributions of $X_1$ is specified with $\mu_1 = 
25$ and $\sigma_1 = 2$, and $X_2$ with $\mu_2 = 5$ and $\sigma_2 = 1.5$, 
resulting in $M = 5$ and $R = 1.9$, and baseline noise of $\sigma_N = 1$ has 
also been added. The resultant fractional linear polarisation distribution is 
shown in Fig.~\ref{fig:siml_pol}(d) and has an isotropic nature peaking around 
60\% level. The distribution resembles that of PSR J1740$-$3015~in 
Fig.~\ref{fig:fracpolC2}(a), that has been classified in the $C_2$ group but 
also shows certain properties of the $C_1$ group (see discussion in section 
\ref{splin}), and can be considered as a transitional object between the two 
groups.
\item In the third example we have considered $X_1$ with $\mu_1 = 13$ and
$\sigma_1 = 7$, and $X_2$ with $\mu_2 = 72$ and $\sigma_2 = 35$, which has $M =
0.2$ and $R = 0.5$, along with baseline noise $\sigma_N = 5$. The distribution 
of the Stokes parameters I, Q and U for the different time samples using this 
setup is shown in Fig.~\ref{fig:siml_pol}(b). The distribution of the 
fractional polarisation as a function of PPA is also shown in 
Fig.~\ref{fig:siml_pol}(f), where the OPM appears resembling the behaviour of 
the $C_2$ class pulsar J2013+3845~in Fig.~\ref{fig:J2013_0.5deg}(a).
\end{enumerate}

The above results show when $M \gg 1$ and $R > 1$, the statistical behaviour 
resemble that of the $C_1$ class of pulsars, while smaller values of $M$ and $R
\lesssim 1$ is the condition for the $C_2$ class. The decrease in $M$ reduces 
the mean level of the fractional linear polarisation, and as $R$ goes below 
unity the distributions starts to overlap such that OPMs are seen in the 
polarisation behaviour. More quantitative estimates of the mode parameters have
been suggested \citep[see][for details]{1998ApJ...502..883M}, like taking into 
account sources of instrumental noise as well as mitigating the effect of 
scintillation, and applied to other single pulse studies 
\citep{2004MNRAS.348.1229J}. However, the primary drawback of this scheme is
the lack of any physical motivation for the origin of the $X_1$ and $X_2$ 
distributions. We have also avoided discussion about the nature of circular 
polarisation, particularly because of the small power and also the lack of any 
significant change of circular polarisation with $\dot{E}$. In the discussion 
below a physical framework for the development of two fully polarised 
orthogonal modes in the pulsar plasma is presented, that can largely explain 
the polarisation behaviour in the $C_1$ and $C_2$ classes of energetic pulsars,
as shown here. We also describe how the circular polarisation can emerge from 
this model.

\section{Discussion} \label{discuss}
In the normal pulsar population several observational studies have clearly 
demonstrated that PPAs of time samples with high levels of linear polarisation 
are distributed along two RVM-like tracks corresponding to the OPMs \citep[see 
e.g.][]{2023MNRAS.521L..34M,2023ApJ...952..151M,2024MNRAS.530.4839J}. This 
result provides conclusive evidence that the radio emission mechanism in normal
pulsars is CCR from charge bunches \citep[see][for a 
review]{2024Univ...10..248M}. The radio emission heights are estimated to be 
below 10\% of the light cylinder radius (see section \ref{avgprop}), where the
plasma waves follow the dispersion relation of strongly magnetized pair plasma.
In this environment the only known mechanism capable of sustaining stable 
charge bunches and emit CCR are charge separated envelope solitons, that 
develop due to nonlinear growth of instability in relativistic, non-stationary,
outflowing plasma \citep{1980Afz....16..161M,2000ApJ...544.1081M,
2018MNRAS.480.4526L,2022MNRAS.516.3715R}. The outflowing plasma in pulsars is 
postulated to emerge from a charge starved inner acceleration region (IAR) 
above the polar cap \citep{1975ApJ...196...51R}. The IAR is dominated by large 
non-dipolar magnetic fields ($\sim10^{13}$ G) and electric fields ($\sim 
10^{12}$ V) parallel to the magnetic fields. Magnetic field assisted 
electron-positron pair production is set off in the IAR from background 
$\gamma$-ray photons, that are accelerated in opposite directions by the 
electric field, and emit additional $\gamma$-ray photons triggering a cascading
effect. This phenomenon is known as sparking discharge and fills up the IAR 
with primary plasma particles till the potential difference along the gap is 
screened. However, it has proved particularly challenging to sustain multiple 
isolated sparking columns within the polar cap area \citep{CR80}. 

The PSG model of the IAR provides a stable setup for sparking discharges and 
serve as a significant advancement towards understanding the nature of 
outflowing plasma. The polar cap surface is heated to high temperature ($\sim 
10^6$ K) by the back streaming particles during sparking, which is around the 
critical temperature for free flow of positively charged ions from the surface.
Abundant supply of these surface ions can completely screen the electric 
potential difference in the IAR and stop the cascading effect. When the 
temperature goes below the critical level at any location within the IAR the 
potential difference reappears, once again triggering a local sparking 
discharge. Thus the sparks serves as a mechanism of thermal regulation of the 
surface temperature and are formed in a tightly packed manner for effective 
regulation \citep{2022ApJ...936...35B}. The PSG model has been successful in 
explaining the observed subpulse drifting behaviour \citep{2020MNRAS.496..465B,
BMM23} and the thermal X-ray emission from pulsar polar caps \citep{SG20}. The 
initiation of the spark in the IAR occurs due to production of primary pairs 
which further undergoes secondary pair cascade and as a result the spark grows 
in size: both along and across the magnetic field lines until the full electric
field in the gap is screened. Across the magnetic field lines at the IAR the 
density of the pair plasma is highest at the center of the spark and decreases 
towards the boundary, where the ionic density becomes comparable. We identify 
the central region of sparks with the superscript ``$sp$'' and the overlapping 
boundary between sparks as ``$isp$''. In normal pulsars the typical Lorentz 
factors of the secondary plasma are $\gamma^{sp}_s = \gamma^{isp}_s \sim 100$, 
while the ion component has values of $\gamma^{sp}_{ion} \sim 10^{3}$ and 
$\gamma^{isp}_{ion} \sim 10$, respectively. The multiplicity of the secondary 
pair plasma density is $\kappa^{sp} \sim 10^{4}$ at the spark center, which 
gradually decreases to $\kappa^{isp} \sim 1$ at the boundary region. The 
sparking process is intermittent in nature and is associated with the IAR gap 
opening and closing times. This results in a spark associated non-stationary 
plasma flow along the magnetic field lines consisting of regions of high and 
low multiplicity pair plasma. The development of two stream instability in this
plasma flow excites Langmuir waves and its modulational instability leads to 
the formation of charge solitions at a distance of about 1\% of the Light 
Cylinder radius. In regions of dense pair plasma the envelope solitons 
excite curvature radiation with the excited waves having characteristic 
frequency $\nu_{cr}$, such that $\nu_{cr} < 2\sqrt{\gamma_s} \nu_p$, here 
$\nu_p$ is the plasma frequency. In the wave generation region, since the pair 
plasma multiplicity changes across the magnetic field line, $\nu_p$ also 
changes since it depends on $\kappa^{0.5}$ and hence  for convenience of 
discussion we use the term high-multiplicity plasma cloud where $\nu_{cr} < 
2\sqrt{\gamma_s} \nu_p$ and low-multiplicity plasma where $\nu_{cr} > 
2\sqrt{\gamma_s} \nu_p$.

A qualitative description of the polarisation properties of the radio emission
from the outflowing plasma generated in a PSG has been presented in 
\cite{2023ApJ...952..151M}, that can be briefly summarized as follows. The 
eigen modes of pair plasma in strong magnetic field is described following
the convention of \cite{2003PhRvE..67b6407S}, where CCR excites the transverse 
electromagnetic ($t$) mode and the sub-luminal longitudinal-transverse ($lt_1$)
mode. The $t$-mode and $lt_1$-mode have electric vectors directed perpendicular
and parallel to the plane containing the propagation vector, $\bm{k}$, and the 
magnetic field, $\bm{B}$, respectively. A charge separated envelope 
soliton excites CCR in high-multiplicity plasma cloud within a narrow angular 
cone specified by 1/$\gamma^{sp}_s$, with the intrinsic power in the 
$lt_1$-mode being 7 times higher than the $t$-mode. The $t$ and $lt_1$ modes 
have different refractive indices in the pulsar plasma, resulting in splitting 
of the waves during propagation in the high-multiplicity plasma clouds and they
travel independently as 100\% linearly polarised modes. The $t$-mode has vacuum
like dispersion property and can freely escape from the high-multiplicity to 
low-multiplicity regions, and eventually detach from the magnetosphere.  The 
$lt_1$-mode on the other hand in the high-multiplicity plasma cloud gets ducted
along magnetic field lines and should be suppressed due to Landau damping 
\citep{2004ApJ...600..872G,2014ApJ...794..105M}). However, the $lt_1$-mode can 
also escape the magnetosphere provided the escape condition $\nu_{cr} > 2 
\sqrt{\gamma_s} \nu_p$ is satisfied before the wave gets damped. The 
low-multiplicity regions naturally exist along the magnetic field lines 
between two spark associated plasma columns, where in-between the columns there 
exists a region where the multiplicity decreases significantly such that 
$\nu_{cr} > 2 \sqrt{\gamma_s} \nu_p$, and this region can serve as an escape 
channel for the waves. Once the $t$ and $lt_1$ modes (also known as the 
extraordinary (X) and ordinary (O) modes) enter the low-multiplicity region, 
the dispersion relation of both modes become vacuum like, and the waves are 
polarised perpendicular or parallel to the magnetic field line planes 
respectively. \cite{2023ApJ...952..151M} also argued that due to preponderance 
of ions in the low-multiplicity plasma the eigen modes become circularly 
polarised, and hence the linearly polarised waves that enter the 
low-multiplicity region can become partially elliptically polarised before they
escape the pulsar magnetosphere.

The incoherent mixing of intensities of $t$ and $lt_1$ modes excited due to 
curvature radiation from a large number of solitons occurs in the high 
multiplicity plasma cloud, and eventually the resultant emission escapes from 
the edge of dense plasma cloud. Further several plasma clouds are averaged 
during the observing time resolution $t_{res}$ which results in additional 
stochasticity arising due to temporal variations of the plasma clouds, and 
hence the observed $t$ and $lt_1$ modes can be identified with two random 
variables with intensities $A$ and 7$A$ drawn from a distribution. 
Additionally, the sub-luminal $lt_1$-mode interacts with the plasma and can get 
damped, which we identify with an average damping factor denoted by a random 
variable $\zeta$, such that the effective intensity of the $lt_1$-mode is 
$7\zeta A$. The statistical scheme of \cite{1998ApJ...502..883M} presented
in \ref{sec:stat} can be interpreted using the curvature radiation mechanism, 
where the two stochastic distributions of the orthogonally polarised modes can 
be identified as the $t$ and the $lt_1$ modes, with $X_1 = A$ and $X_2 = 7\zeta
A$. \\

\subsection{Statistical model of $C_1$ class of pulsars using CCR characteristics} 
\label{d1}
\begin{figure}
\gridline{\fig{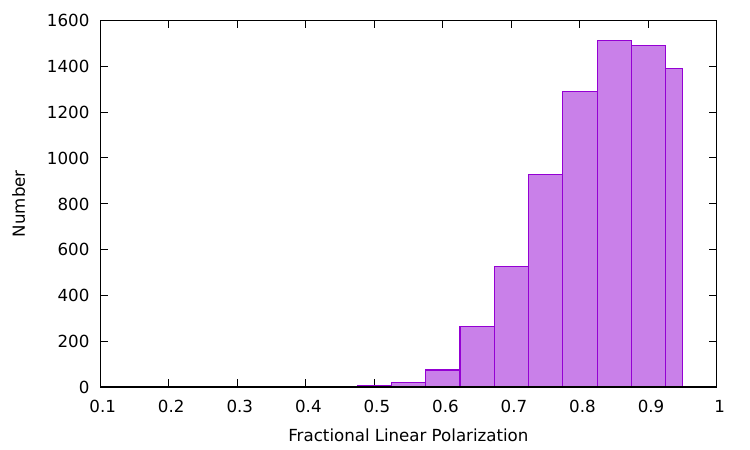}{0.42\textwidth}{(a)}
          \fig{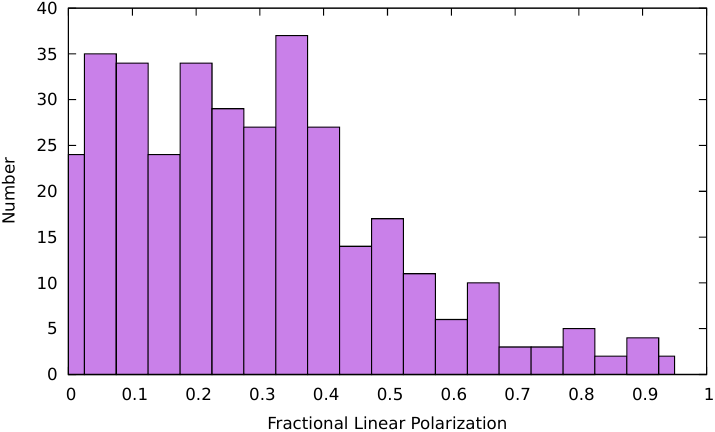}{0.42\textwidth}{(b)}}
\gridline{\fig{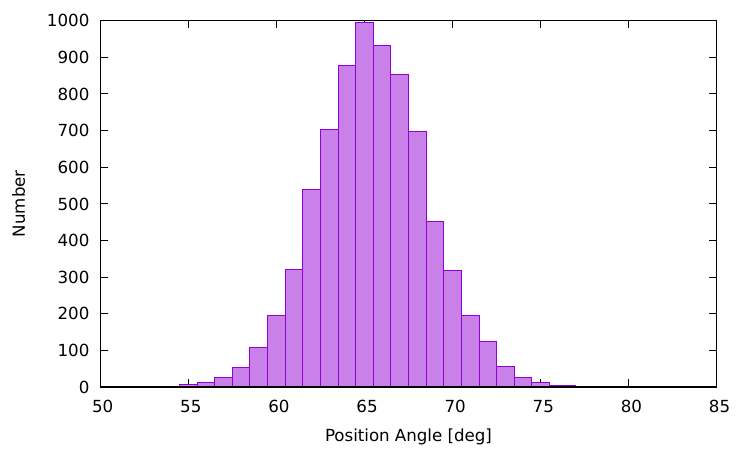}{0.42\textwidth}{(c)}
          \fig{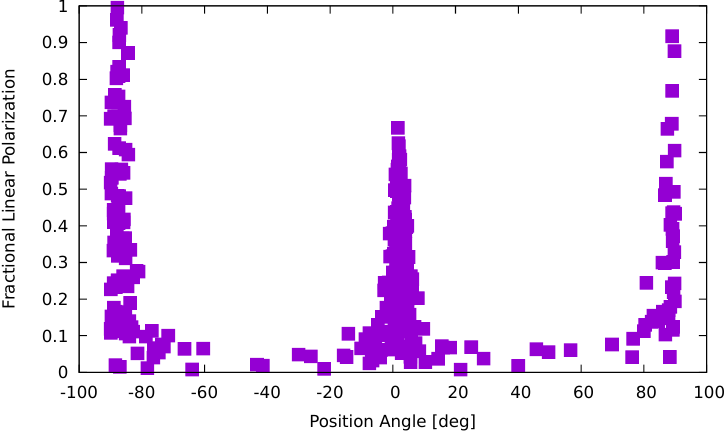}{0.42\textwidth}{(d)}}
\caption{The top panels show the simulated fractional linear polarisation 
distribution for the $C_1$ class (left panel) and $C_2$ class of pulsars, using
the expected OPM from CCR (see text for details). The bottom left panel shows 
the distribution of the PPA in the $C_1$ case, while the bottom right panel
shows the distribution of fractional linear polarisation as a function of the
PPA, exhibiting the presence of OPM seen in the PPA of the $C_2$ class of 
pulsars.}
\label{fig:ccr_modl}
\end{figure}

The analysis in section \ref{sec:stat} shows that in the $C_1$ class of pulsars
with high levels of linear polarisation, the statistical relationship between
the two OPMs is characterised by $M \gg 1$ and $R > 1$. Using the $t$ and 
$lt_1$ modes as model for the two OPMs, and considering an average damping 
factor of the $lt_1$ mode as $\mu_{\zeta}$, we obtain the parameter $M = 
1/7\mu_{\zeta}$. If we now assume a reasonable value of $M \sim 20$ then 
$\mu_{\zeta} = 0.007$ for the $C_1$ class. To simulate the polarisation 
statistics we use a Gaussian distributed amplitude of the of the $t$-mode, $A$,
with mean $\mu_A = 1000$ and standard deviation $\sigma_A = 100$. The inherent 
amplitudes of the $lt_1$ mode distribution is scaled by a factor of 7. For the 
damping factor we also assume a normal distribution with mean $\mu_{\zeta} = 
0.007$ and $\sigma_{\zeta} = 0.0005$, while instrumental noise parameters added
to the distributions have zero mean and standard deviation $\sigma_N = 60$. An 
additional redistribution of the Stokes parameters U and Q has been carried out
by rotating the linear polarisation by PPA phase of $\psi_{\circ} = 66\degr$ in
order to match the observed behaviour of PSR J0138+5814. 
Fig.~\ref{fig:ccr_modl}(a) shows the fractional linear polarisation 
distribution of the simulated time samples which peaks around 85\% level and 
has a stretched leading side, symptomatic of the $C_1$ class. The PPA 
distribution is shown in Fig.~\ref{fig:ccr_modl}(c) and follows the expected 
behaviour of PSR J0138+5814~in Fig.~\ref{fig:B0136_band4}(c). \\

\noindent
\subsection{Statistical model of $C_2$ class of pulsars using CCR characteristics} 
\label{d2} 
In case of $C_2$ class of pulsars the statistical dependence between the two
OPM distributions is specified by relatively smaller values of $M$ and $R < 1$.
For example, if $M = 1$ then the mean value of the damping factor is 
significantly higher with $\mu_{\zeta} = 0.14$. We have simulated the 
polarisation statistics of a $C_2$ type pulsar using distribution of the 
$t$-mode with mean and standard deviations, $\mu_A = 2000$, $\sigma_A = 50$, 
the mean and standard distribution of the damping factor specified by 
$\mu_{\zeta} = 0.11$ and $\sigma_{\zeta} \sim 0.18$, and baseline noise levels
with zero mean and standard deviation $\sigma_N = 100$ added to the Stokes 
parameters. Fig.~\ref{fig:ccr_modl}(b) shows the distribution of the fractional
linear polarisation which has irregular shape with a tail, where the majority 
of time samples have low polarisation levels below 50\%. The variation of the
fractional polarisation as a function of the PPA is shown in 
Fig.~\ref{fig:ccr_modl}(d) where the two OPMs are clearly visible, the primary 
mode bunched around $\psi \sim -85\degr$ and the less dominant mode around 
$\psi \sim 5\degr$, and replicates the observed behaviour of PSR J2013+3845~in 
Fig.~\ref{fig:J2013_0.5deg}(b). \\

From the above discussion it is clear that the $lt_1$-mode gets significantly 
damped in $C_1$ class of pulsars compared to the $C_2$ sources, with more than 
one order of magnitude difference in the the intensities of the emergent modes 
from these two cases. The mechanism of escape of $lt_1$-mode from 
high-multiplicity plasma cloud to reach the low-multiplicity region is not well
established, however it is clear that this depends on the details of the 
inhomogeneities present in the outflowing plasma. One condition for the escape 
of $lt_1$-mode is that $\kappa$ in the low-multiplicity region should be such 
that the plasma frequency should be smaller than frequency of the generated 
radio wave, which in turn is determined by how the multiplicity profile changes
from high to low multiplicity region in the transverse direction. An additional
requirement for the escape of $lt_1$-mode is that the length scale over which 
$\kappa$ changes from high to low multiplicity region should be comparable to 
the wavelength of the radio emission in the plasma frame of reference. 
\cite{2023ApJ...952..151M} estimated the variation in the number of sparks in a
PSG with pulsar energetics and found the number to increase with increasing 
value of $\dot{E}$, while the characteristic size of the spark decreases. Using
suitable parameters describing the nature of PSG, like the strength of surface
magnetic field, the mean screening factor of the gap potential difference due
to ions, rotation period and its derivative, etc., \cite{2023ApJ...952..151M}
gives estimates of the $\dot{E}$ dependence of typical half of the transverse 
(across the open field lines) distance between the center of two neighbouring 
sparks in the wave generation region and the number of sparks in the IAR. For 
$\dot{E} \sim 10^{34}$ erg~s$^{-1}$, the transverse size is about 10$^6$ cm and
there are about 100 sparks. When such high number of sparks exist in the IAR, a
large number of high-multiplicity plasma clouds forms in the emission region 
and hence the likelihood of the $lt_1$-mode escaping the medium goes down, due
to several likely reasons. For example, as the $lt_1$-mode reaches the boundary
there is higher probability of encountering an adjacent cloud that will damp 
the mode. Another possibility is that the multiplicity of plasma in the 
boundary remains high as they have contribution from more adjacent sparks, such
that the escape condition for the $lt_1$-mode cannot be satisfied. 
\cite{2023ApJ...952..151M} showed that $\dot{E} \sim 5\times 10^{33}$ 
erg~s$^{-1}$ acts as a rough limit below which the $lt_1$ mode will not be 
completely damped, which can also be considered the boundary above which the 
$C_1$ class of pulsars are observed.

\section{Conclusion}
We report detailed measurements of the polarisation behaviour in the radio 
emission of energetic pulsars at the relatively unexplored frequency range of 
300--750 MHz. We have estimated the average profile as well as single pulse 
properties of 35 pulsars with $\dot{E} > 5 \times 10^{33}$ erg~s$^{-1}$. At 
meterwavelengths the mean levels of average linear polarisation fractions in 
the energetic pulsars are higher than the less energetic cases, similar to 
higher frequencies, although there is considerable spread of polarisation 
levels within this group. The emission behaviour follow two distinct 
categories, in the first case the average linear polarisation fraction exceeds 
70\% level and the PPA distribution is usually bunched around a single 
RVM-like track. The second group has lower polarisation level and the PPA 
behaviour is more scattered, sometimes showing two OPMs with RVM-like tracks 
separated by 90$\degr$ phase. We also find evidence that even in the highly 
polarised pulsars there exist time samples, with low levels of polarisation, 
that show presence of the second OPM. 

The observed polarisation features can be reproduced from incoherent mixing of 
two fully polarised, orthogonal modes having stochastic distributions. In the 
strongly magnetized pair plasma medium in pulsars, two such modes, the $t$-mode
and the $lt_1$-mode, polarised perpendicular and parallel to magnetic field 
line planes, are excited by CCR from charge separated envelope solitons.
The two modes have different refractive indices in the pulsar medium and hence 
split during propagation and travel independently as fully polarised modes. 
Although the $lt_1$-mode has seven times more intensity than the $t$-mode, they
are ducted along the magnetic field lines and are expected to get damped in the 
medium. The electromagnetic $t$-mode is the viable source of the single 
polarisation mode seen in one category of energetic pulsars. However, the 
$lt_1$-mode can enter the low-multiplicity region between plasma clouds and 
escape as an electromagnetic mode that emerges as the second OPM seen in 
emission of the other group of pulsars. We find that the CCR from charge 
separated envelope solitons developing in the pulsar plasma is able to 
explain the variations in the observed polarisation behaviour of highly 
energetic pulsars studied in this work.

\begin{acknowledgments}
We thank the referee for comments that helped to improve the
manuscript. We thank the staff of the GMRT that made these observations presented in this paper possible. GMRT is run by the National Centre for Radio Astrophysics of the Tata Institute of Fundamental Research.
D.M. acknowledges the support of the Department of Atomic Energy, Government of
India, under project No. 12-R\&D-TFR-5.02-0700. This work was supported by the
grant 2020/37/B/ST9/02215 of the National Science Center, Poland.
\end{acknowledgments}




\bibliography{References}{}
\bibliographystyle{aasjournal}

\appendix

\section{Average profile features of energetic pulsars} \label{sec:app}

\noindent
{\bf PSR J0117+5914 (B0114+58) :} The pulsar has a single component profile 
that gets scattered at frequencies below 300 MHz \citep{2016A&A...586A..92P}. 
Polarization observations from this source has been reported between the 
frequencies of 400 MHz and 1.6 GHz, and no significant frequency evolution of 
polarisation properties is apparent \citep{1998MNRAS.301..235G,
2009ApJS..181..557H}. Although the value of $\dot{E} \sim 2 \times 10^{35}$ 
erg~s$^{-1}$ is high, the fractional linear polarisation is around 20\% which 
is relatively low compared to other energetic pulsars. The sensitive GMRT 
observations have revealed the PPA traverse of this source that exhibit a 
prominent single RVM-like track. \\

\noindent
{\bf PSR J0139+5814 (B0136+57) :} The pulsar profile is scattered below 300 MHz 
\citep{2015A&A...576A..62N}, while at higher frequencies the profile has a 
bright leading component and hint of a merged trailing component. The 
polarisation behaviour has been reported over a wide frequency range between
150 MHz and 4.8 GHz \citep{1998MNRAS.301..235G,1997A&AS..126..121V,
2015A&A...576A..62N}. Above 300 MHz the pulsar shows high levels of linear 
polarisation, with polarisation fraction $> 70$\%, with the PPAs distributed 
along a single polarisation mode. The GMRT measurements between 300--750 MHz 
have significantly higher sensitivity and largely follows the previous results.
The average PPA distribution shows deviations from RVM behaviour in the 
trailing parts of the profile which may be due to presence of OPMs, however 
rest of the emission window follows the RVM nature. \\

\noindent
{\bf PSR J1637$-$4553 :} There is a weak inter-pulse emission in the average 
profile and the polarisation properties have been previously reported at 
frequencies above 1 GHz \citep[see e.g.][]{2005MNRAS.359..481K,
2008MNRAS.391.1210W,2018MNRAS.474.4629J}. The main pulse has a single component
and shows high levels of linear polarisation, while the inter-pulse is weakly 
polarised. The underlying PPA traverse remains relatively flat across the 
main pulse. At both the observing frequency bands of GMRT the profile shows
scattering, although the high levels of linear polarisation and flat PPA 
traverse across main pulse is still evident. We find moderately high levels
of around 50\% fractional linear polarisation in the inter-pulse at the 550--750
MHz frequency band. \\

\noindent
{\bf PSR J1705$-$3950 :} The 1.4 GHz polarisation observation of this pulsar 
show two clear components in the profile that are close to 100\% linearly 
polarised \citep{2018MNRAS.474.4629J}. We have detected the pulsar only at the 
550-750 MHz frequency band of GMRT, where the profile is affected due to 
scattering. Although, the two components and high levels of linear polarisation
are also clear from our measurements. \\

\noindent
{\bf PSR J1709$-$4429 (B1706$-$44) :} The average polarisation behaviour of 
this pulsar have been studied over wide frequencies ranging from 950 MHz to 8.4
GHz \citep{2021MNRAS.504..228S,2006MNRAS.369.1916J,2018MNRAS.474.4629J}. The 
average profile has a single component with high levels of close to 100\% 
linear polarisation across the entire frequency range. At both frequency bands
of GMRT the pulsar remains highly polarised and the PPA traverse closely 
follows the RVM tracks. \\

\noindent
{\bf PSR J1718$-$3825 :} The 1.4 GHz profile reported in 
\citet{2018MNRAS.474.4629J} shows that the pulsar has a complex profile with 
three identifiable components and has close to 100\% linear polarisation.
In the GMRT observations at the 550-750 MHz band the leading and trailing 
components of the profile could be distinguished despite low detection 
sensitivity. There is high levels of linear polarisation and the underlying PPA
traverse closely follow the RVM nature. \\

\noindent
{\bf PSR J1733$-$3716 (B1730$-$37) :} The pulsar profile has two well separated 
components in the average profile, measured at 618 MHz and 1.4 GHz, that 
exhibit high levels of linear polarisation \citep{2016ApJ...833...28M,
2018MNRAS.474.4629J}. However, the linear polarisation levels diminish at the 
higher frequency of 3.1 GHz \citep{2005MNRAS.359..481K}. The 300-500 MHz GMRT
measurements show scattering in the profile despite the two components being
clearly visible with high levels of linear polarisation. At the higher 
frequency band, 550-750 MHz, the leading component is brighter and has close to
100\% linear polarisation. The RVM nature is also evident in the PPA traverse 
across the profile. \\

\noindent
{\bf PSR J1739$-$2903 (B1736$-$29) :} The pulsar has been observed at 610 MHz 
and 1.4 GHz, and has a prominent inter-pulse emission in the average profile
\citep{1998MNRAS.301..235G,1998MNRAS.295..280M,2010MNRAS.402..745K,
2016ApJ...833...28M,2018MNRAS.474.4629J}. Both the main pulse and inter-pulse 
has a single component, and while the main pulse is significantly de-polarised 
the inter-pulse has moderate levels of linearly polarisation. The detection 
sensitivity at the lower frequency band of GMRT is pretty low and the profile
is significantly scattered. The pulsar is prominently detected at 550-750 MHz 
frequency band, but the PPA do not show any clear pattern despite reports of
RVM fits at previous 1.4 GHz studies \citep{2010MNRAS.402..745K}. \\

\noindent
{\bf PSR J1739$-$3023 :} The pulsar has a single component profile and is 
highly polarised at 1.4 GHz \citep{2018MNRAS.474.4629J}. We have detected this 
object at 550--750 MHz band of GMRT showing similar emission properties, and 
the PPA traverse closely following the RVM nature. \\

\noindent
{\bf J1740+1000 :} The pulsar has been extensively observed over a wide 
frequency range between 300 MHz and 4.8 GHz \citep{2019MNRAS.489.1543O,
2018MNRAS.474.4629J}. The average profile has three merged components which 
evolves with frequency, and shows close to 100\% linear polarisation across the
frequency range. The GMRT observations between 550--750 MHz is consistent with
other frequencies and the underlying PPA has RVM nature. \\

\noindent
{\bf PSR J1740$-$3015 (B1737$-$30) :} The average polarisation properties of 
the pulsar has been measured between 400 MHz and 8.3 GHz and has single 
component profile with moderately high levels of linear polarisation over this 
range \citep{1998MNRAS.301..235G,1998A&A...334..571V,2006MNRAS.369.1916J,
1995MNRAS.274..572Q,2018MNRAS.474.4629J}. The pulsar has been measured at the 
550--750 MHz frequency range of GMRT and exhibit comparable features with
other measurements. There is hint of scattering tail near the trailing edge of the profile and the PPA traverse deviates from the RVM nature. \\

\noindent
{\bf PSR J1757$-$2421 (B1754$-$24) :} The average polarisation profile at 1.4 
GHz has three distinct components with low levels of linear polarisation 
\citep{2018MNRAS.474.4629J}. At lower frequencies the archival measurements
have poor detection sensitivity \citep{1998MNRAS.301..235G} compared to the 
GMRT observations at the two frequency bands. The pulse profiles, particularly
at the 300-500 MHz band, are scattered with low fractional linear polarisation
and there is no clear structure in the PPA traverse. \\

\noindent
{\bf PSR J1803$-$2137 (B1800-21) :} The polarised profile at 1.4 GHz has three
distinct components with the trailing component possibly merged 
\citep{2018MNRAS.474.4629J}. The linear polarisation shows variation across the
emission window with the leading component having higher fractional 
polarisation. The pulsar profile transforms into a double component structure
with high levels of linear polarisation at 4.8 GHz \citep{1998A&A...334..571V}.
The lower frequency of observations with GMRT, between 300-500 MHz, has less
detection sensitivity, and after further averaging across the longitudes the 
trailing component is visible along with the weak leading component, also seen 
in the previous measurements at 400 MHz \citep{1998MNRAS.301..235G}. The higher 
detection sensitivity at the 550-750 MHz range reveals all three components of 
the profile, where the PPA traverse also shows RVM like characteristics. \\

\noindent
{\bf PSR J1809$-$1917 :} The observations of the average polarisation 
properties at 1.4 GHz show a two component profile with high levels of linear 
polarisation \citep{2018MNRAS.474.4629J}. The GMRT observations detected the 
pulsar at the 550-750 MHz frequency band but with lower signal to noise ratio. 
Averaging across the longitude range revealed the double component structure in
the profile. The pulsar also has high levels of linear polarisation at lower 
frequencies and the PPA traverse exhibit clear RVM nature. \\

\noindent
{\bf PSR J1824$-$1945 (B1821$-$19) :} The average polarisation profiles around 
1 GHz show that the pulsar has a bright component with hint of a merged 
emission feature on the leading side \citep{2018MNRAS.474.4629J,
2015MNRAS.453.4485F}. The average polarisation have negligible intensity in 
these measurements. The GMRT observations at the two frequency bands show the 
profiles to be highly scattered with low levels of linear polarisation and PPA 
being similarly affected by scattering. \\

\noindent
{\bf PSR J1826$-$1334 :} The higher frequency measurements of the pulsar 
emission between 1.4 GHz and 4.8 GHz show the average profiles to have two well 
separated components that are highly polarised \citep{1998MNRAS.301..235G,
1998A&A...334..571V,2018MNRAS.474.4629J}. The pulsar is measured at the 
550--750 MHz frequency band of GMRT but has low detection sensitivity and 
the profile is scattered. But despite the scattering effects the high levels of
linear polarisation is also observed at low frequencies. \\

\noindent
{\bf PSR J1835$-$1020 :} The pulsar average profile at 1.4 GHz has a single 
component structure with low polarisation levels \citep{2018MNRAS.474.4629J}.
The measured profile at 550--750 MHz also has a prominent single component
feature with relatively lower levels of linear polarisation. The PPA traverse 
across the profile follows a single RVM like track. \\

\noindent
{\bf PSR J1841$-$0345 :} The average polarisation profile at 1.4 GHz shows a 
single broad component with close to 100\% linear polarisation 
\citep{2018MNRAS.474.4629J}. The profile measurements at 550-750 MHz has low 
signal to noise ratio and has signatures of scattering. After averaging across
the longitudinal phase the high linear polarisation levels is detected, but 
no clear pattern is visible in the PPA traverse. \\

\noindent
{\bf PSR J1847$-$0402 (B1844$-$04) :} Observations at 400 MHz 
\citep{1998MNRAS.301..235G} show the pulsar profile to be scattered but at 
frequencies above 600 MHz the effect of scattering decreases, revealing two 
merged components \citep[also see][]{2016ApJ...833...28M}. The higher 
resolution observations of \cite{2018MNRAS.474.4629J} at 1.4 GHz reveal 
additional bifurcation in the trailing component. The pulsar has low levels of 
linear polarisation across all frequencies. The GMRT measurements at the 
550-750 MHz frequency band has high detection sensitivity with the profile 
showing the merged double component feature. The linear polarisation level is 
low, consistent with previous studies, while the PPA traverse show several 
orthogonal jumps and no clear fits for the RVM could be found. \\

\noindent
{\bf PSR J1856+0113 (B1853+01) :} Average polarisation observations at 330 MHz
show a single component scattered profile with very low levels of linear 
polarisation, while at 1 GHz the scattering effects are less prominent and  
the average profile has two components with moderate levels of polarisation 
\citep{2023MNRAS.520..314W}. There is significant evolution in the emission
characteristics at higher frequencies of 4.5 GHz, where the pulsar shows a 
single component profile with close to 100\% linear polarisation levels 
\citep{2019MNRAS.489.1543O}. The pulsar is detected at the at the 550-750 MHz
frequency range of GMRT, showing a double component profile with moderate
levels of linear polarisation and a relatively flat PPA traverse. \\

\noindent
{\bf PSR J1913+0904 :} The high sensitivity observations at 1.25 GHz with FAST 
radio telescope show the pulsar profile to have three components and high 
levels of linear polarisation \citep{2023RAA....23j4002W}. The GMRT 
observations detected the pulsar at 550--750 MHz frequency band, where a broad 
single component profile is seen with possible signatures of scattering. The 
profile shows high levels of linear polarisation and although the PPA traverse 
has been modelled using the RVM fits, there is possibility of scattering 
affecting these estimates. \\

\noindent
{\bf PSR J1915+1009 (B1913+10) :} The pulsar has been observed over a wide 
frequency range between 400 MHz and 4.8 GHz \citep{1998A&A...334..571V,
1998MNRAS.301..235G,2018MNRAS.474.4629J,2023MNRAS.519.3872R}. The average 
profile has a single component structure that gets scattered at frequencies 
below 600 MHz. The pulsar has moderate levels of linear polarisation which 
decreases with increasing frequency. In the GMRT observations at 300--500 MHz
frequency range the pulsar profile shows significant scattering, with moderate
levels of linear polarisation fraction where the PPA is affected by scattering. \\

\noindent
{\bf PSR J1921+0812 :} The pulsar has been observed at 1.25 GHz where a single 
component profile is seen with high levels of linear polarisation 
\citep{2023RAA....23j4002W}. The GMRT observations detected the pulsar at the 
550--750 MHz frequency band and the emission behaviour is similar to the higher 
frequency features. The underlying PPA traverse across the profile can be 
modelled using RVM. \\

\noindent
{\bf PSR J1922+1733 :} The FAST radio telescope observations of this pulsar at 
1.25 GHz show a double component profile, where the trailing component is
brighter, and moderately high levels of linear polarisation 
\citep{2023RAA....23j4002W}. The GMRT measurements at 550--750 MHz frequencies
has comparatively lower detection sensitivity and exhibit a single broad 
component profile shape with moderately high linear polarisation fractions. The
PPA traverse across the profile can be clearly estimated using RVM. \\

\noindent
{\bf PSR J1932+2020 (B1929+20) :} In this pulsar the average polarisation 
observations have been carried out at frequencies ranging from 400 MHz to 1.6 
GHz \citep{1998MNRAS.301..235G,2018MNRAS.474.4629J}. The average profile has a 
dominant single component structure, across all frequencies, while a weaker 
leading component is seen at the sensitive 1.25 GHz measurements using FAST 
\citep{2023RAA....23j4002W}. The profiles have low levels of fractional linear 
polarisation at these frequencies. The GMRT observations at 300--500 MHz 
frequencies show significant scattering, while at the higher frequency range, 
between 550--750 MHz, the profile shows some effect of scattering as well as 
the weak leading component. The pulsar exhibits low levels of polarisation in 
the GMRT measurements as well, and no clear pattern is seen in the PPA traverse
across the profile. \\ 

\noindent
{\bf PSR J1932+2220 (B1930+22) :} The average polarisation properties of this  
pulsar have been reported over a frequency range of 400 MHz to 1.6 GHz
\citep{1998MNRAS.301..235G,1999ApJS..121..171W,2023MNRAS.519.3872R}. The 
average profiles at most frequencies show a single bright leading component, 
while an extended weak trailing emission feature can also be seen in the high 
sensitivity FAST measurements at 1.25 GHz \citep{2023RAA....23j4002W}. The 
300-500 MHz observations with GMRT also show the single component structure as 
well as high levels of linear polarisation, which is also seen in earlier 
studies, and the corresponding PPA traverse has RVM characteristics. \\

\noindent
{\bf PSR J1935+2025 :} The pulsar profile has an inter-pulse emission and around
1 GHz both the main pulse and the inter-pulse have single component features 
with high linear polarisation \citep{2006MNRAS.369.1916J,2023MNRAS.519.3872R,
2023RAA....23j4002W}. The pulsar was detected at the 550-750 MHz frequency 
range of GMRT with similar physical characteristics, and the PPA traverse 
corresponding to the main pulse closely follow the RVM track. \\

\noindent
{\bf PSR J1938+2213 :} The pulsar has been observed at 1.25 GHz using the FAST
radio telescope and the profile has a bright central component surrounded by 
two weaker outer components with moderate levels of linear polarisation 
\citep{2023RAA....23j4002W}. The GMRT observations at the higher frequency 
range of 550--750 MHz also show the profile to comprise of a bright central 
component and weaker surrounding emission. The polarisation level is similar to
the earlier study and the PPA has prominent OPMs that cannot be accurately 
modelled using RVM fits. \\

\noindent
{\bf PSR J2002+3217 (B2000+32) :} Average polarisation observations from this 
source have been carried out between 300 MHz and 1.6 GHz, and show single 
component profiles with moderate levels of linear polarisation 
\citep{1998MNRAS.301..235G,2023MNRAS.519.3872R}. The low frequency measurement
has signature of scattering which is also apparent at the GMRT lower frequency 
band. At higher frequency band between 550-750 MHz the profile has similar 
characteristics to the archival studies, but the PPA is disordered near the 
trailing edge and deviates from RVM nature. \\

\noindent
{\bf PSR J2006+3102 :} The observation at 1.25 GHz shows a three component 
profile with high linear polarisation levels \citep{2023RAA....23j4002W}. The 
observations at the two GMRT frequency bands exhibit the previously reported 
nature as well as PPA traverse with RVM characteristics. \\

\noindent
{\bf PSR J2013+3845 (B2011+38) :} The average profile measurements from this
pulsar exist between 400 MHz to 4.8 GHz and has a single component with 
moderate levels of polarisation \citep{1998MNRAS.301..235G,1999PhDT........18H,
2023RAA....23j4002W}. The GMRT measurements at the two frequency bands have
similar properties with the PPA following RVM nature. \\

\noindent    
{\bf PSR J2043+2740 :} The profiles at 320 MHz and 1.25 GHz have two merged 
components with high levels of linear polarisation \citep{2023MNRAS.520..314W,
2023RAA....23j4002W}. The GMRT observations at the two frequency bands has 
bright single component profiles with possibly a merged trailing component. The
profiles are highly polarised and exhibit RVM like PPA traverse. \\

\noindent
{\bf PSR J2150+5247 (B2148+52) :} The average polarised profiles have been
measured between 400 MHz and 1.6 GHz, and show single component structure at
the lower frequencies that transforms into merged double structure at higher 
frequencies, with low levels of linear polarisation throughout 
\citep{1998MNRAS.301..235G}. The GMRT measurements show complex profile shapes
at both observing bands with several merged components, as well as low levels
of linear polarisation. The underlying PPA traverse shows presence of OPM at 
the 550-750 MHz band, that exabit RVM characteristics after taking into account
the deviations due to OPM. \\

\noindent
{\bf PSR J2229+6114 :} The GMRT observations at the 550-750 MHz frequency band
show a single component profile with low levels of linear polarisation and the 
underlying PPA closely aligned with the RVM nature. \\

\noindent
{\bf PSR J2337+6151 (B2334+61) :} Measurements of the average polarisation 
behaviour of this pulsar has been carried out between 400 MHz and 4.8 GHz 
\citep{1998MNRAS.301..235G,1998A&A...334..571V}. The profile shape evolves from
a merged double at the lowest frequencies, changing to well resolved double 
components and finally into a single component structure at the higher 
frequency end. The polarisation fraction also changes from low to moderate 
levels up to 1.6 GHz, but becomes highly polarised at 4.8 GHz. Our measurements
at the two GMRT frequency bands show merged triple structures with low levels
of linear polarisation. The PPA traverse shows evidence of OPM at the 300--500 
MHz band and at both frequencies follow the RVM tracks. 

\begin{figure}
\gridline{\fig{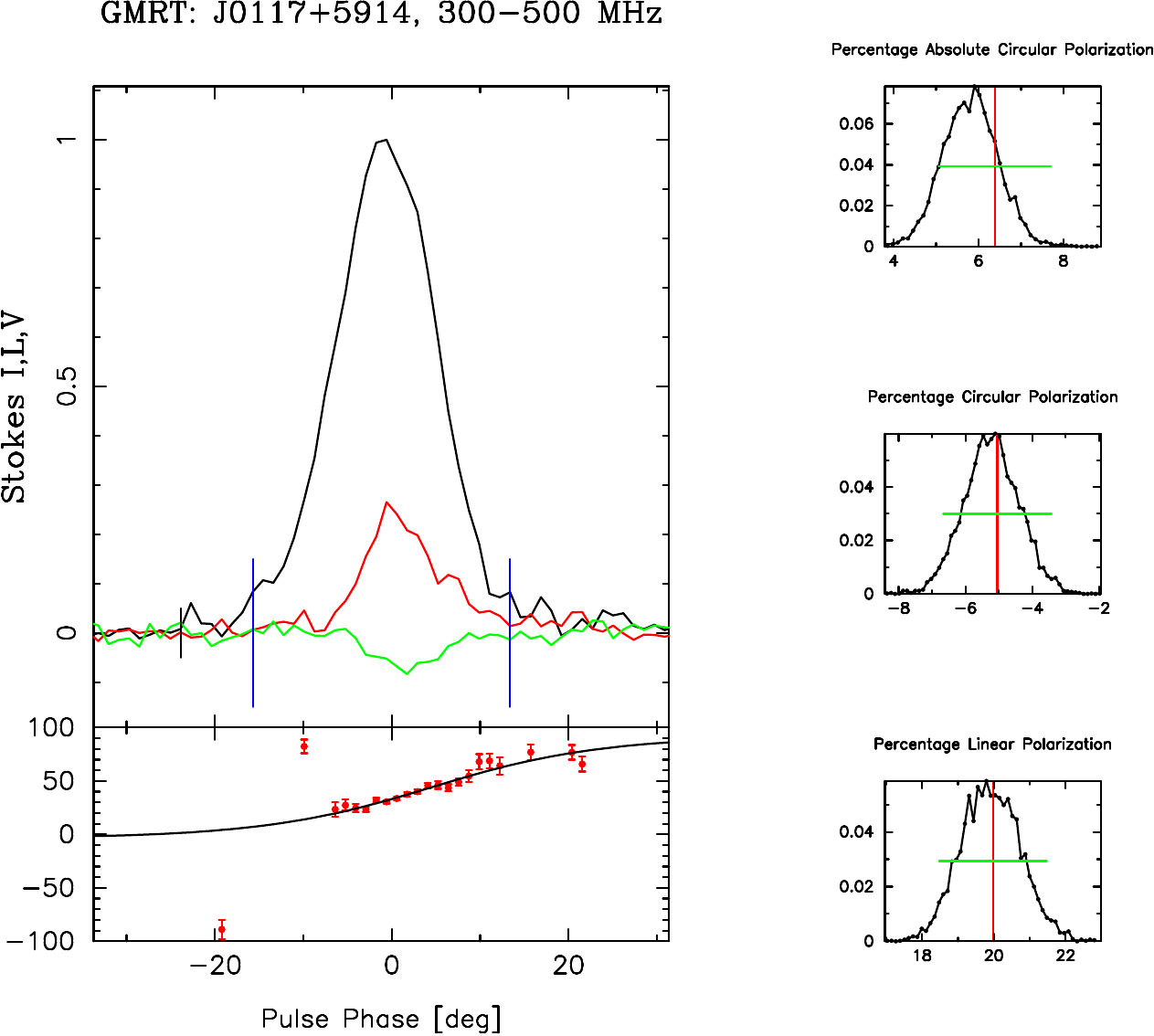}{0.3\textwidth}{(a) PSR J0117+5914 (300--500 MHz)}
          \fig{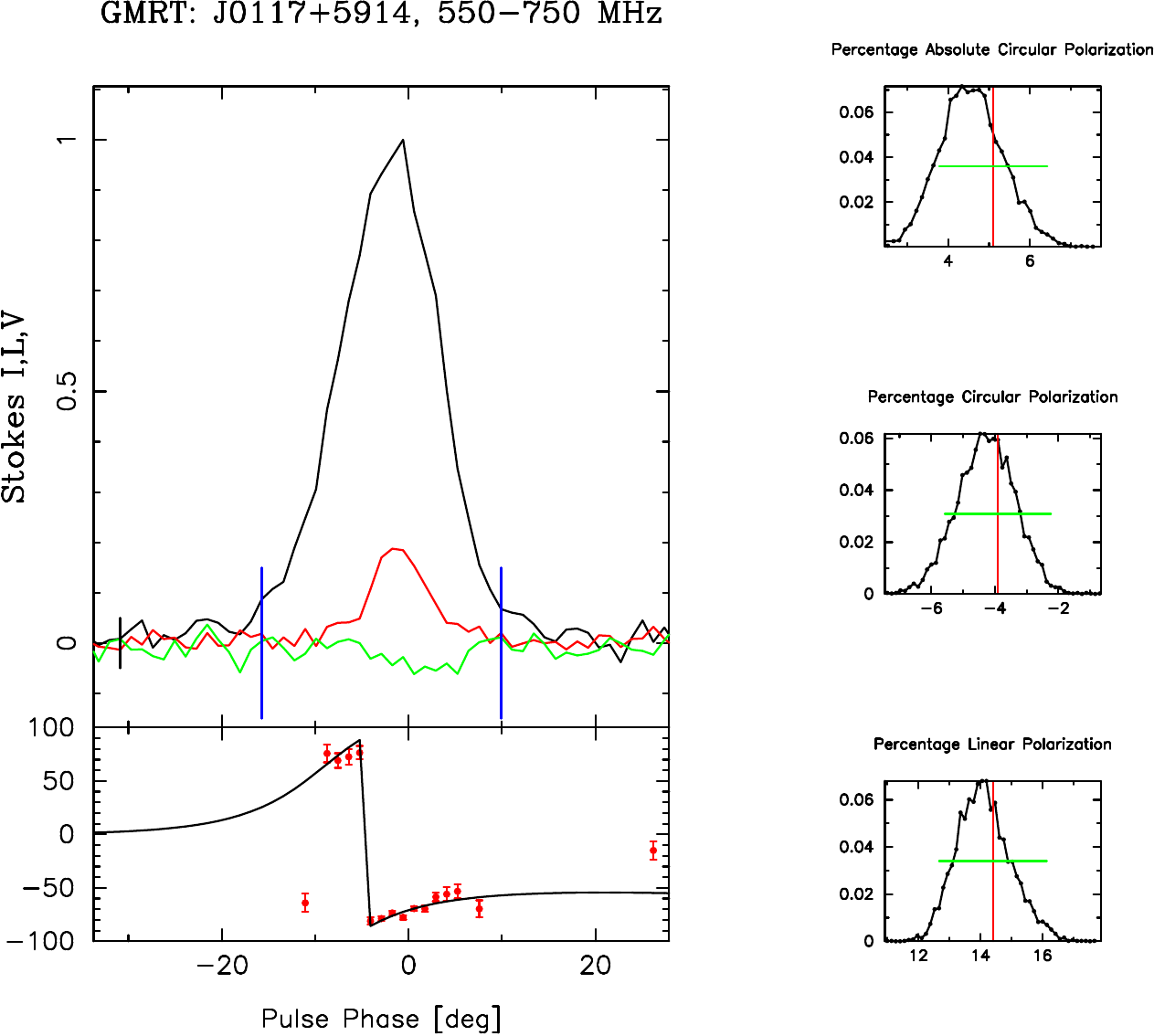}{0.3\textwidth}{(b) PSR J0117+5914 (550--750 MHz)}
          \fig{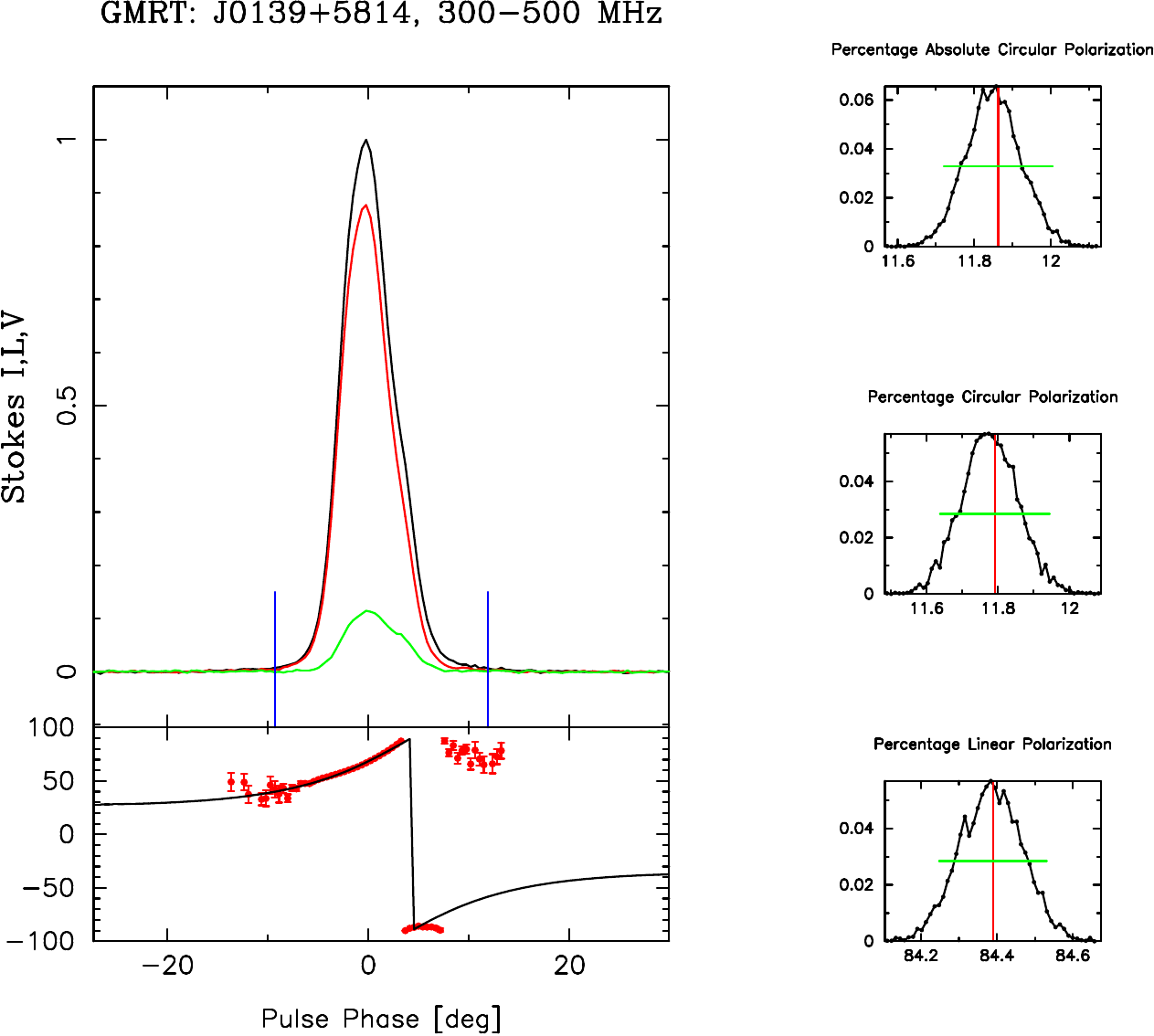}{0.3\textwidth}{(c) PSR J0139+5814 (300--500 MHz)}}
\gridline{\fig{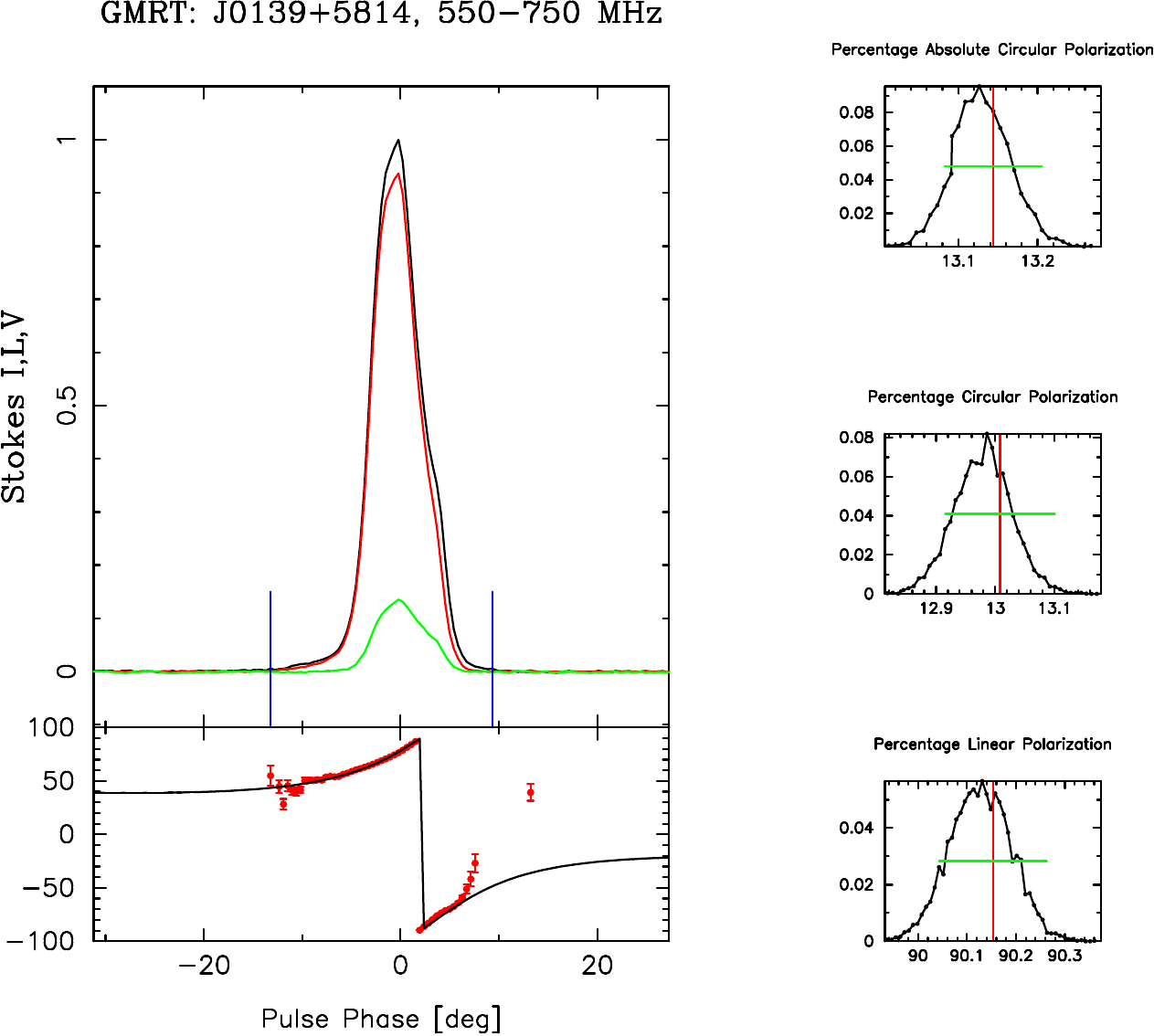}{0.3\textwidth}{(d) PSR J0139+5814 (550--750 MHz)}
          \fig{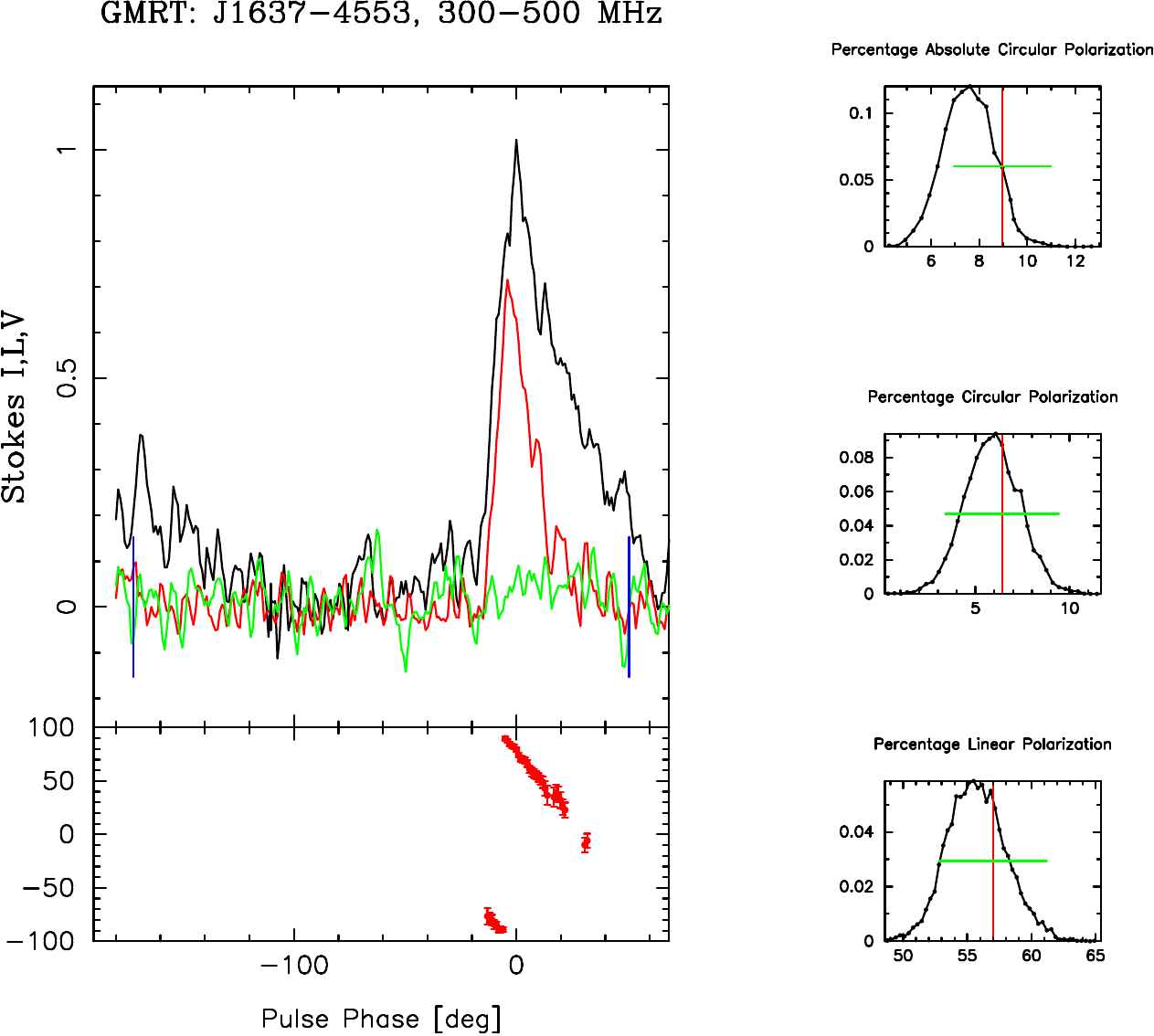}{0.3\textwidth}{(e) PSR J1637$-$4553 (300--500 MHz)}
          \fig{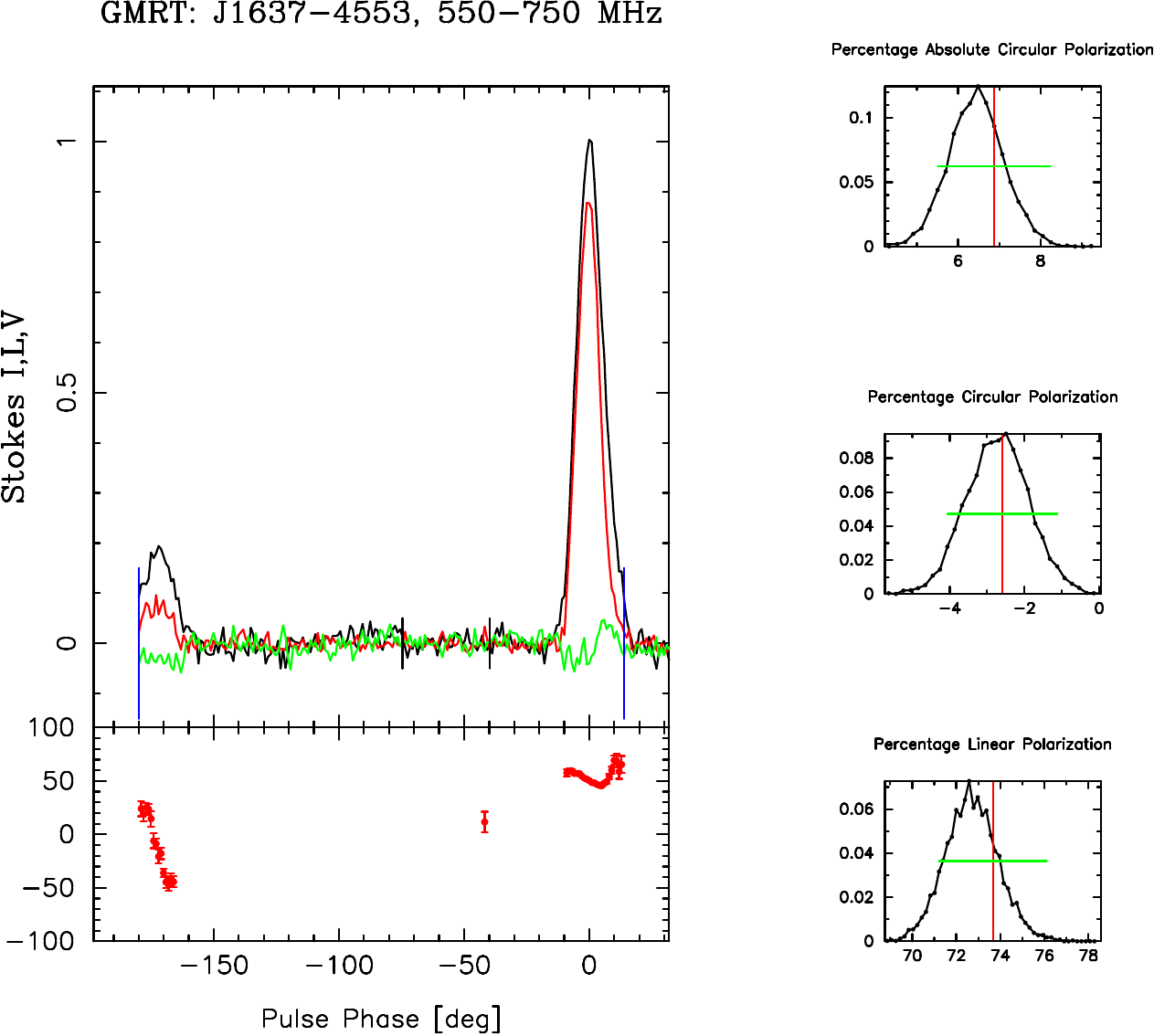}{0.3\textwidth}{(f) PSR J1637$-$4553 (550--750 MHz)}}
\gridline{\fig{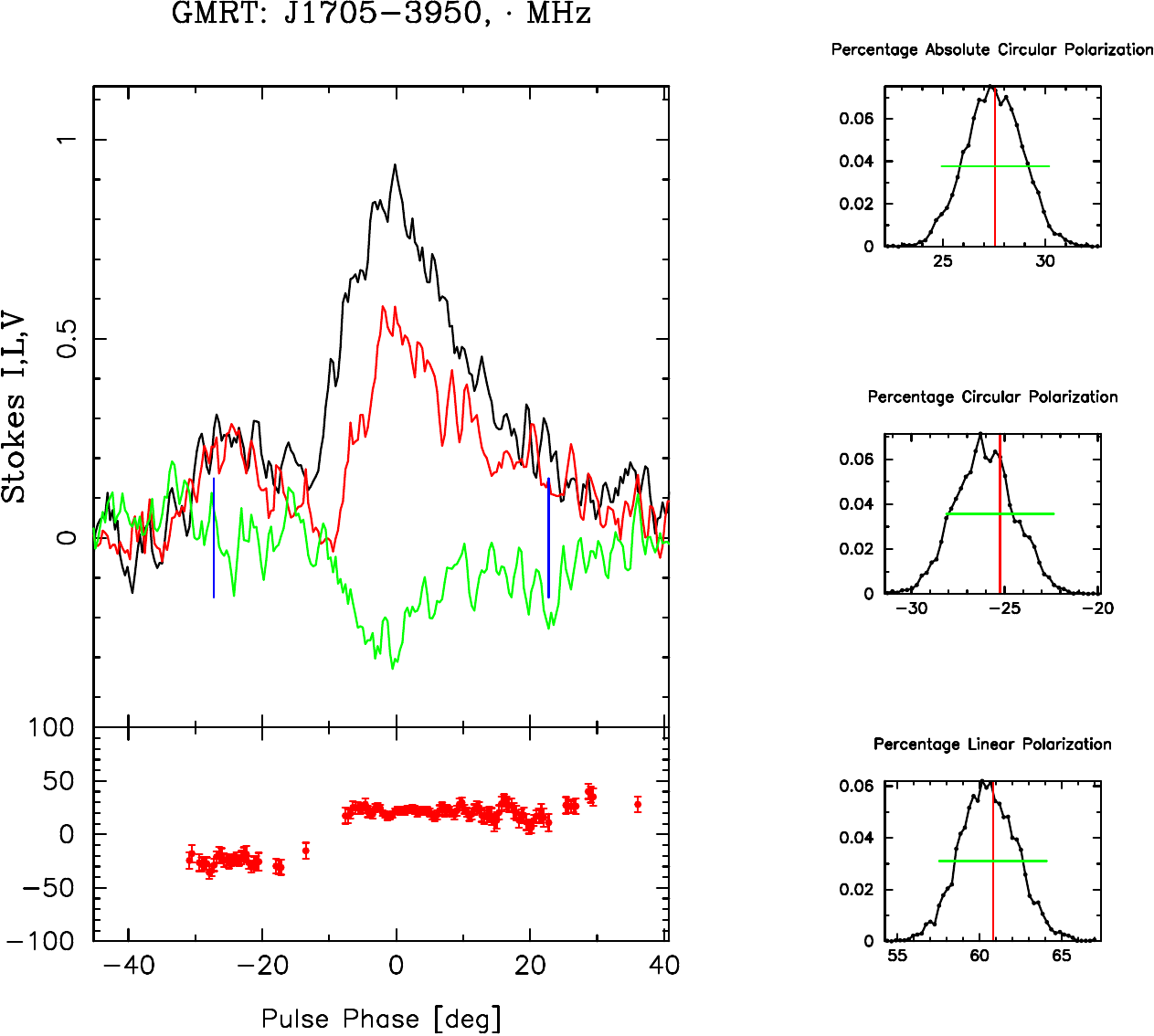}{0.3\textwidth}{(g) PSR J1705$-$3950 (550--750 MHz)}
          \fig{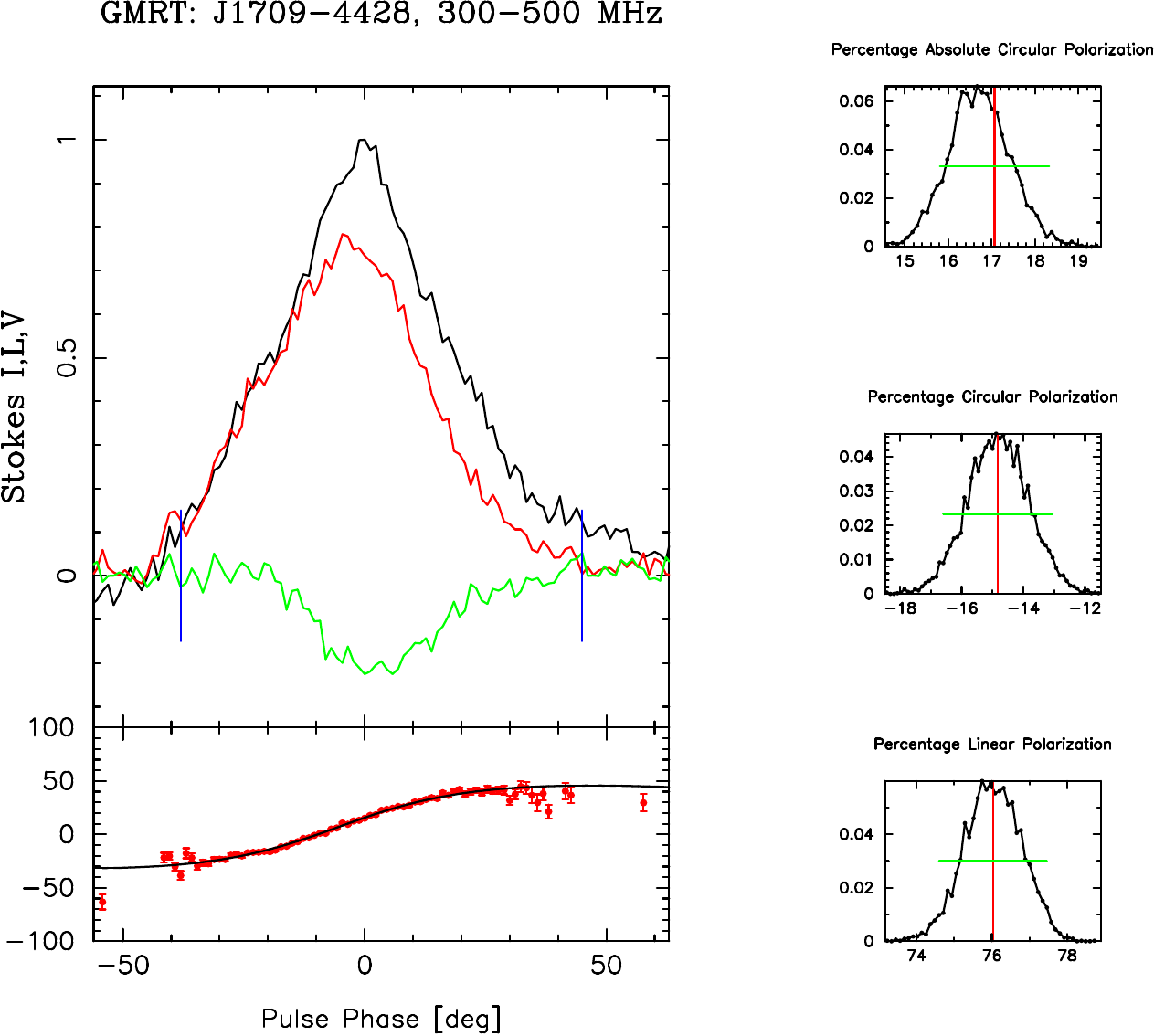}{0.3\textwidth}{(h) PSR J1709$-$4428 (300--500 MHz)}
          \fig{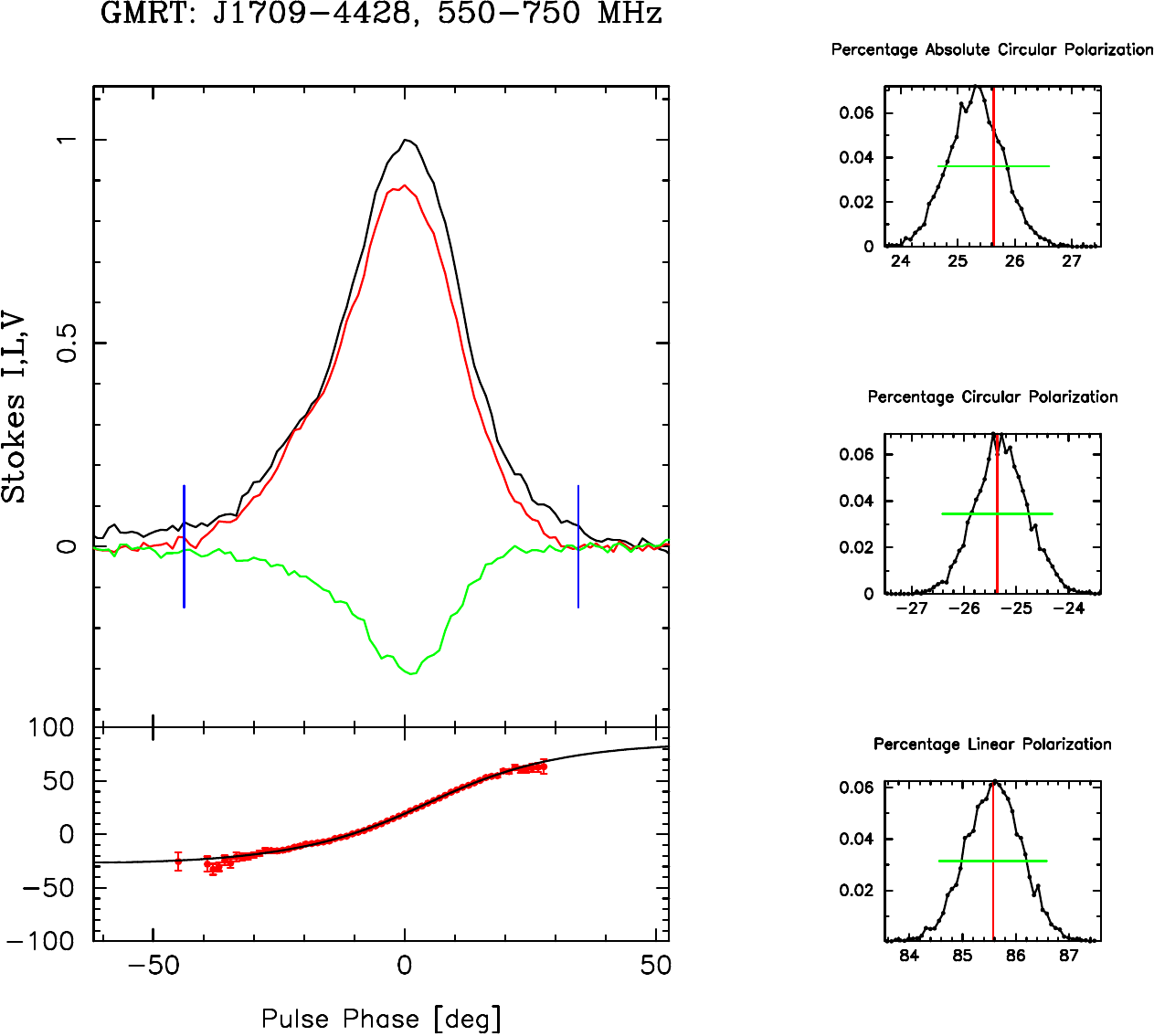}{0.3\textwidth}{(i) PSR J1709$-$4428 (550--750 MHz)}}
\gridline{\fig{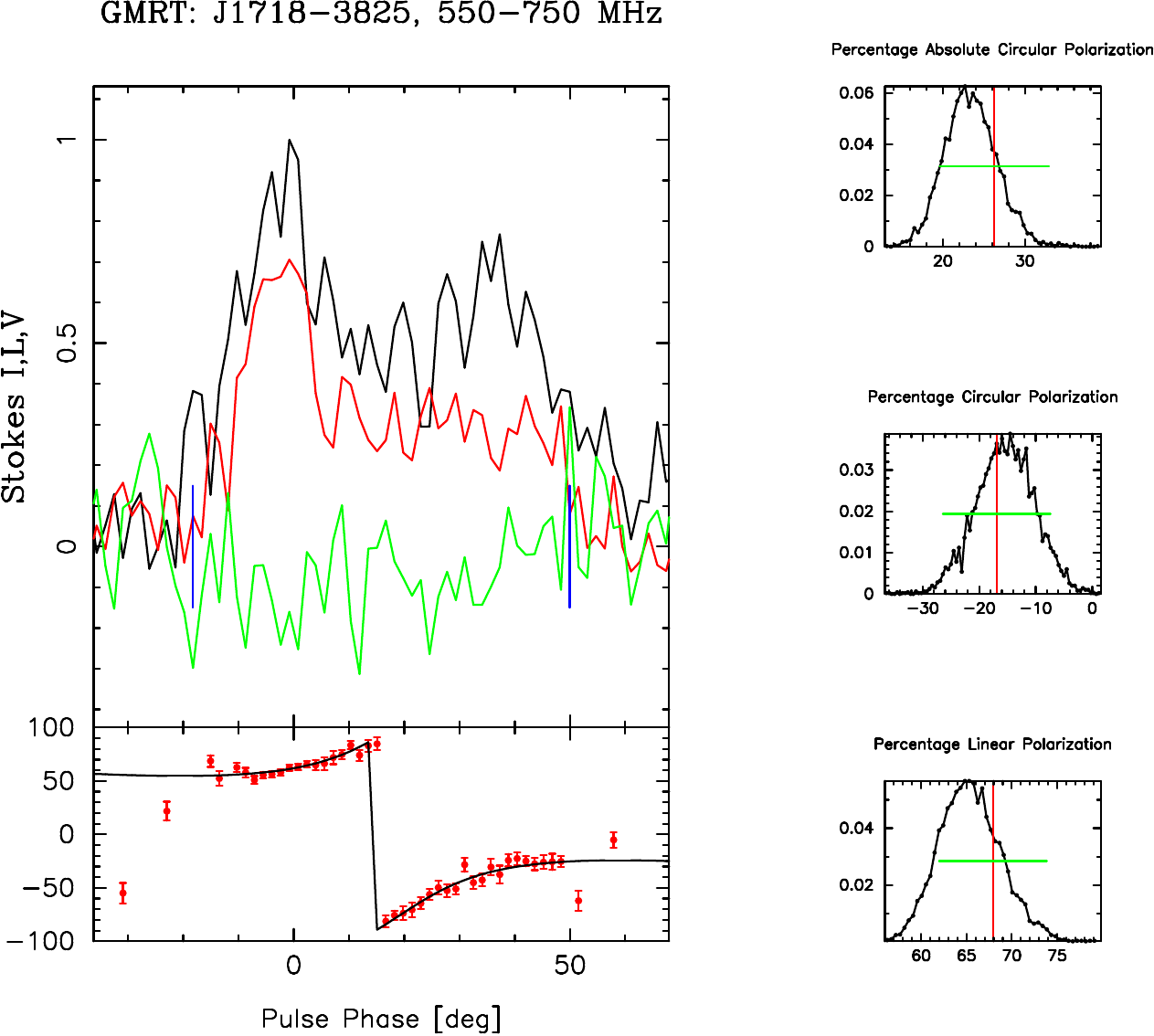}{0.3\textwidth}{(j) PSR J1718$-$3825 (550--750 MHz)}
          \fig{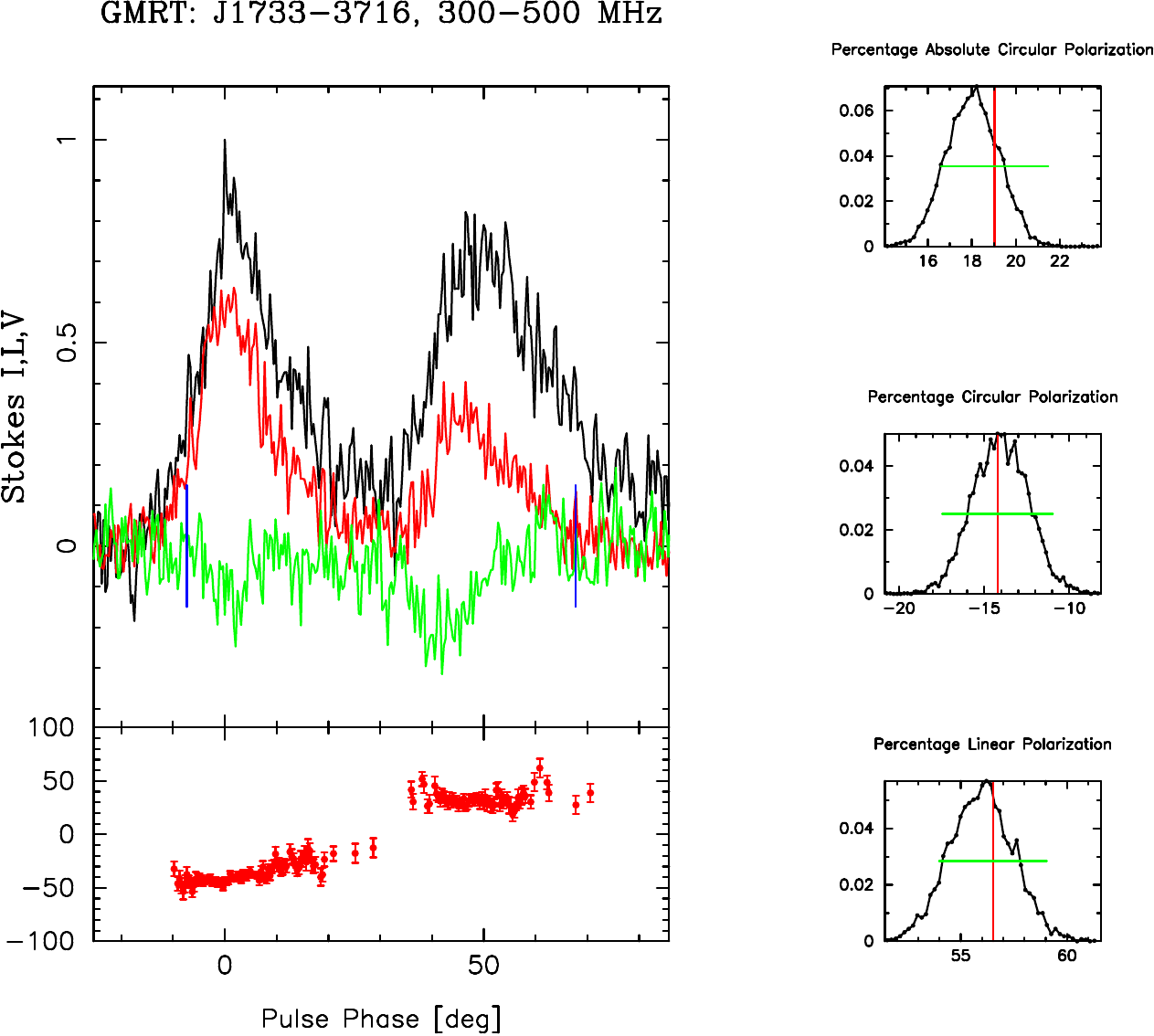}{0.3\textwidth}{(k) PSR J1733$-$3716 (300--500 MHz)}
          \fig{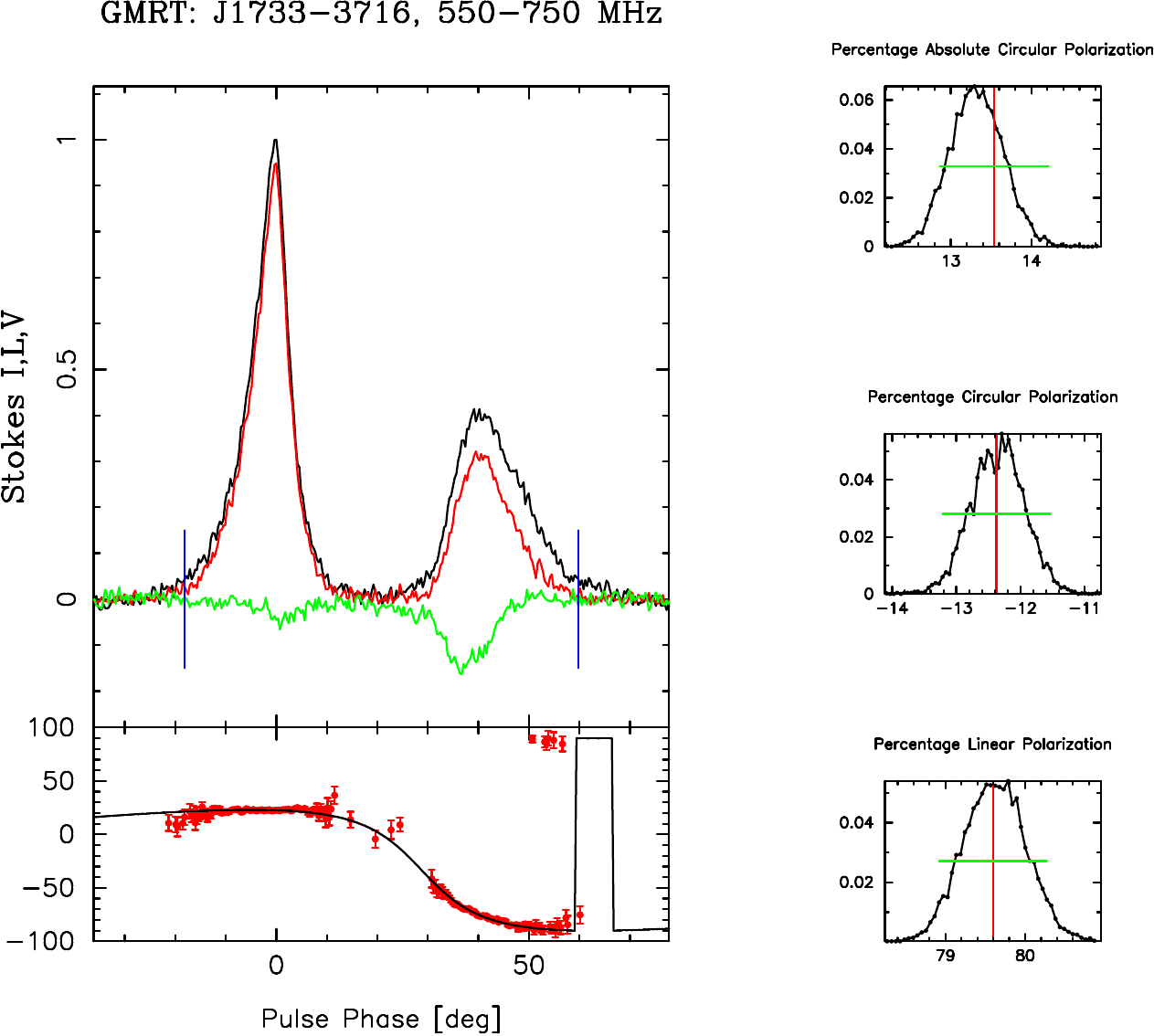}{0.3\textwidth}{(l) PSR J1733$-$3716 (550--750 MHz)}}
\label{avgp1}
\caption{See caption in Fig.~\ref{fig:avgpol}.}
\end{figure}

\begin{figure}
\gridline{\fig{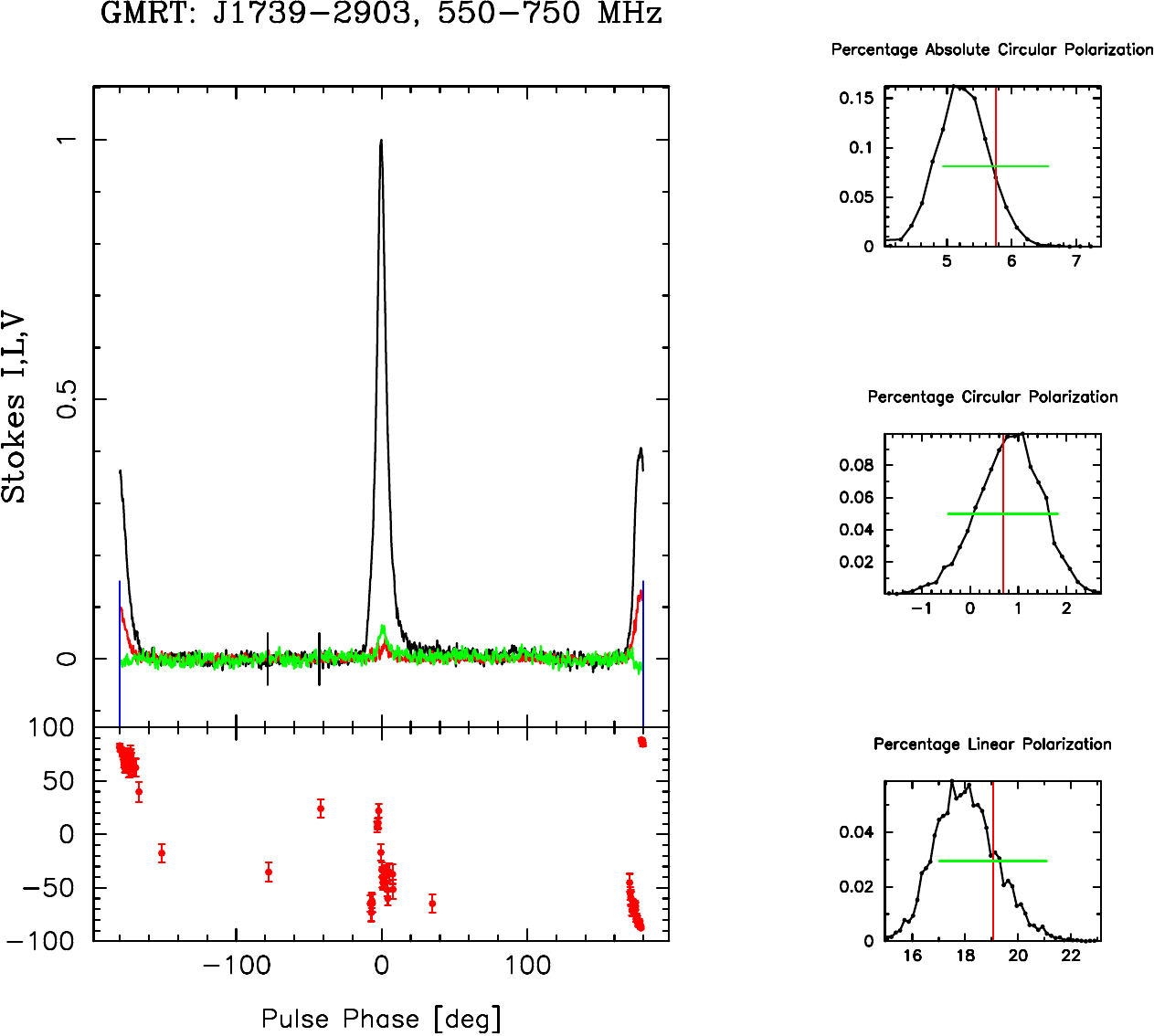}{0.3\textwidth}{(a) PSR J1739$-$2903 (550--750 MHz)}
          \fig{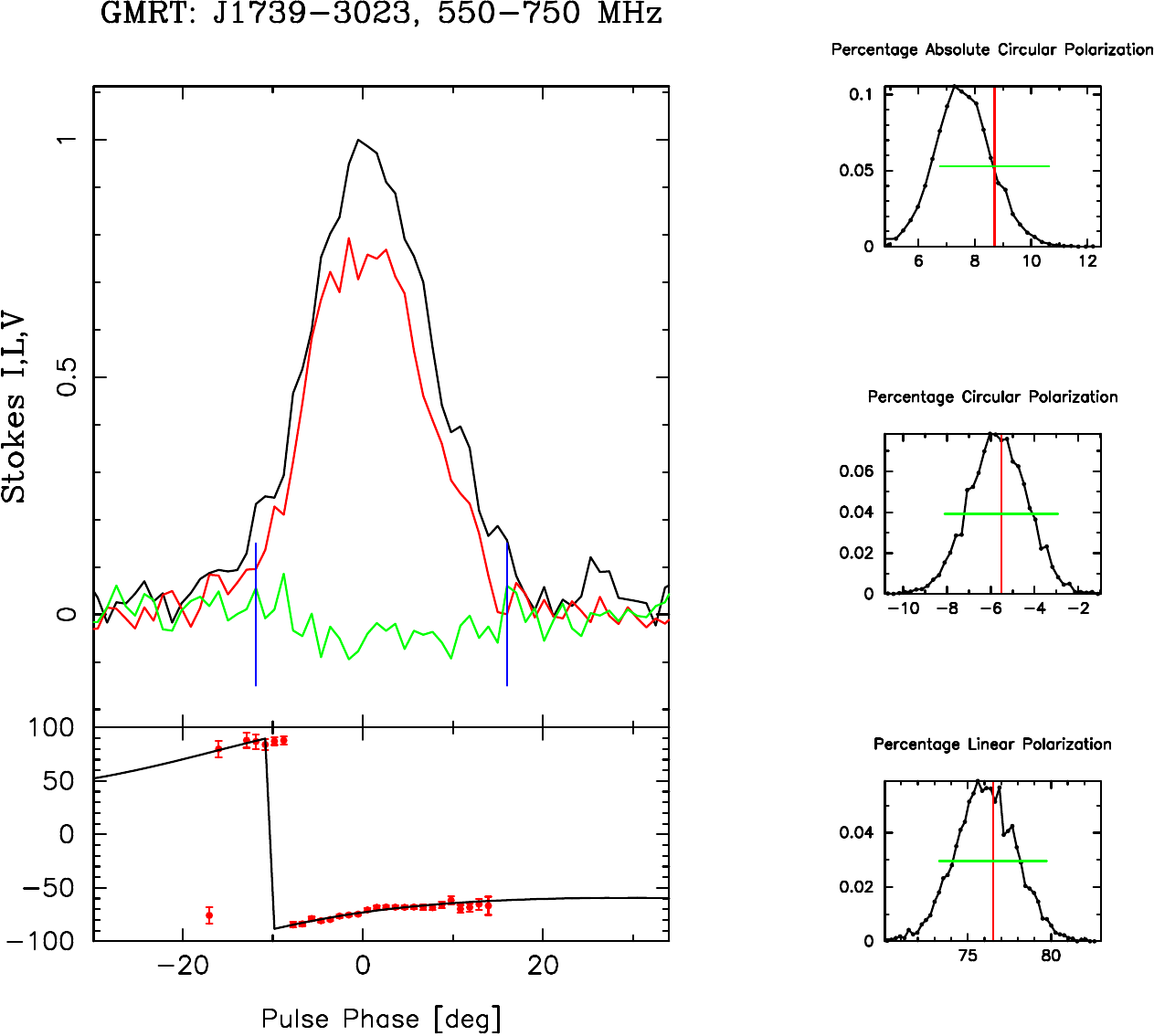}{0.3\textwidth}{(b) PSR J1739$-$3023 (550--750 MHz)}
          \fig{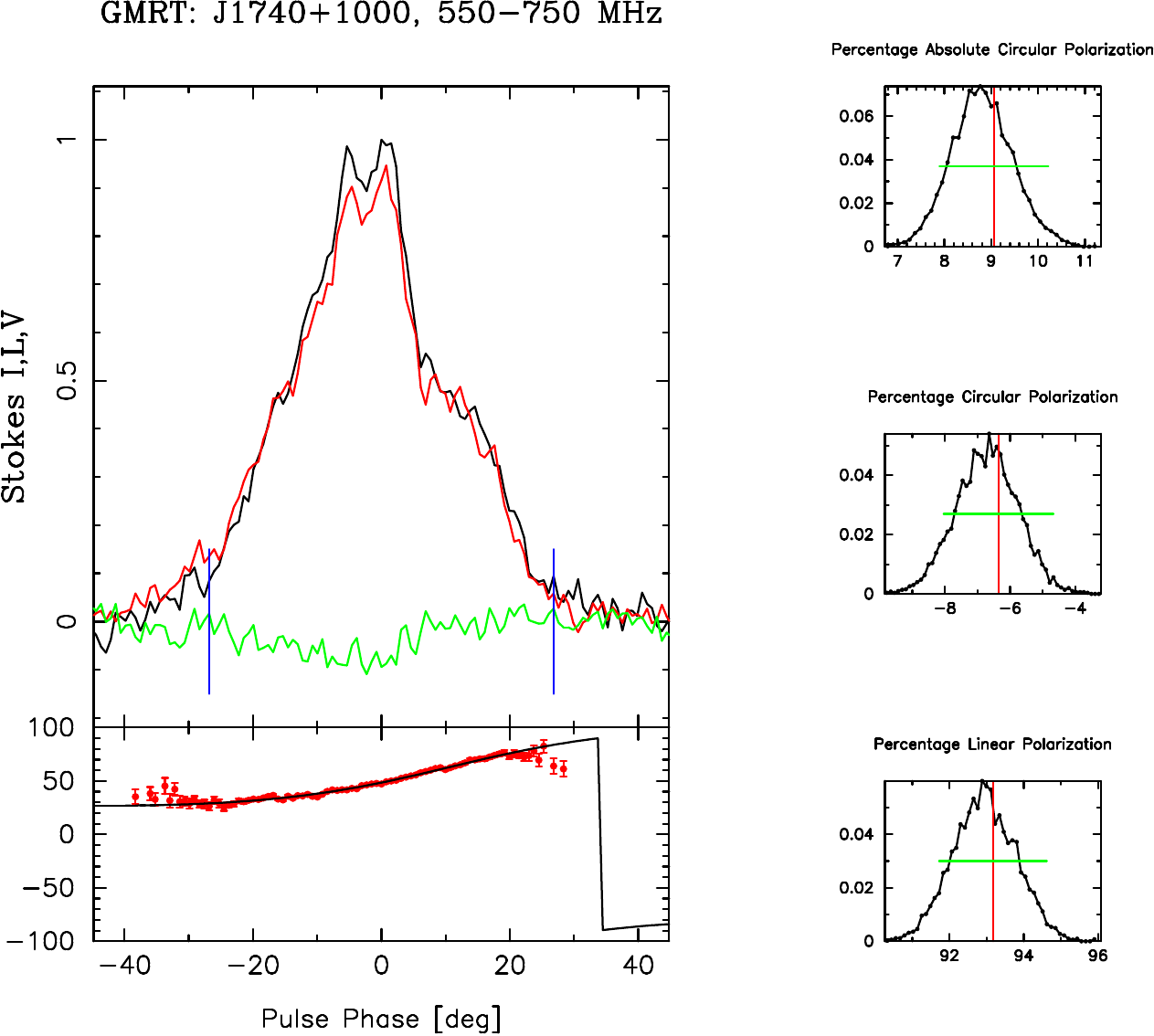}{0.3\textwidth}{(c) PSR J1740+1000 (550--750 MHz)}}
\gridline{\fig{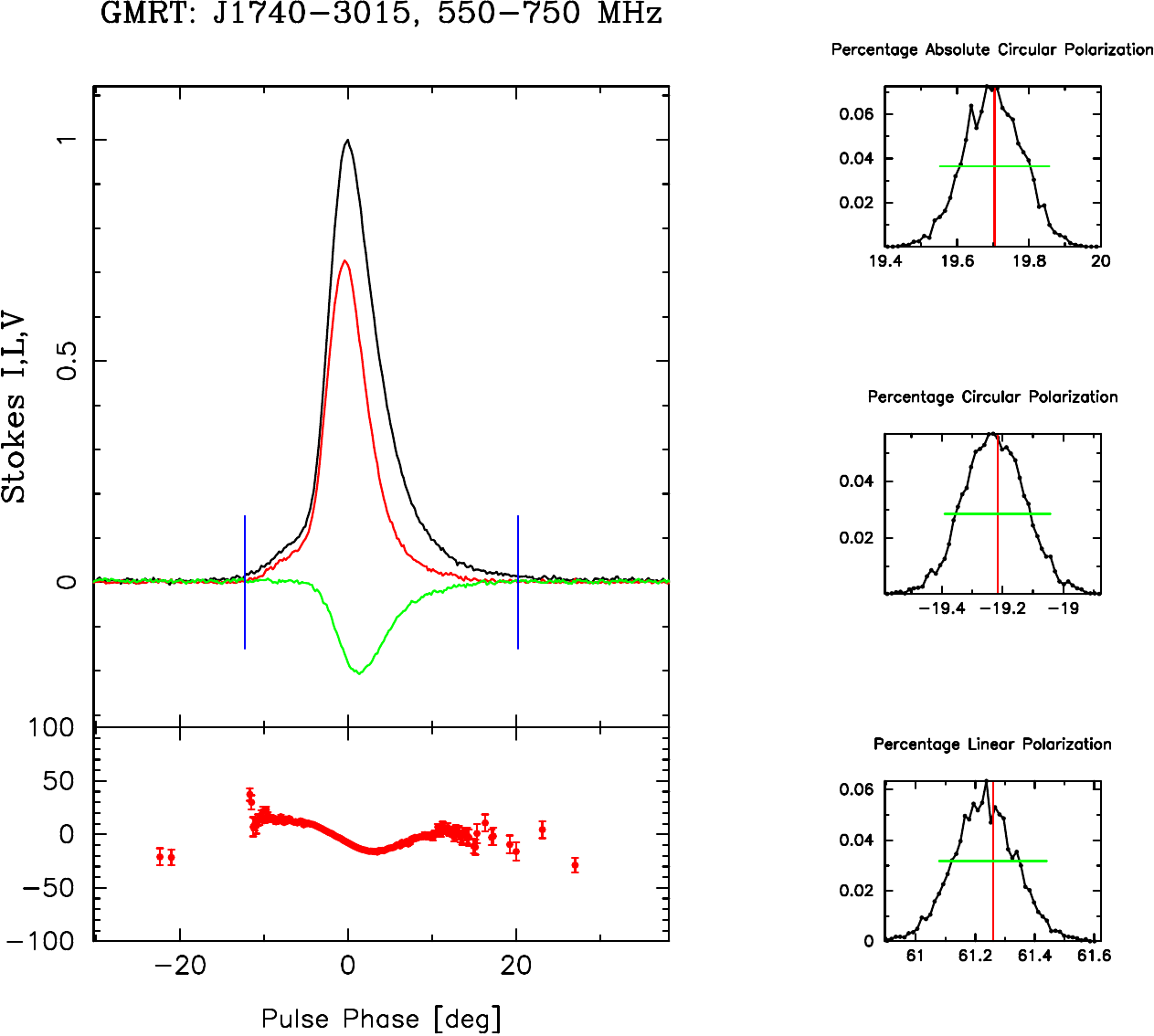}{0.3\textwidth}{(d) PSR J1740$-$3015 (550--750 MHz)}
          \fig{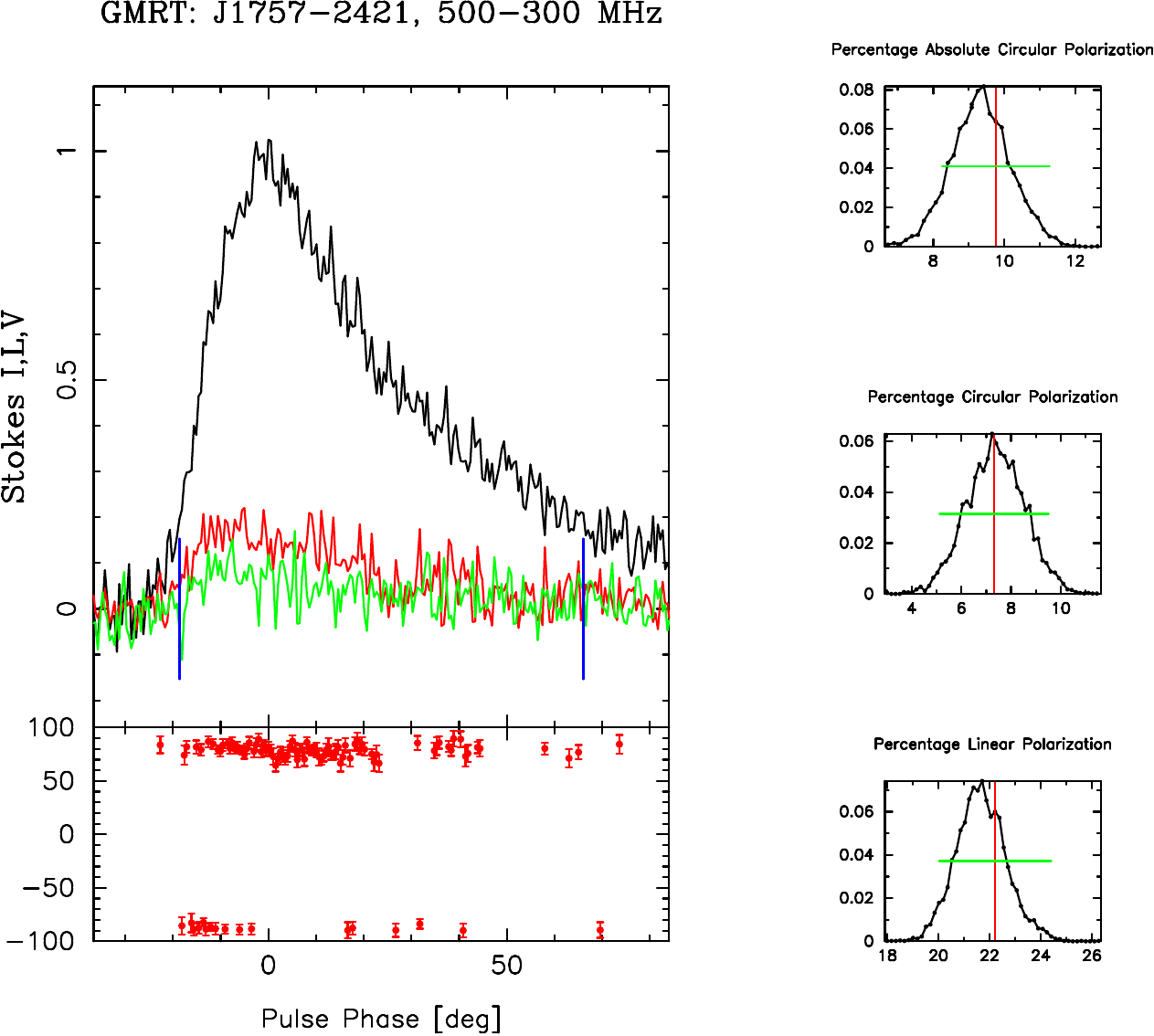}{0.3\textwidth}{(e) PSR J1757$-$2421 (300--500 MHz)}
          \fig{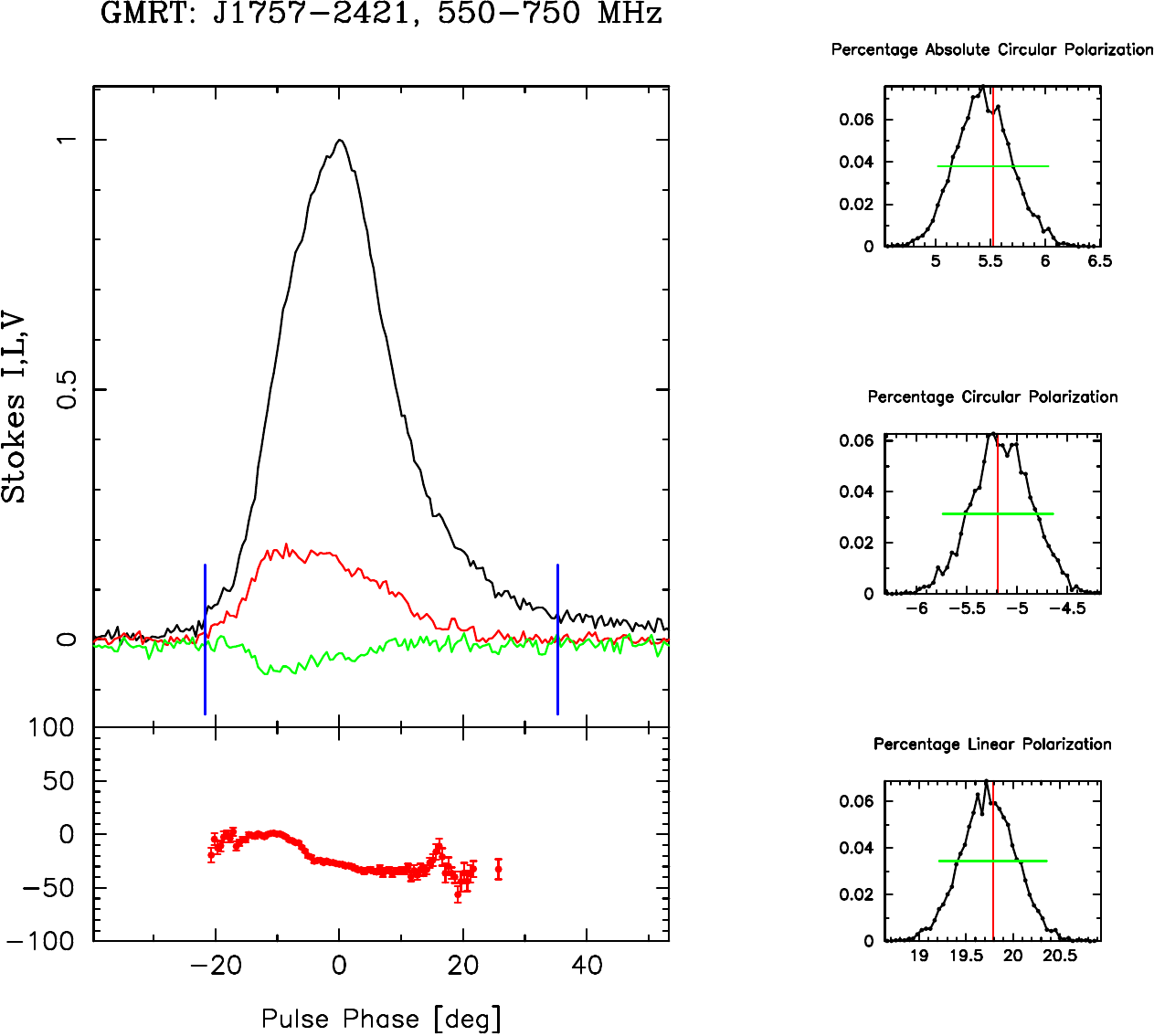}{0.3\textwidth}{(f) PSR J1757$-$2421 (550--750 MHz)}}
\gridline{\fig{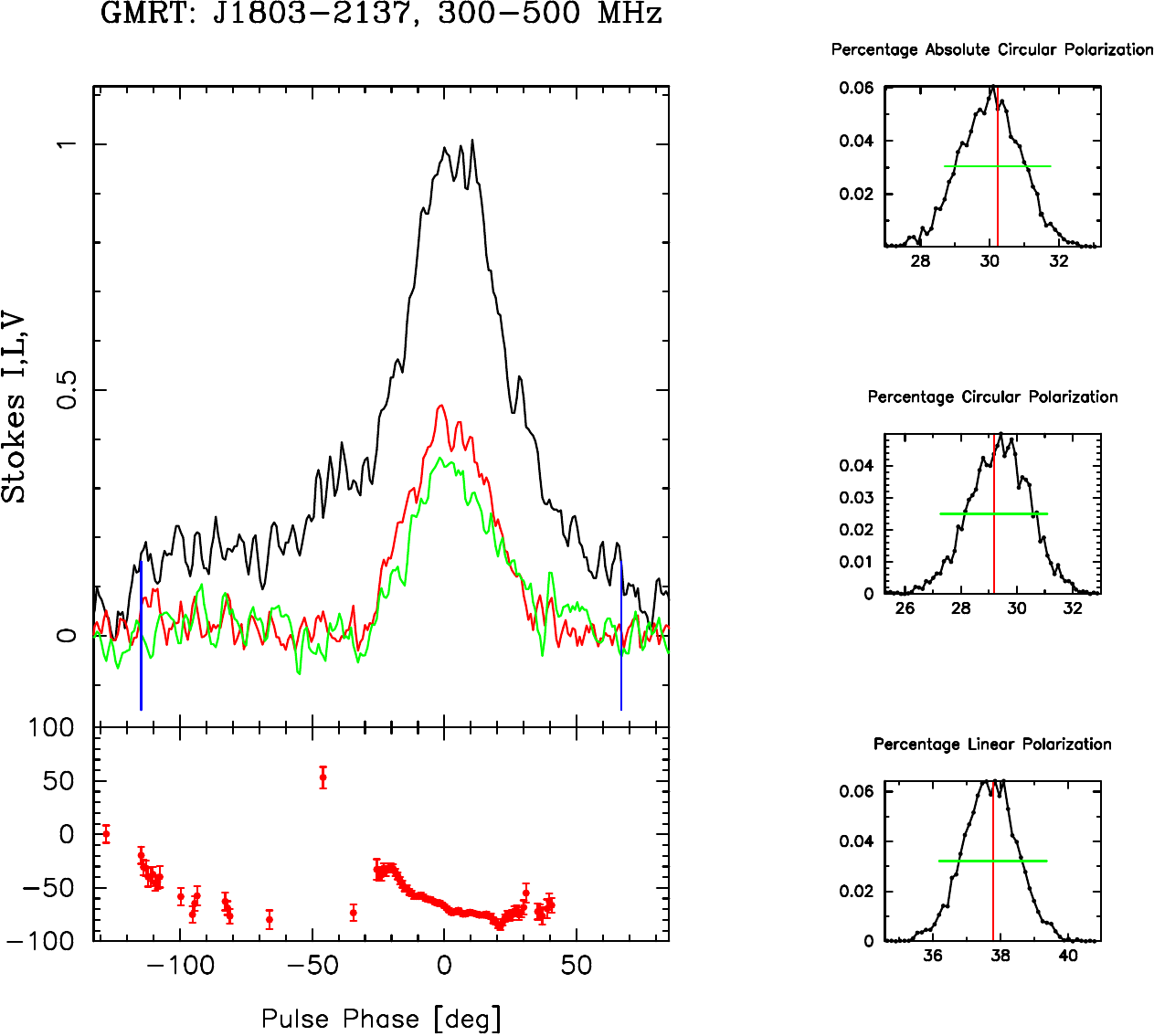}{0.3\textwidth}{(g) PSR J1803$-$2137 (300--500 MHz)}
          \fig{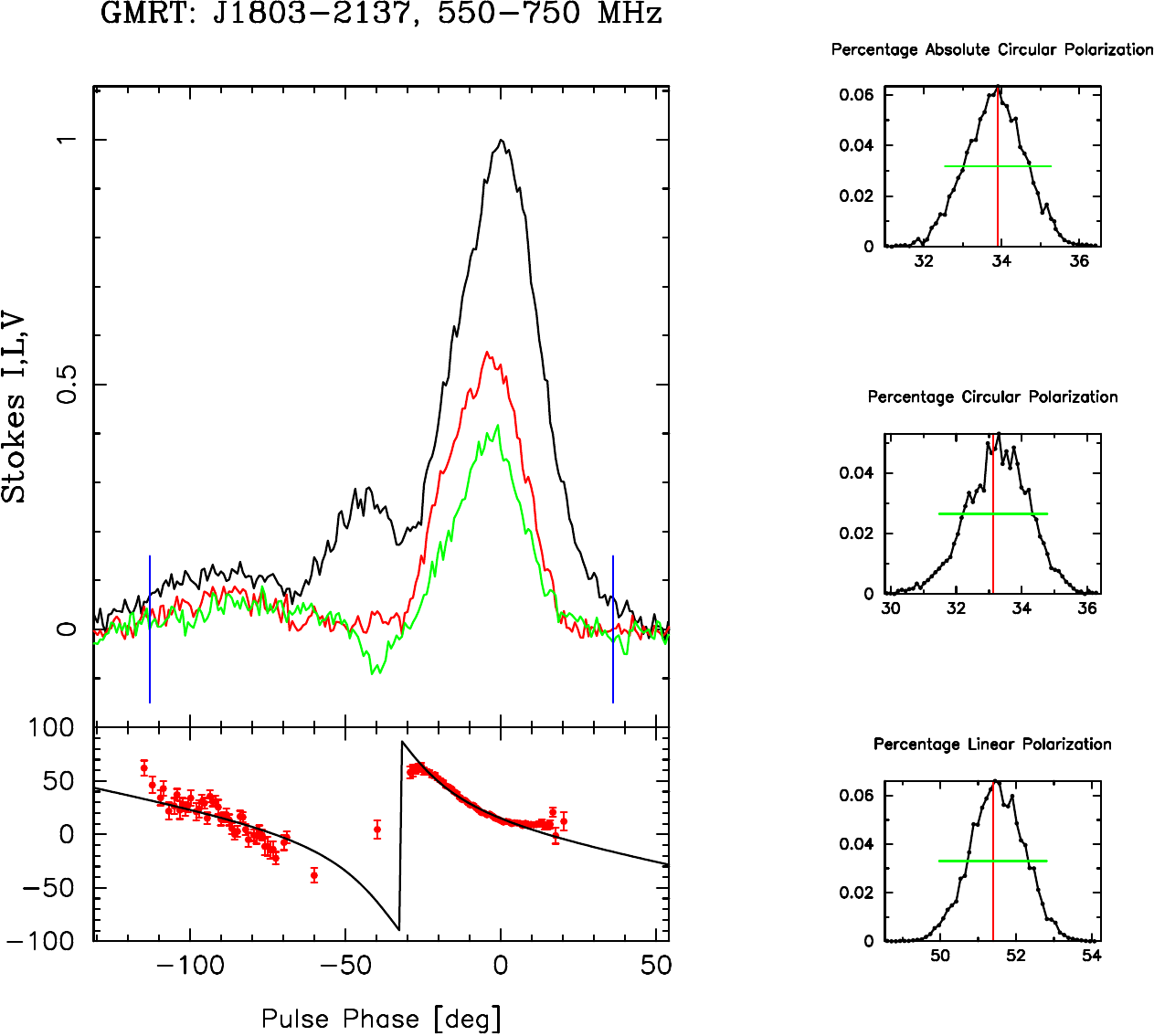}{0.3\textwidth}{(h) PSR J1803$-$2137 (550--750 MHz)}
          \fig{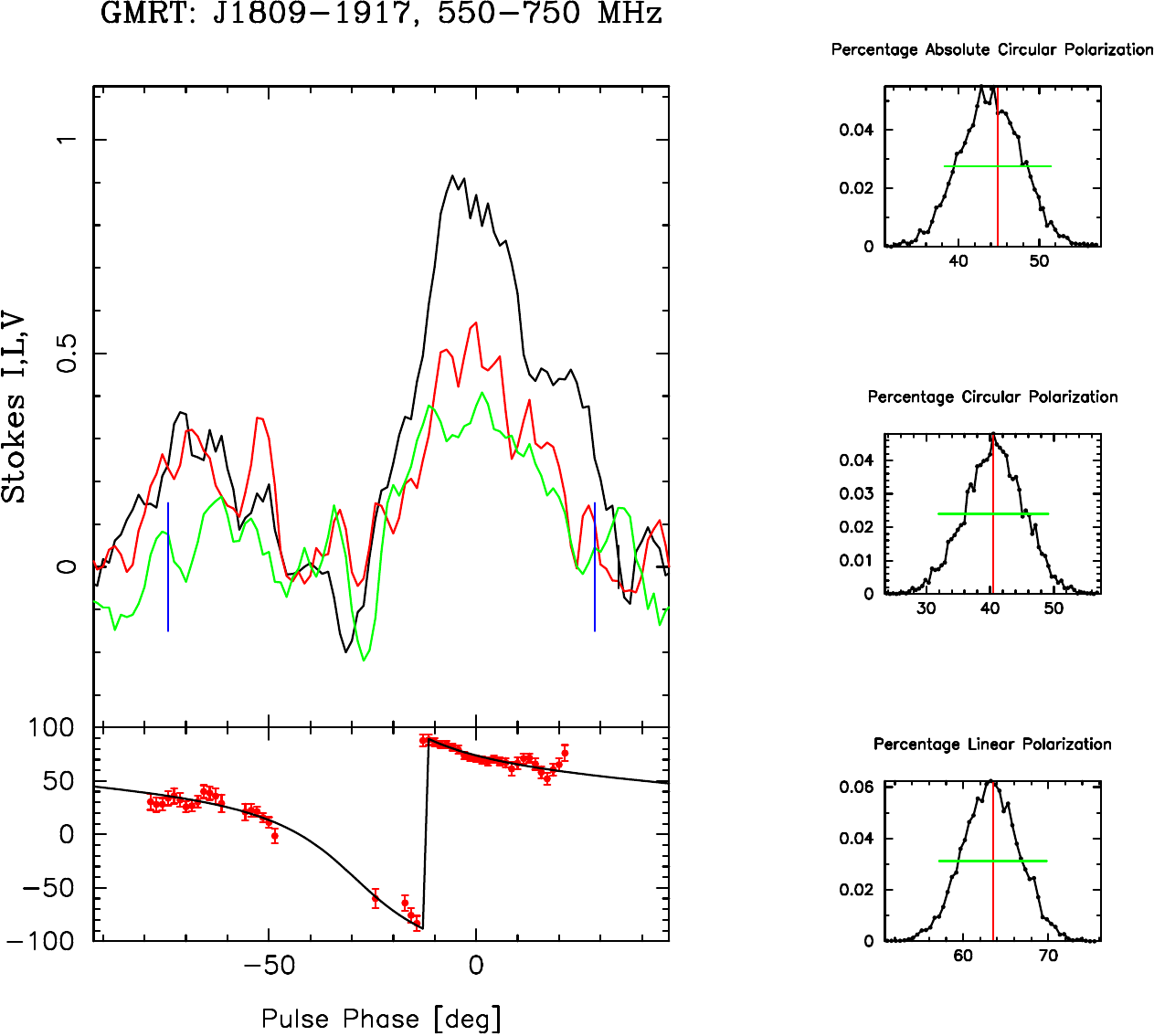}{0.3\textwidth}{(i) PSR J1809$-$1917 (550--750 MHz)}}
\gridline{\fig{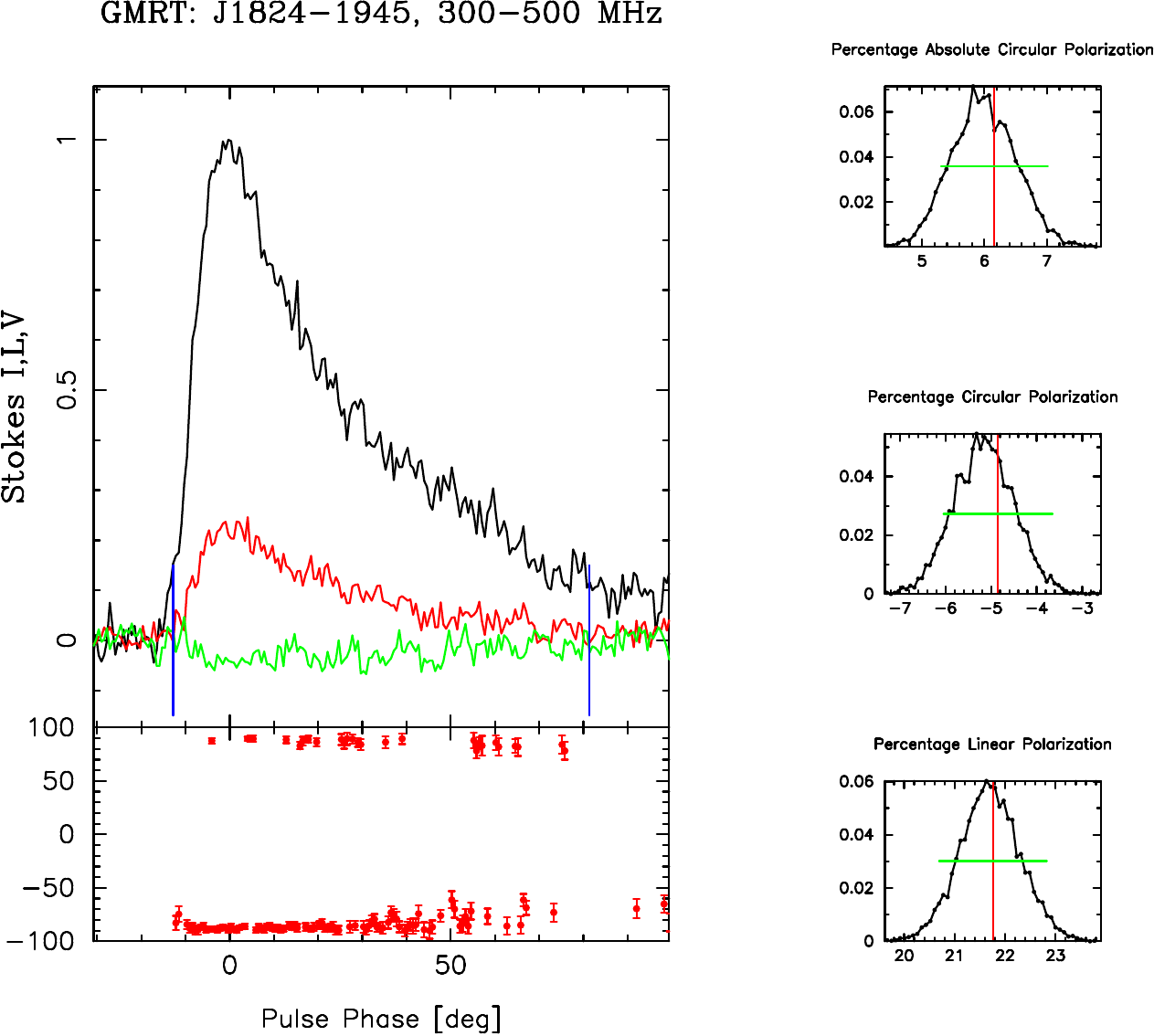}{0.3\textwidth}{(j) PSR J1824$-$1945 (300--500 MHz)}
          \fig{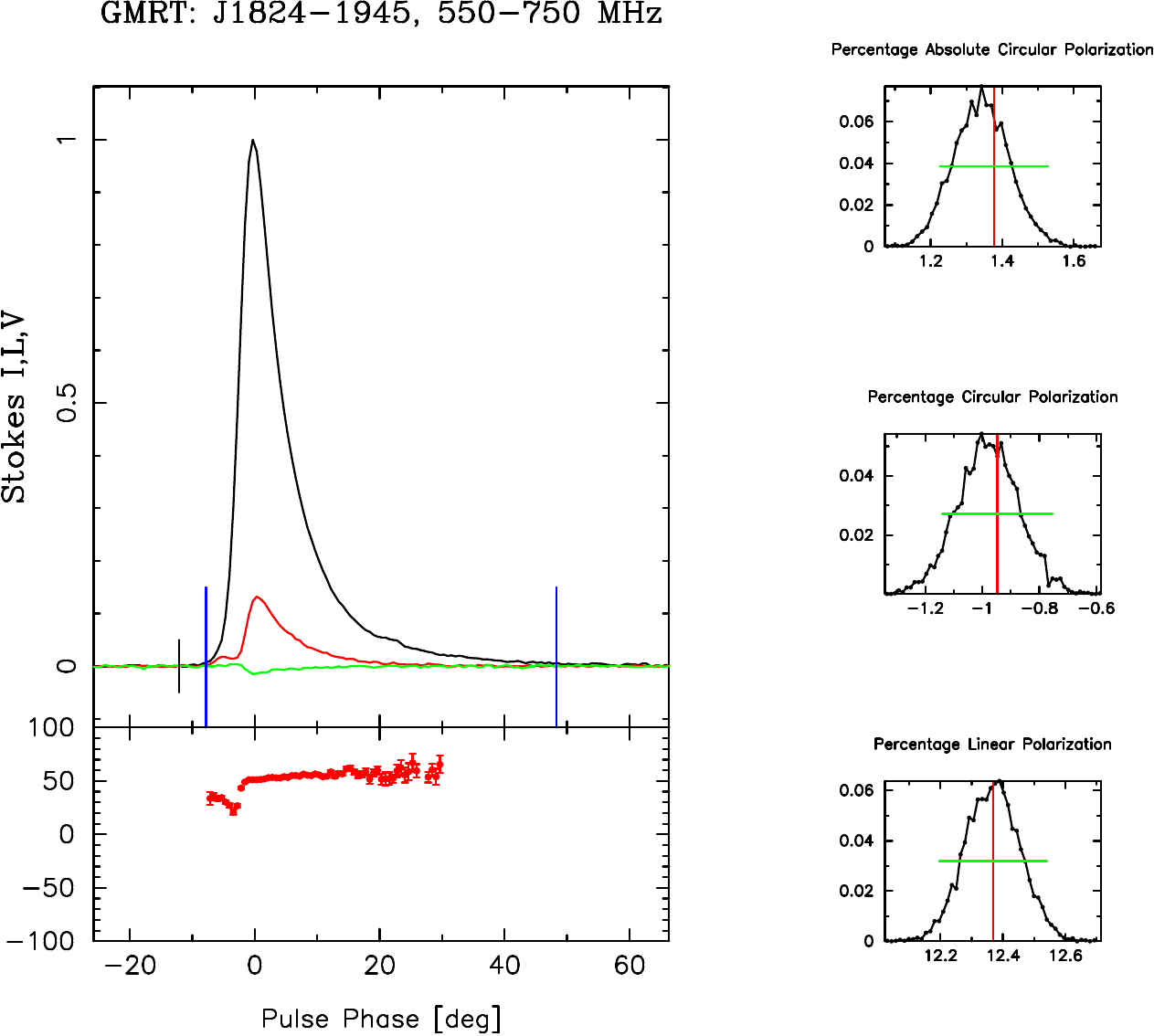}{0.3\textwidth}{(k) PSR J1824$-$1945 (550--750 MHz)}
          \fig{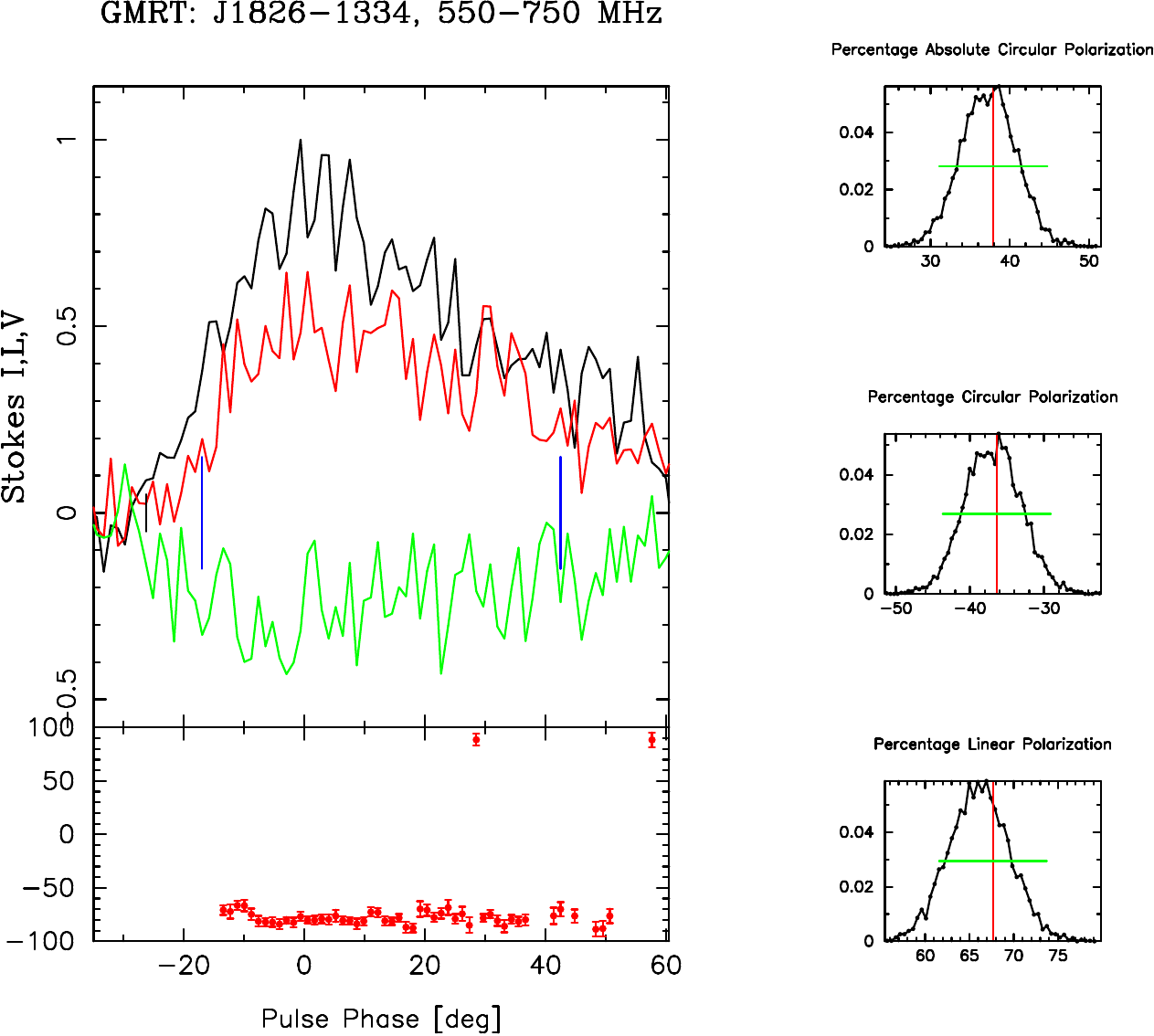}{0.3\textwidth}{(l) PSR J1826$-$1334 (550--750 MHz)}}
\label{avgp2}
\caption{See caption in Fig.~\ref{fig:avgpol}.}
\end{figure}

\begin{figure}
\gridline{\fig{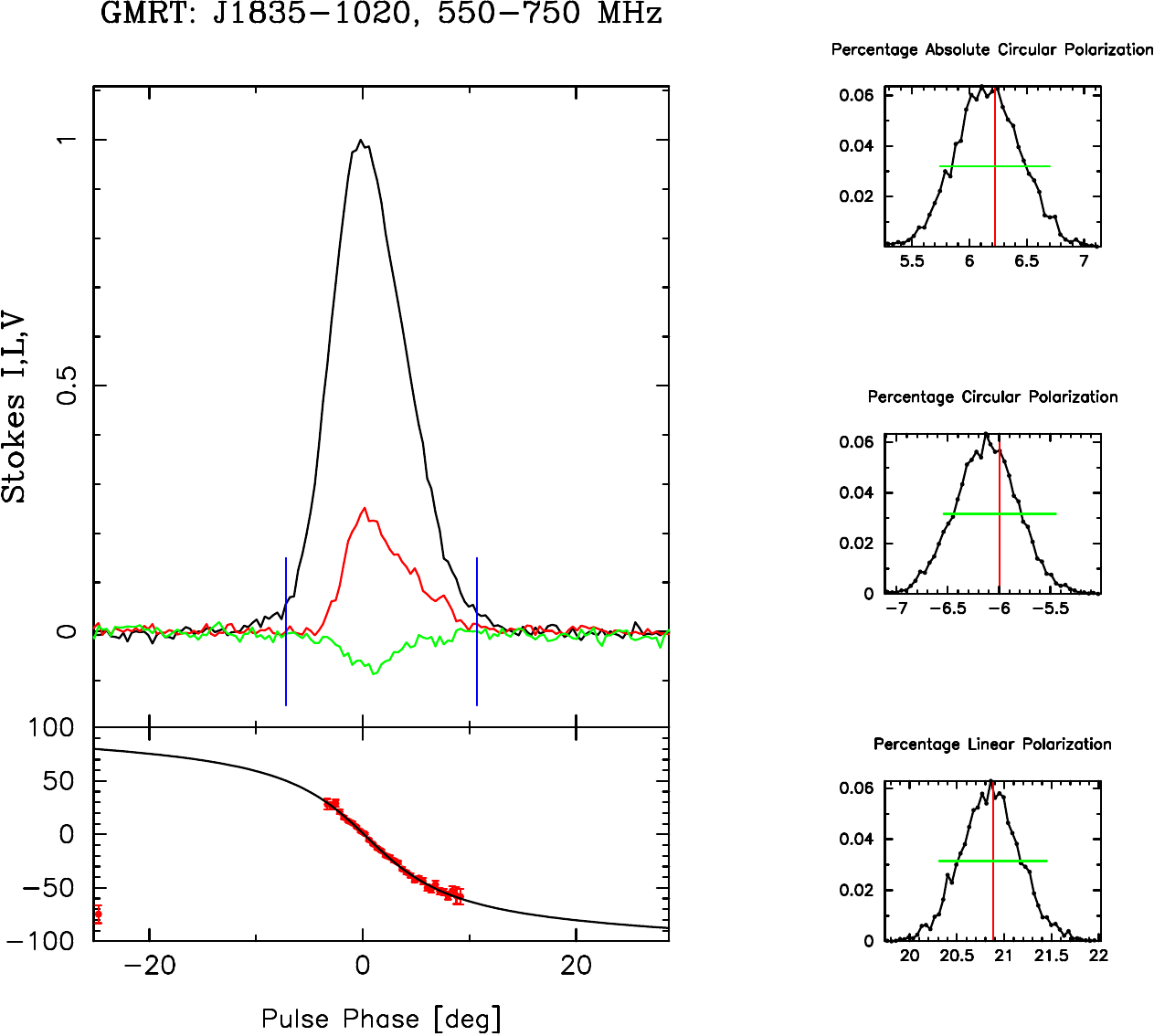}{0.3\textwidth}{(a) PSR J1835$-$1020 (550--750 MHz)}
          \fig{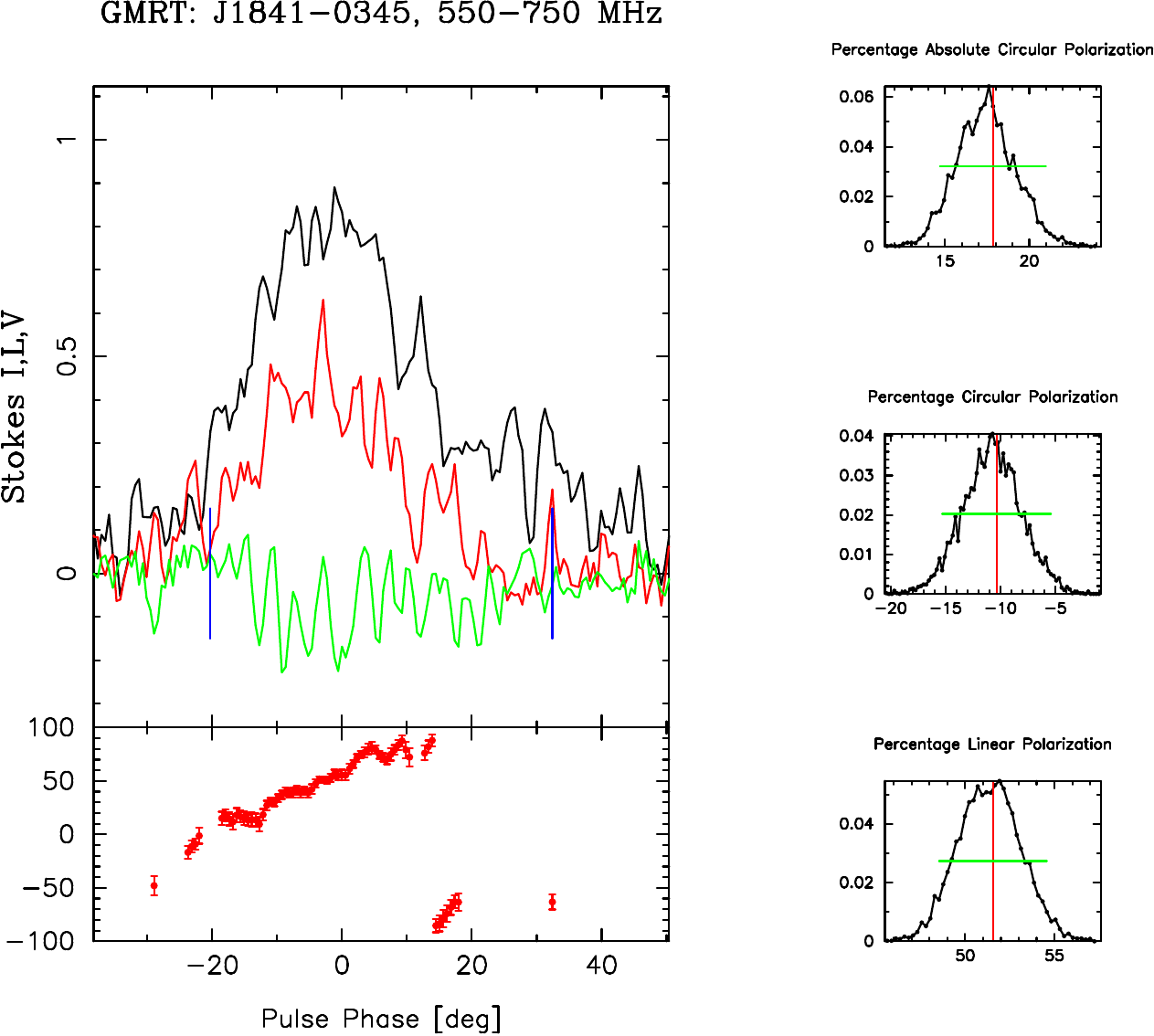}{0.3\textwidth}{(b) PSR J1841$-$0345 (550--750 MHz)}
          \fig{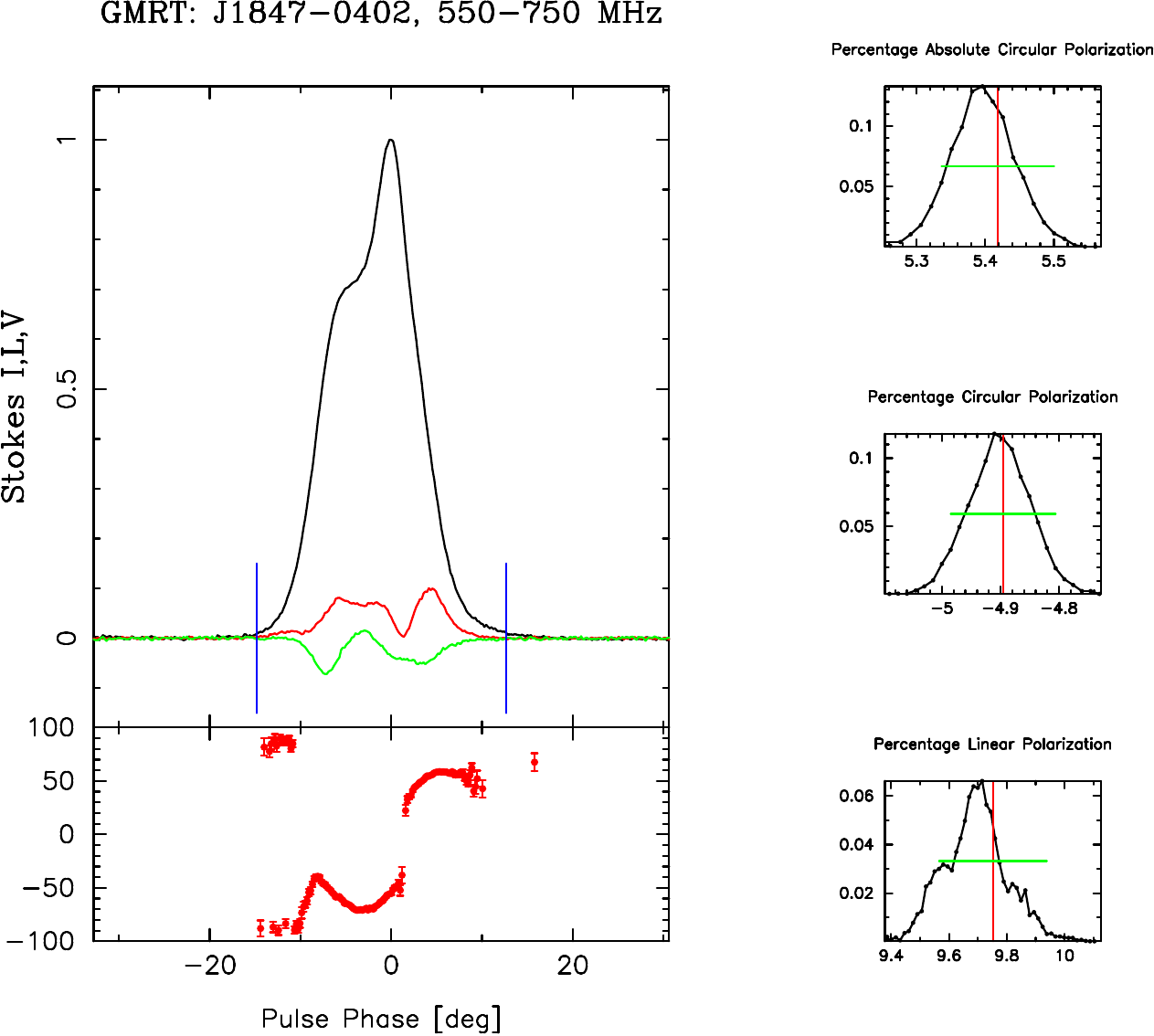}{0.3\textwidth}{(c) PSR J1847$-$0402 (550--750 MHz)}}
\gridline{\fig{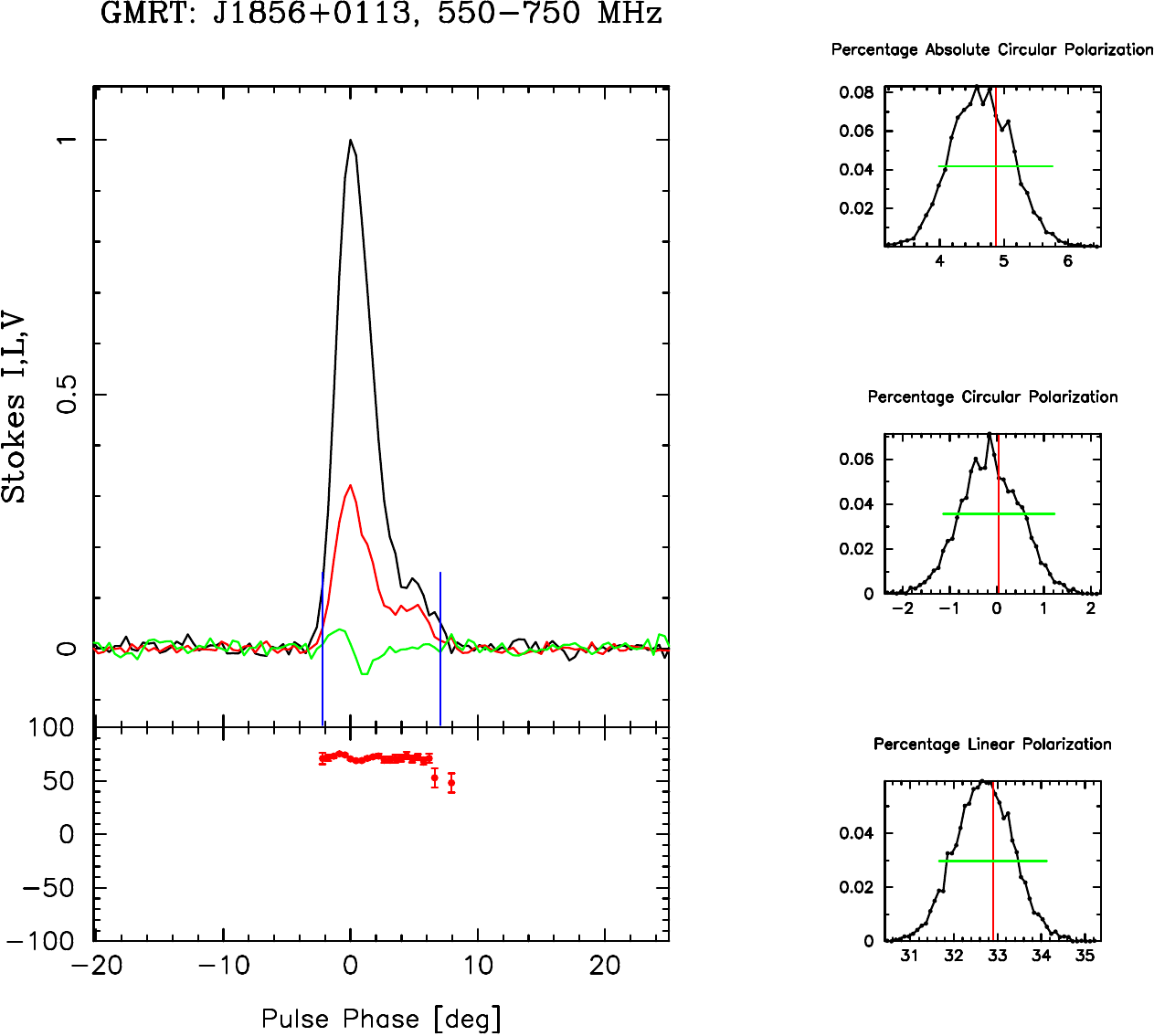}{0.3\textwidth}{(d) PSR J1856+0113 (550--750 MHz)}
          \fig{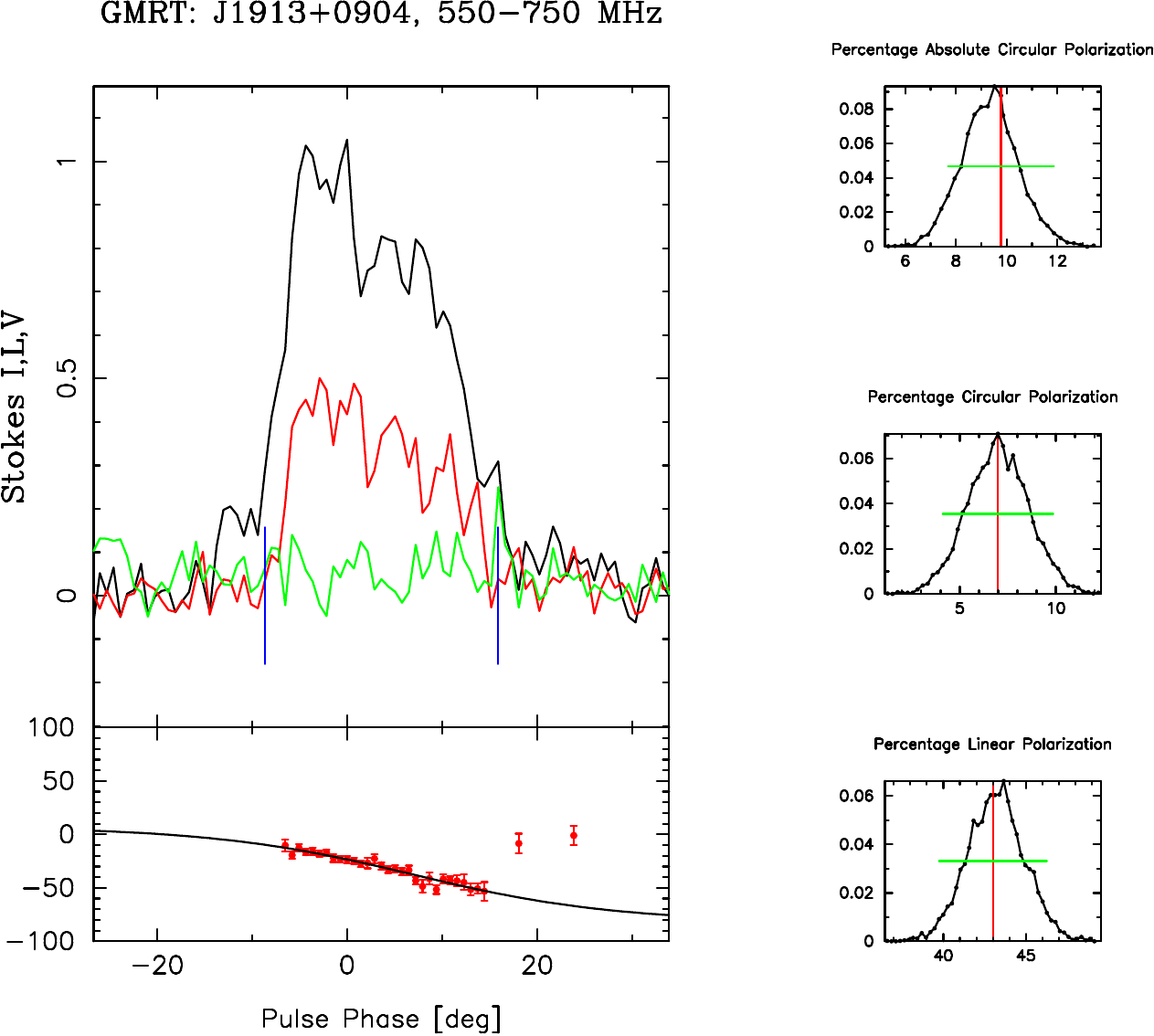}{0.3\textwidth}{(e) PSR J1913+0904 (550--750 MHz)}
          \fig{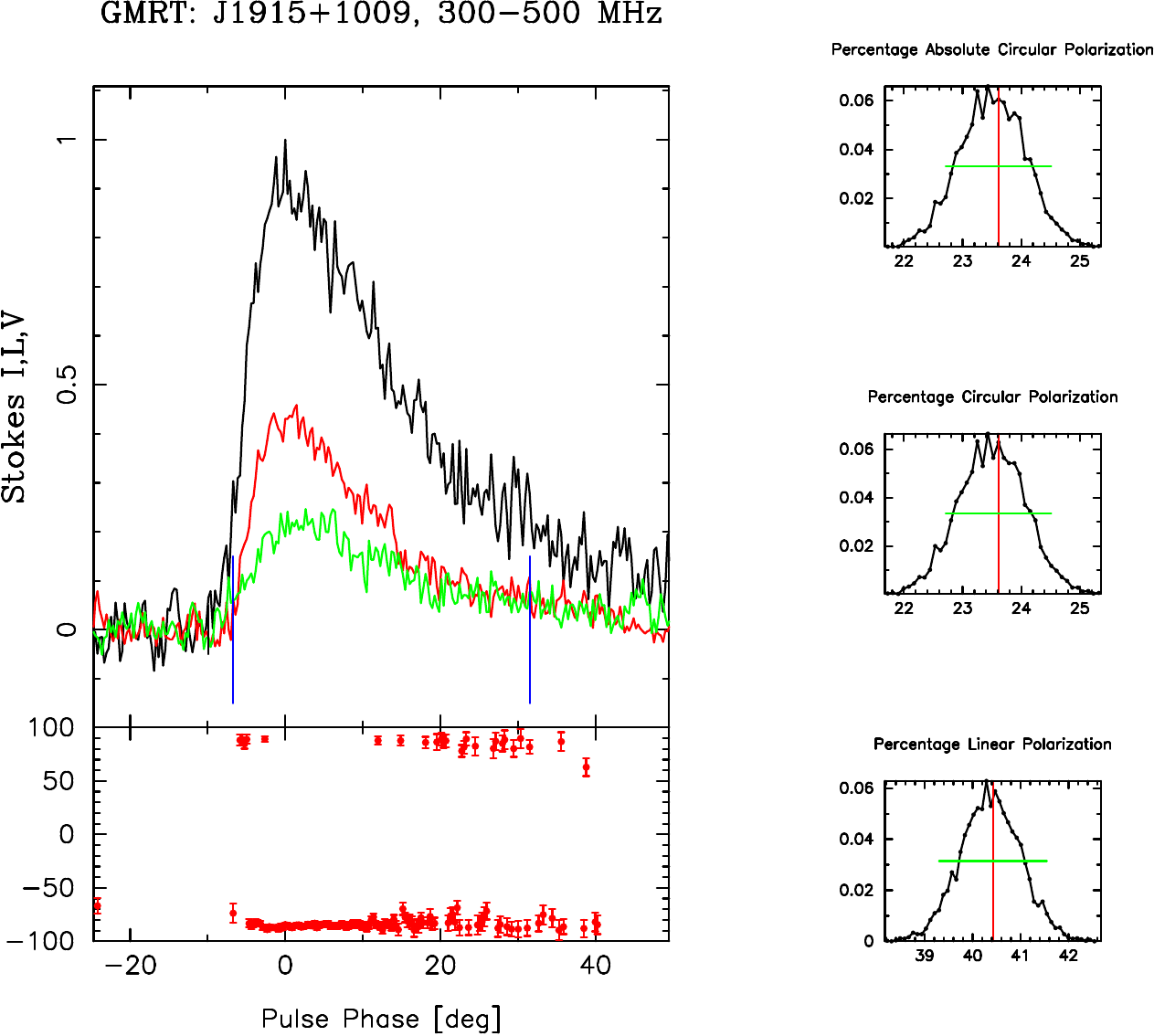}{0.3\textwidth}{(f) PSR J1915+1009 (300--500 MHz)}}
\gridline{\fig{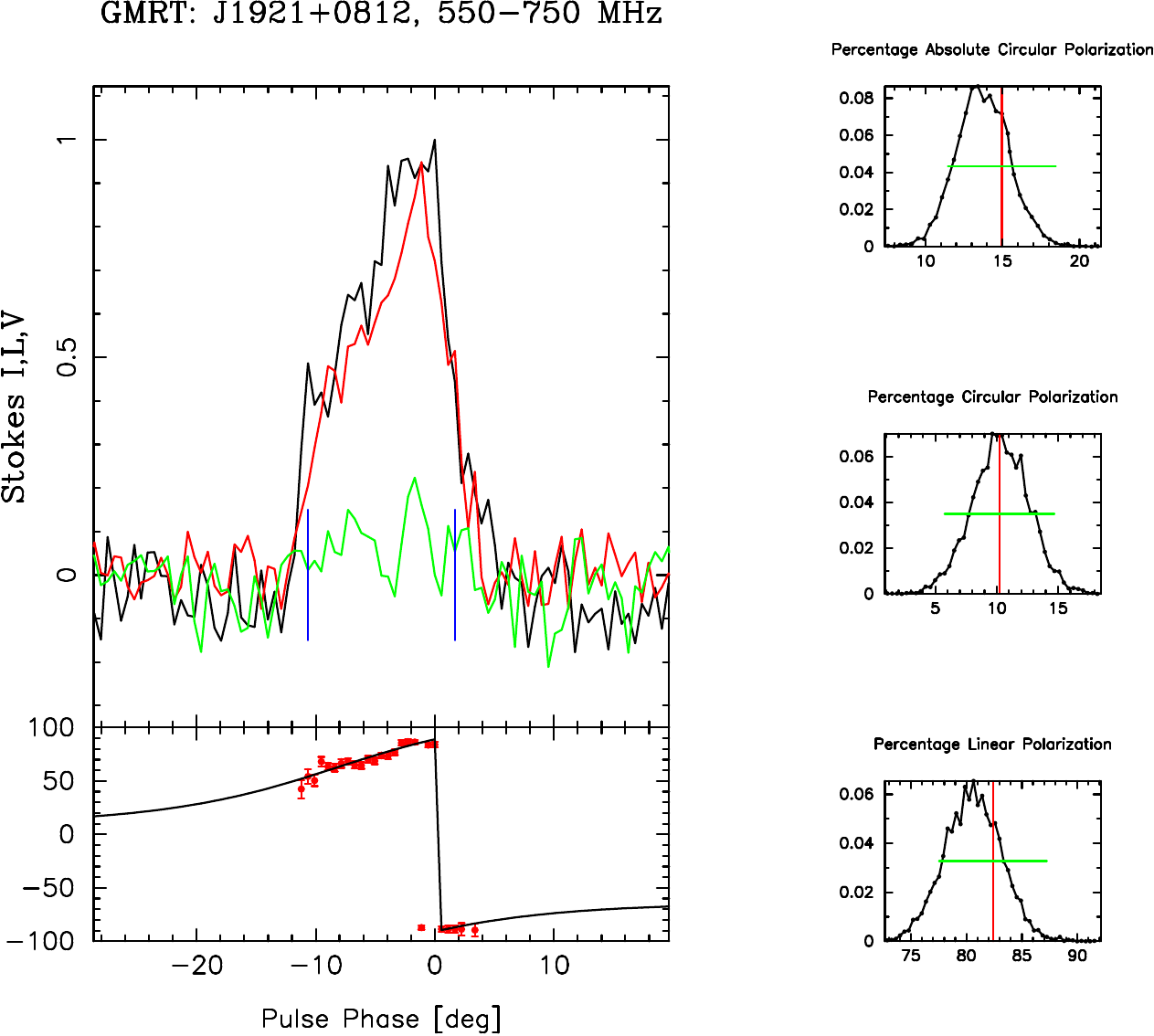}{0.3\textwidth}{(g) PSR J1921+0812 (550--750 MHz)}
          \fig{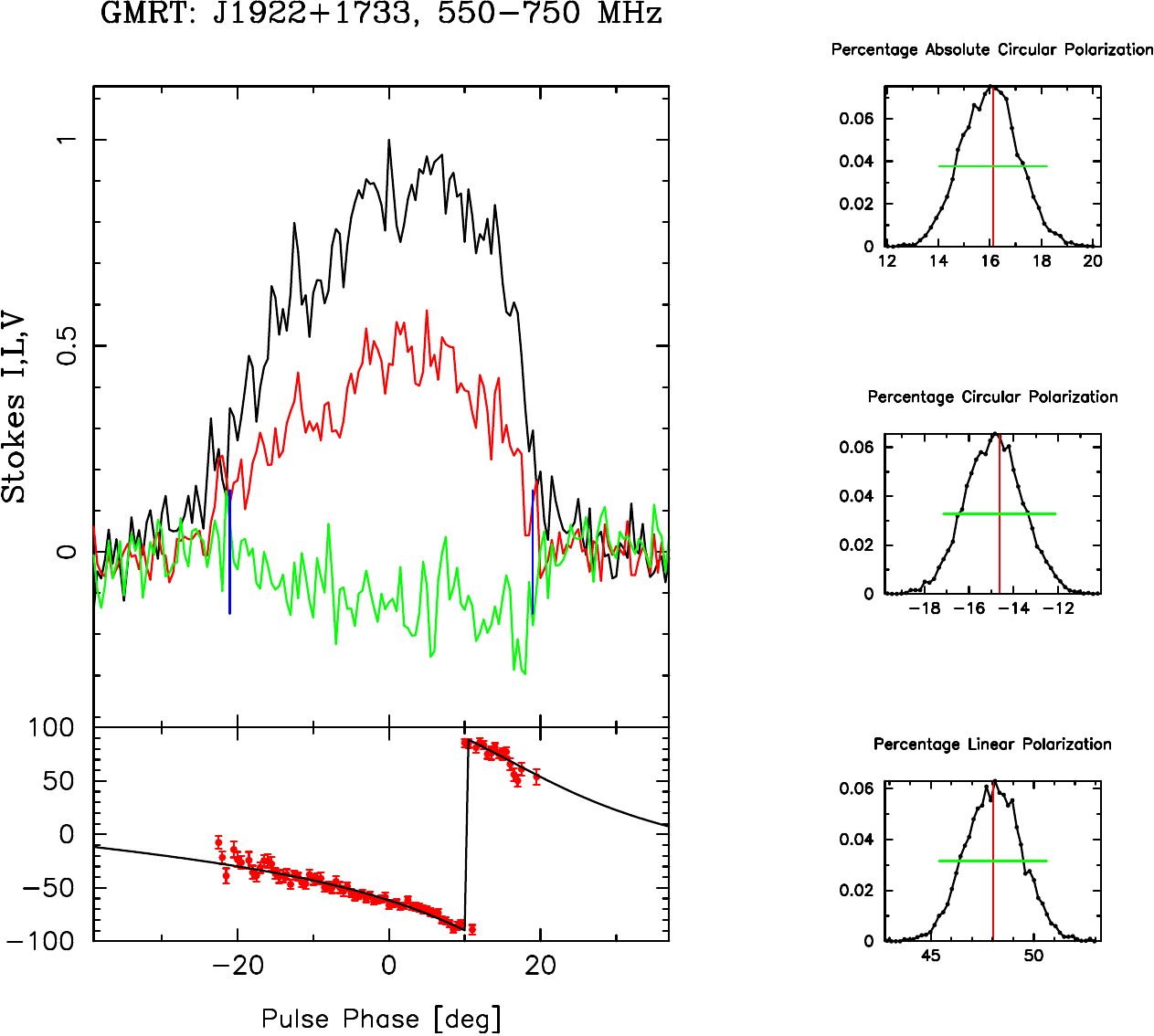}{0.3\textwidth}{(h) PSR J1922+1733 (550--750 MHz)}
          \fig{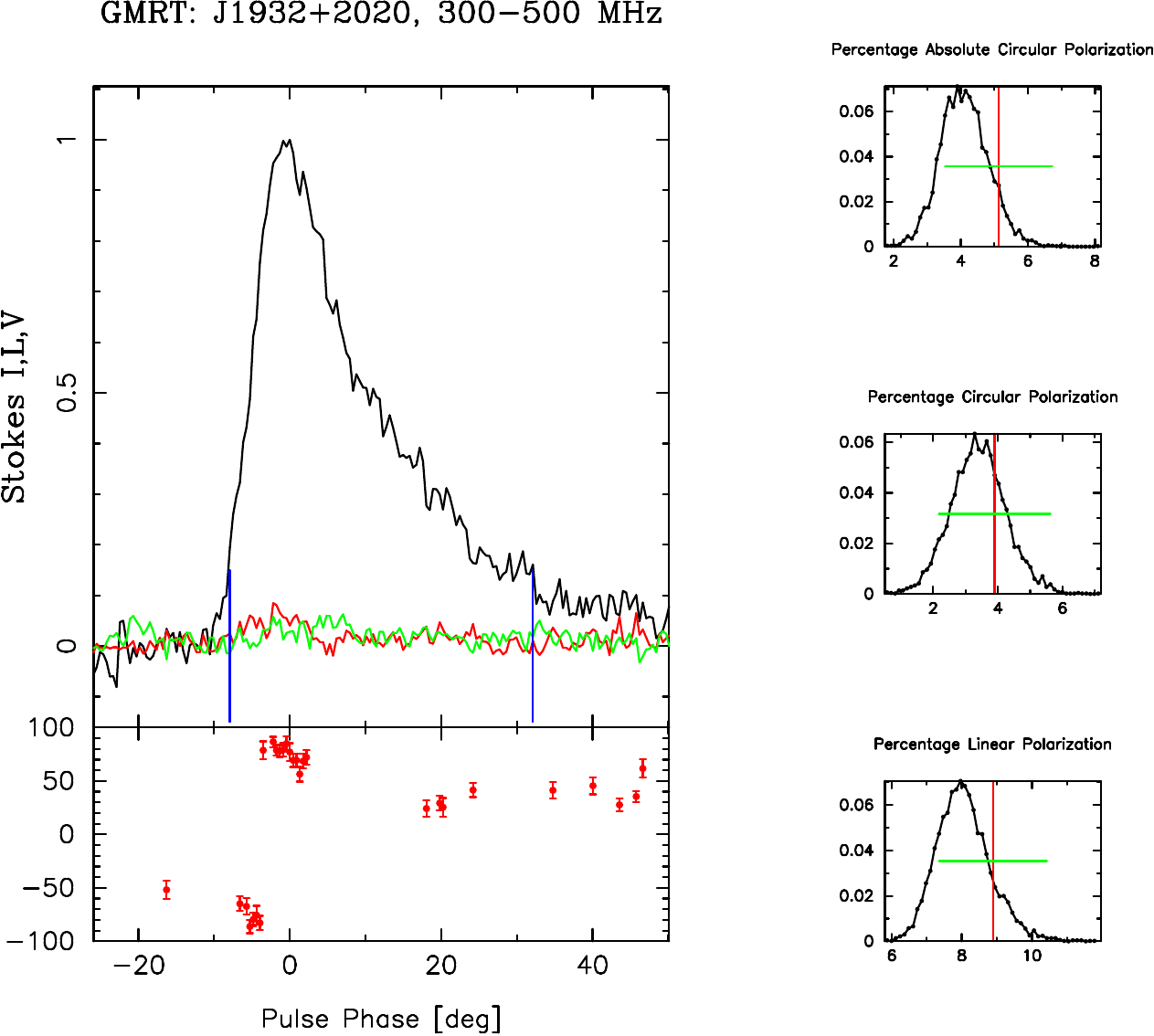}{0.3\textwidth}{(i) PSR J1932+2020 (300--500 MHz)}}
\gridline{\fig{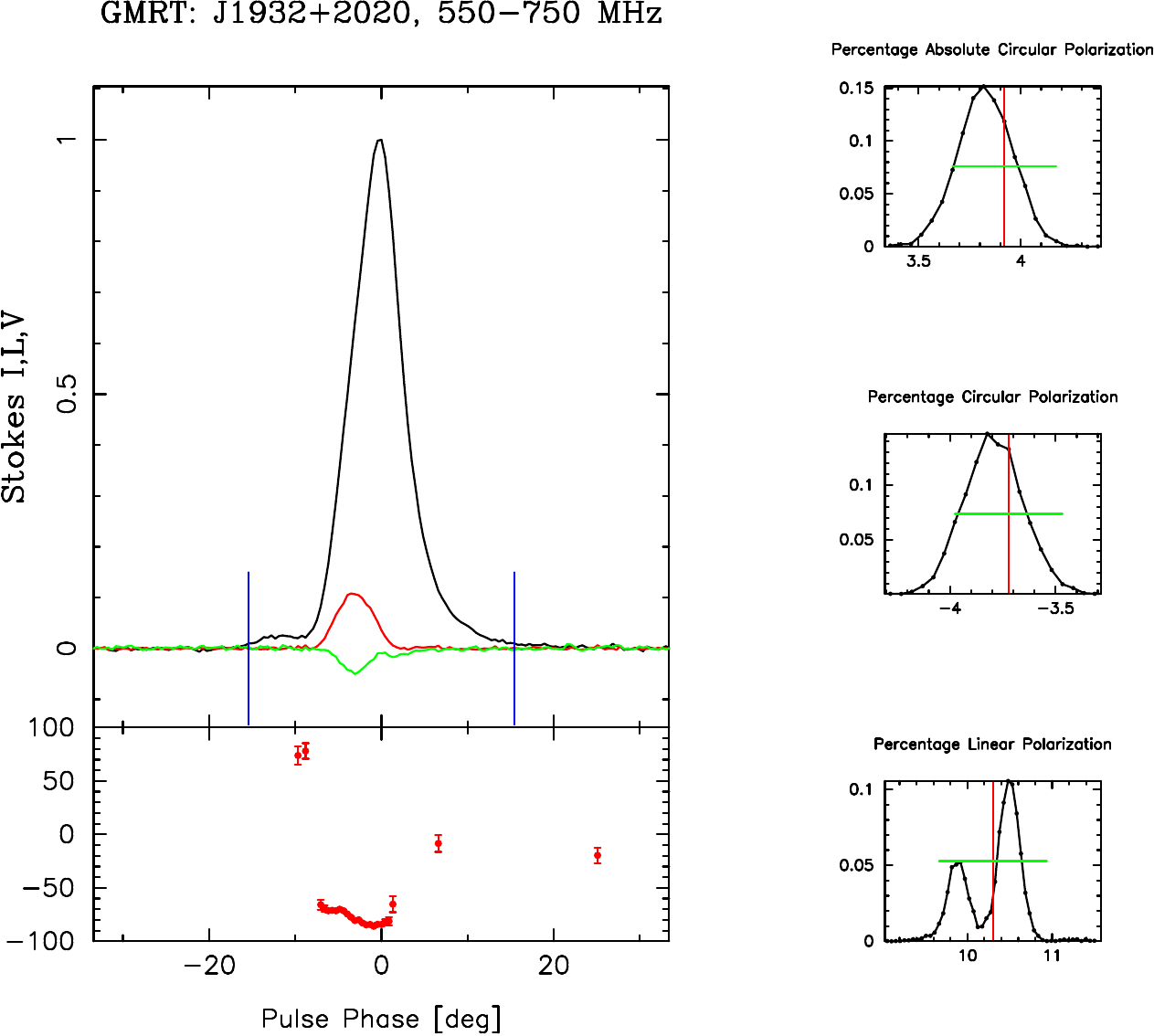}{0.3\textwidth}{(j) PSR J1932+2020 (550--750 MHz)}
          \fig{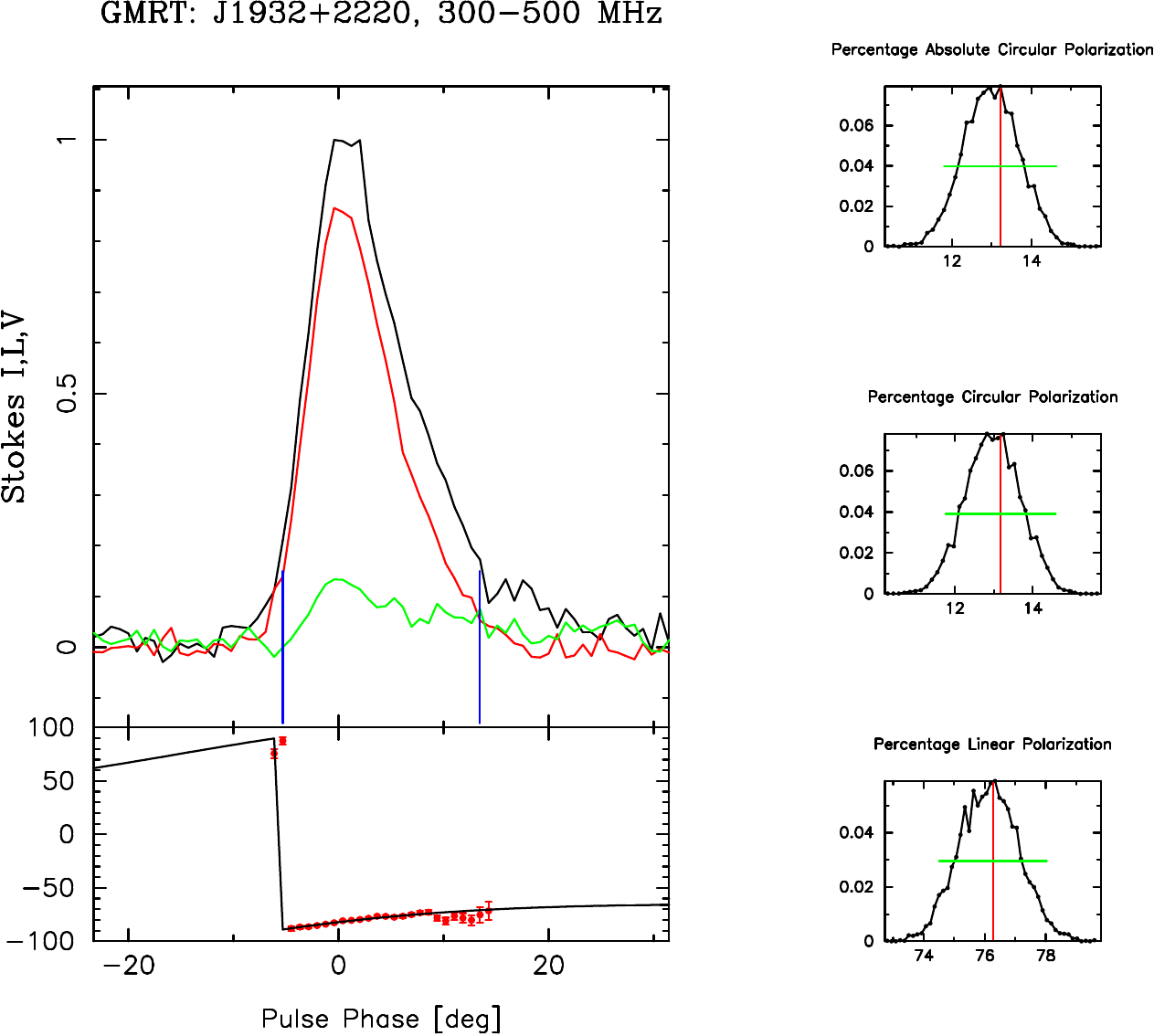}{0.3\textwidth}{(k) PSR J1932+2220 (300--500 MHz)}
          \fig{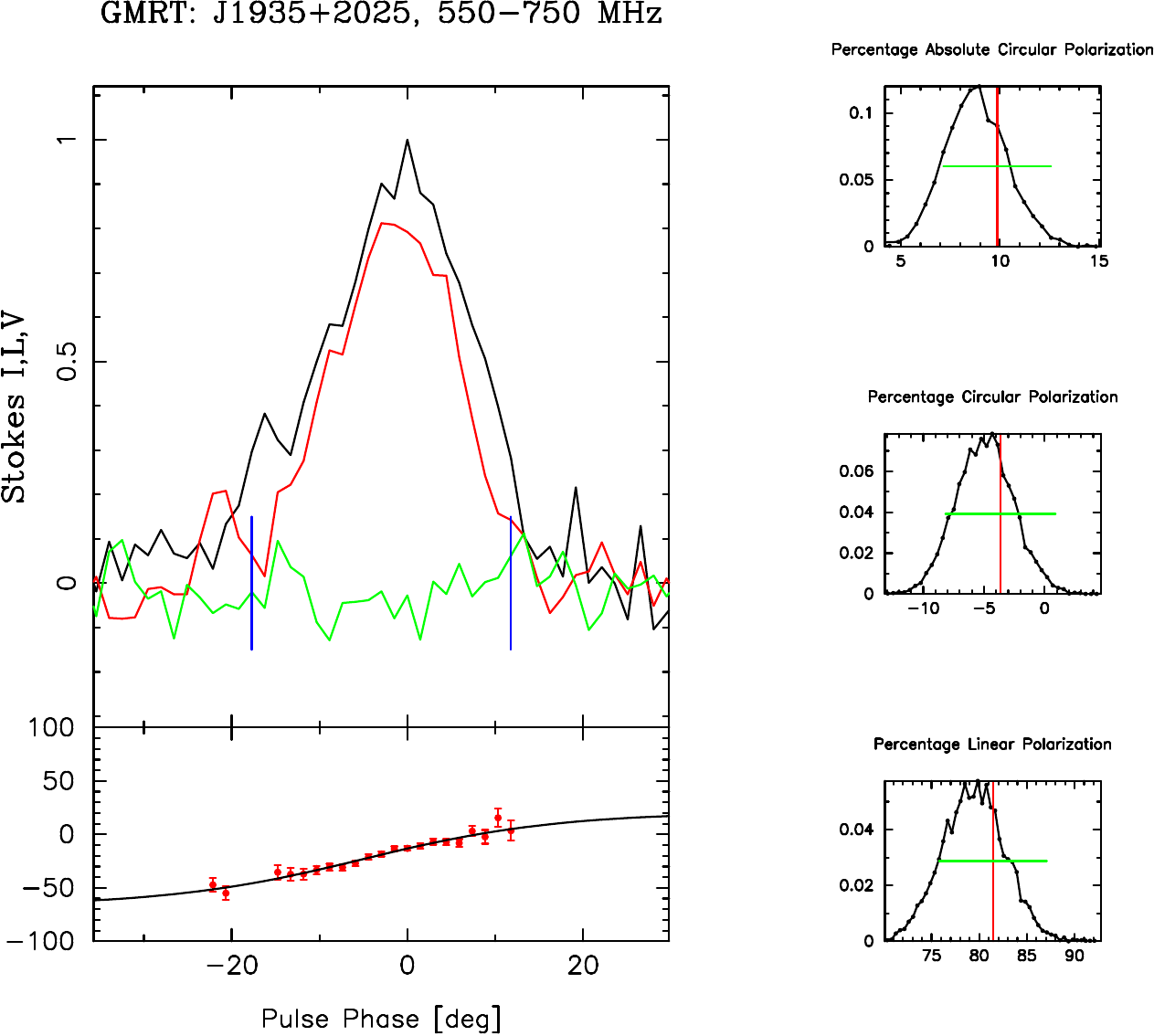}{0.3\textwidth}{(l) PSR J1935+2025 (550--750 MHz)}}
\label{avgp3}
\caption{See caption in Fig.~\ref{fig:avgpol}.}
\end{figure}

\begin{figure}
\gridline{\fig{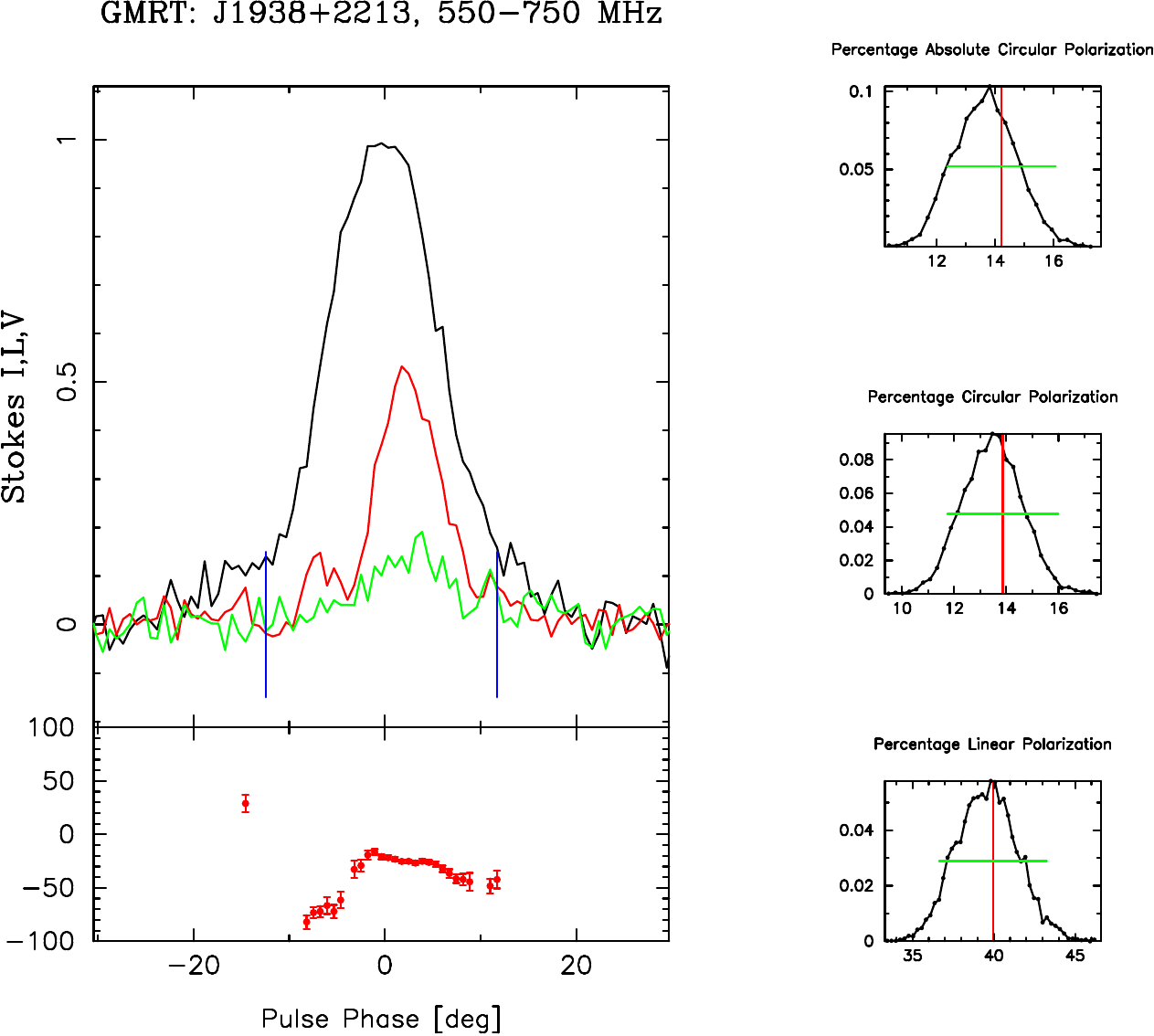}{0.3\textwidth}{(a) PSR J1938+2213 (550--750 MHz)}
          \fig{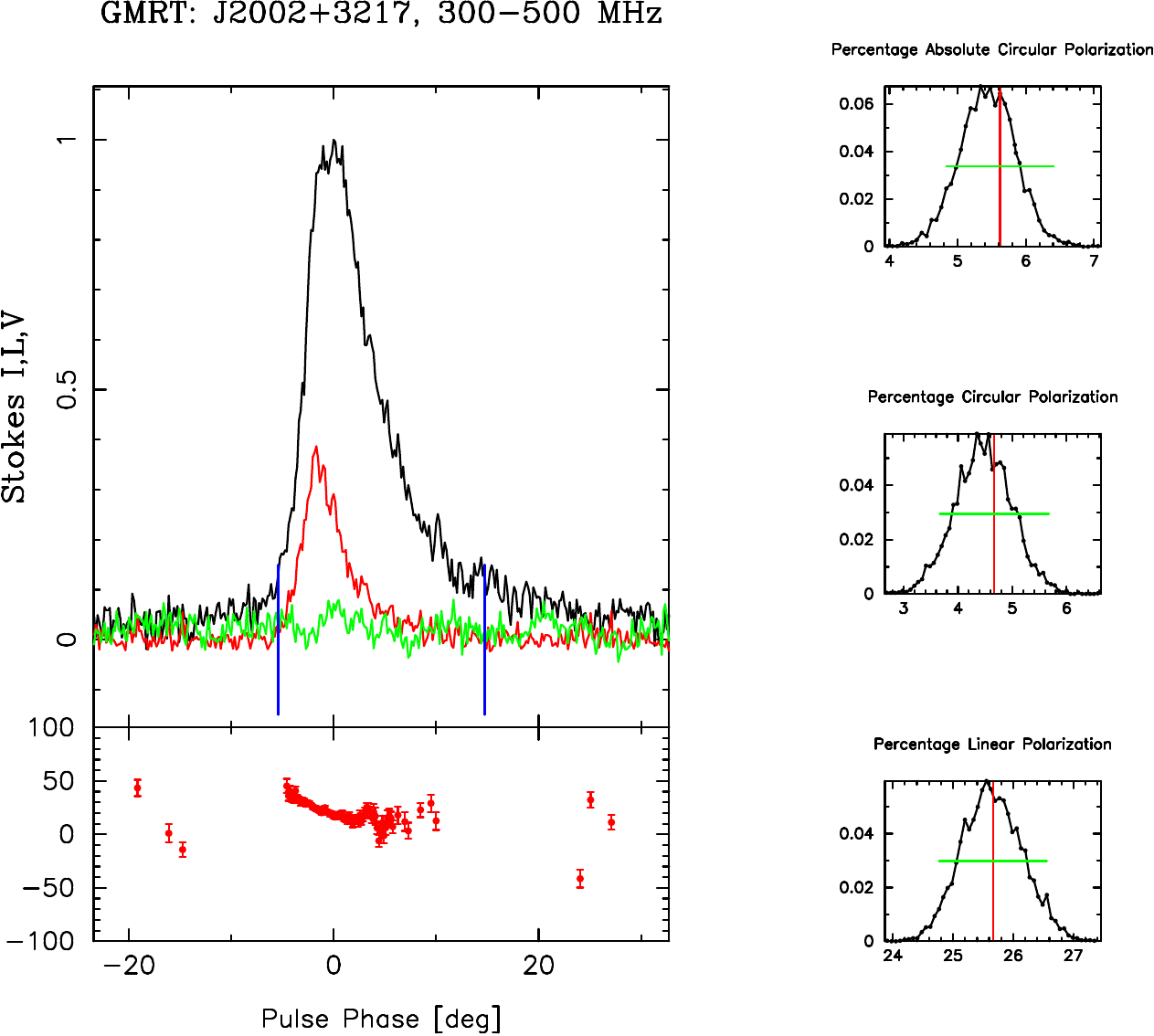}{0.3\textwidth}{(b) PSR J2002+3217 (300--500 MHz)}
          \fig{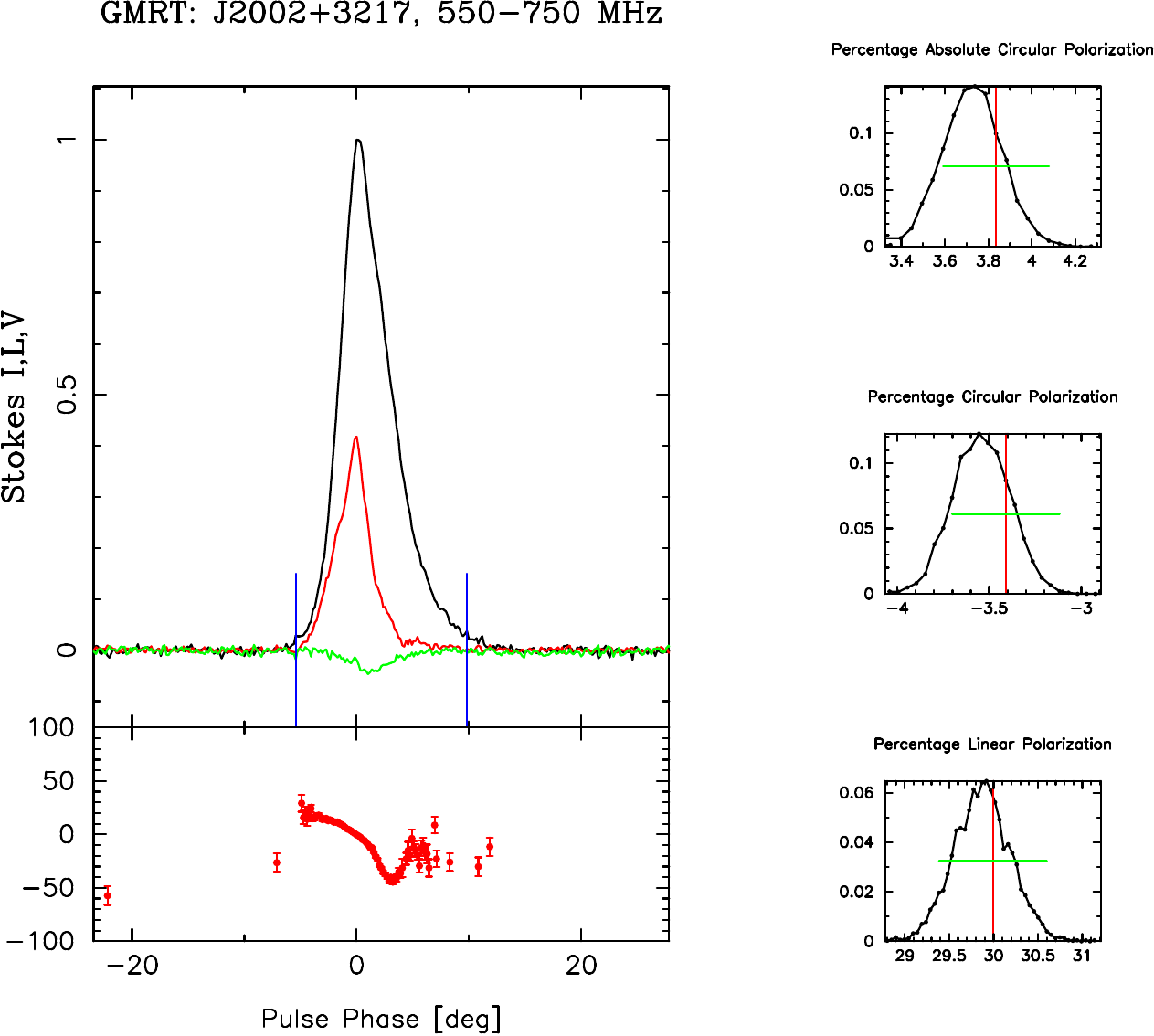}{0.3\textwidth}{(c) PSR J2002+3217 (550--750 MHz)}}
\gridline{\fig{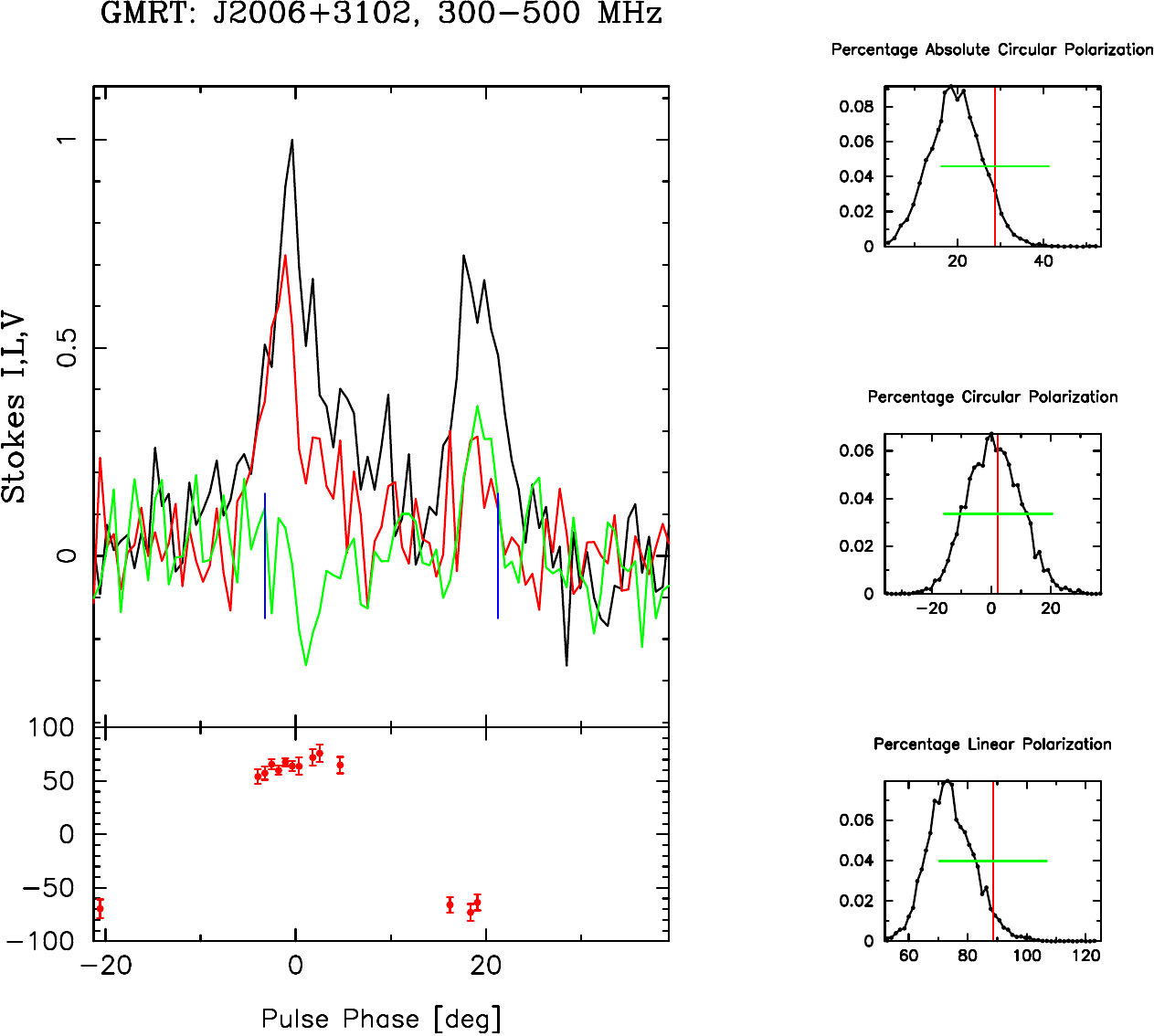}{0.3\textwidth}{(d) PSR J2006+3102 (300--500 MHz)}
          \fig{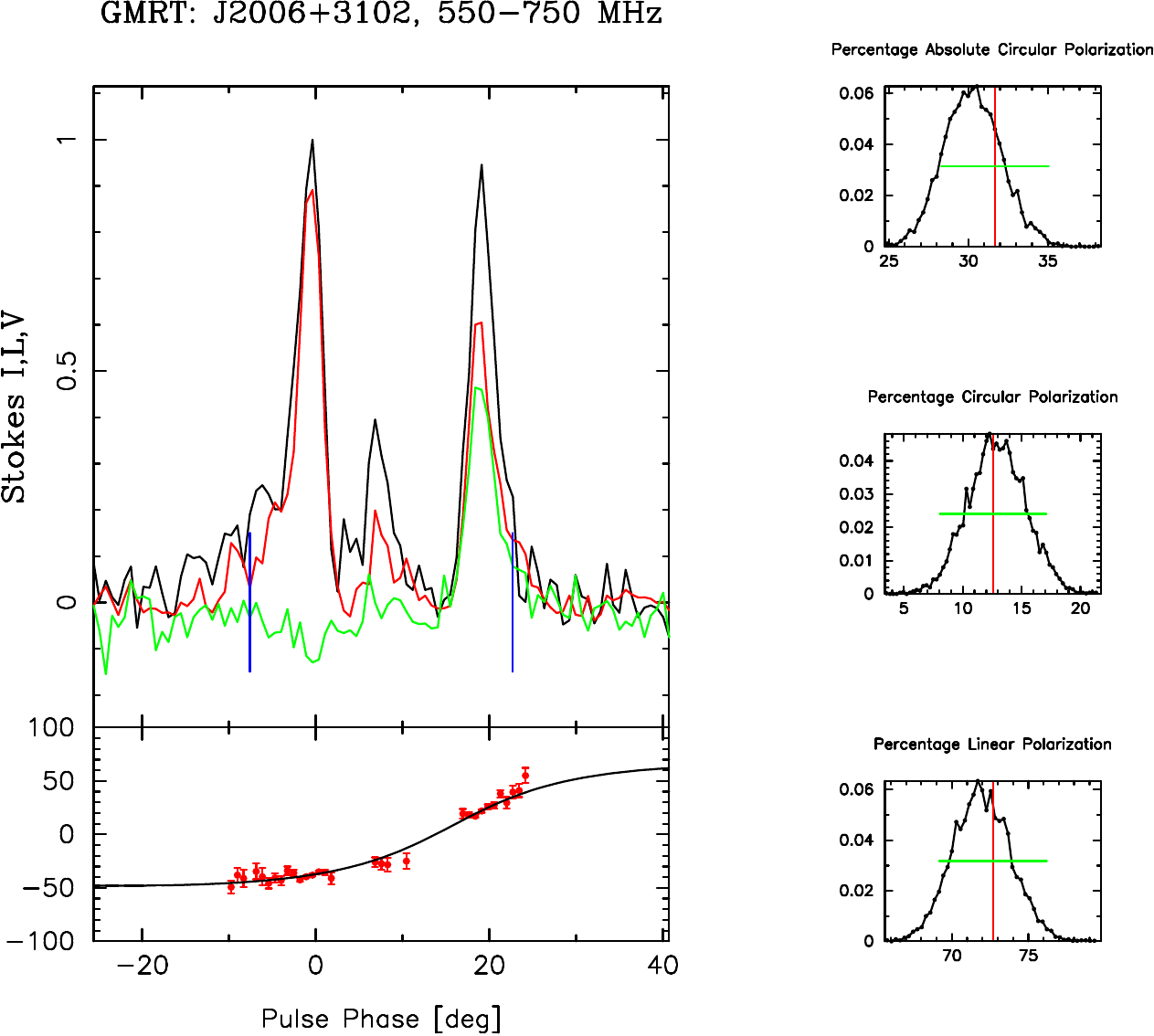}{0.3\textwidth}{(e) PSR J2006+3102 (550--750 MHz)}
          \fig{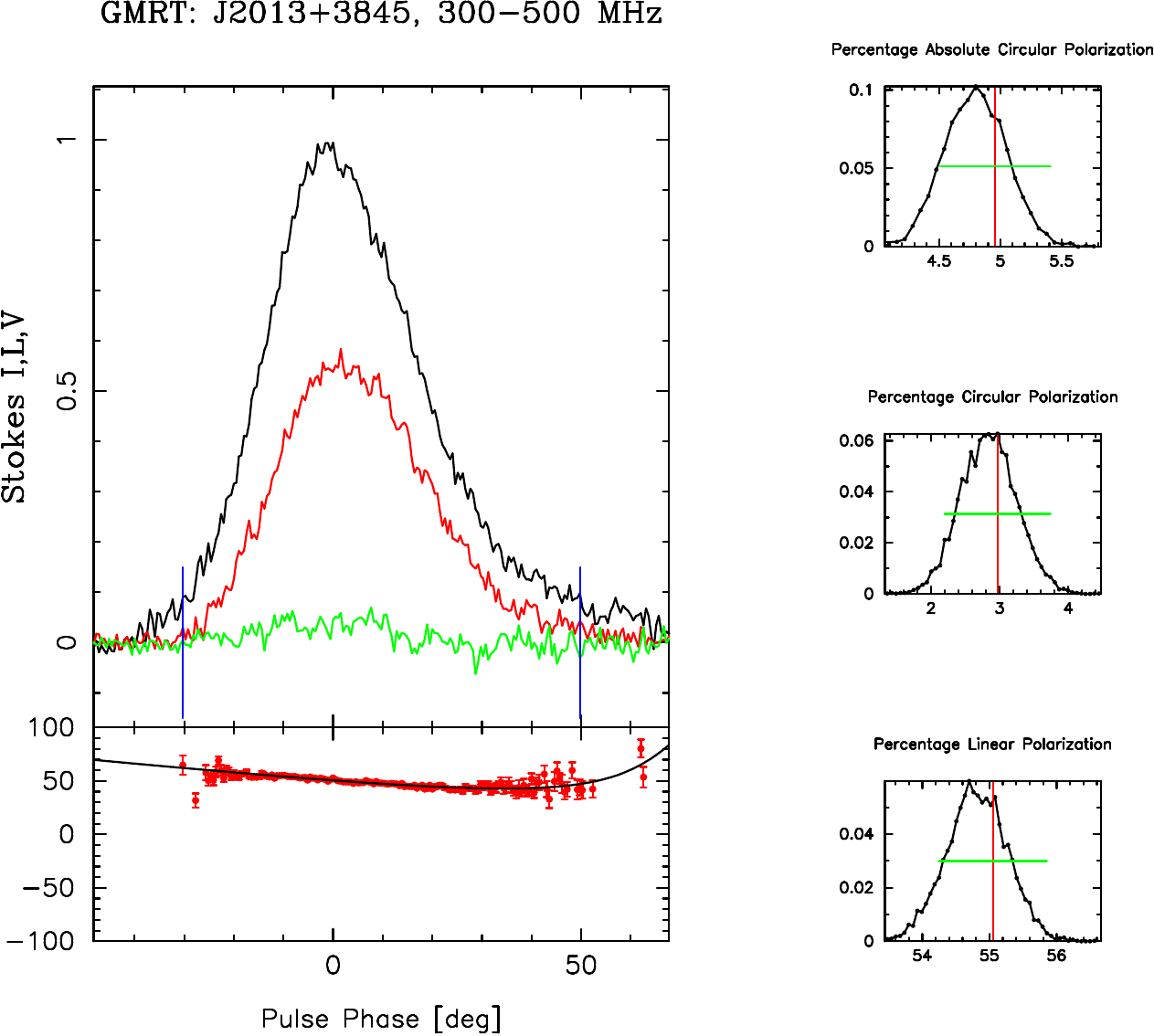}{0.3\textwidth}{(f) PSR J2013+3845 (300--500 MHz)}}
\gridline{\fig{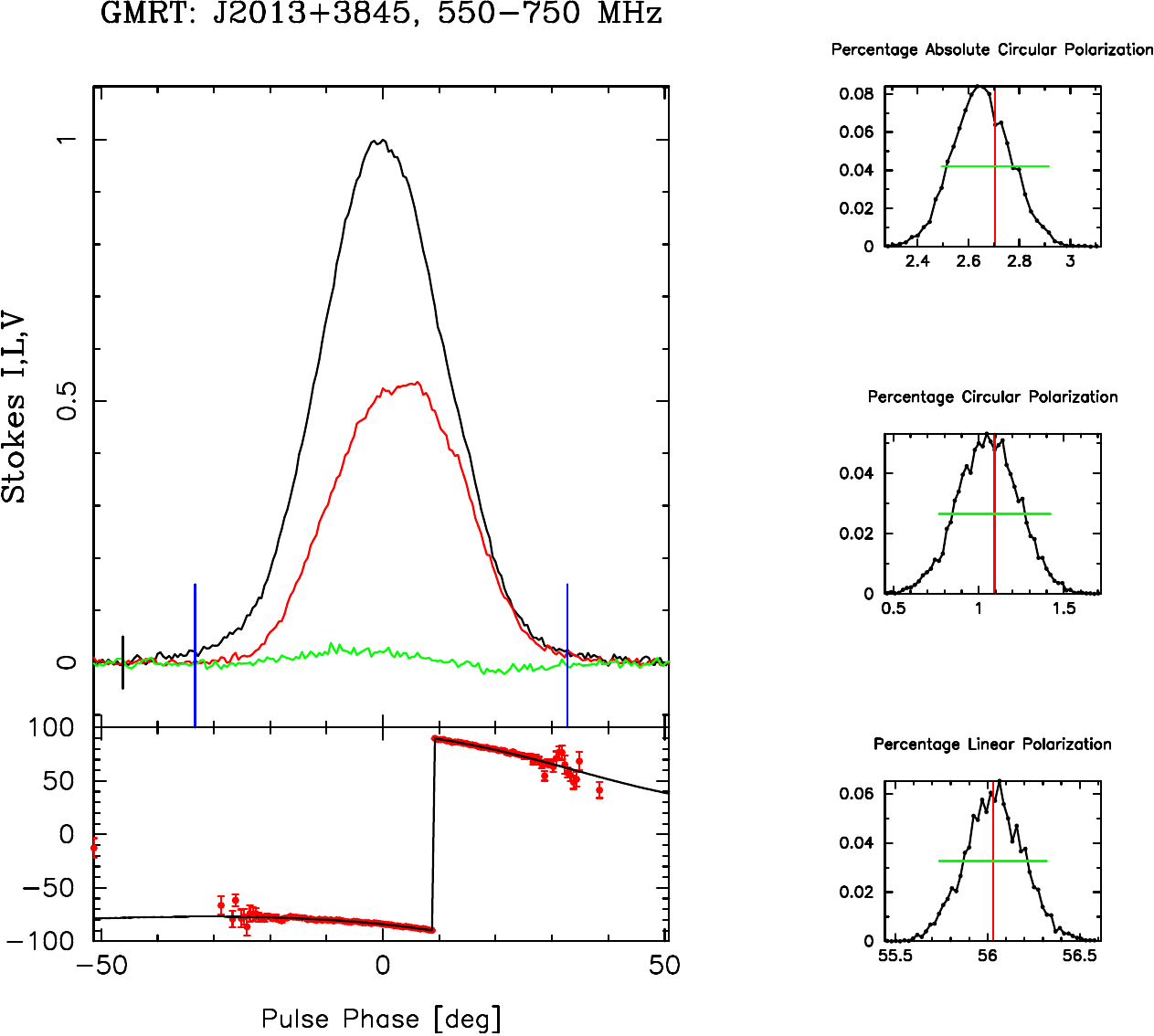}{0.3\textwidth}{(g) PSR J2013+3845 (550--750 MHz)}
          \fig{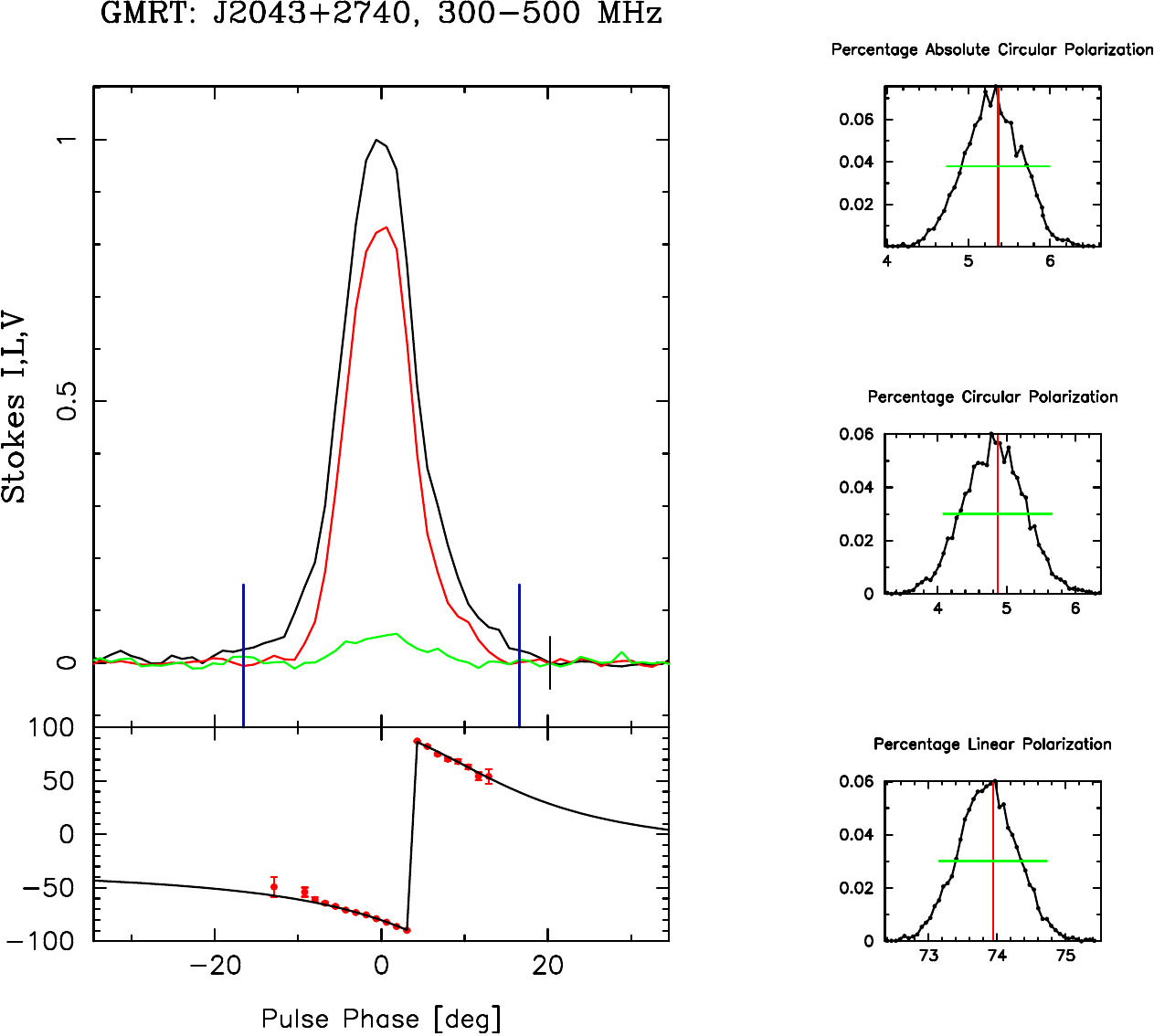}{0.3\textwidth}{(h) PSR J2043+2740 (300--500 MHz)}
          \fig{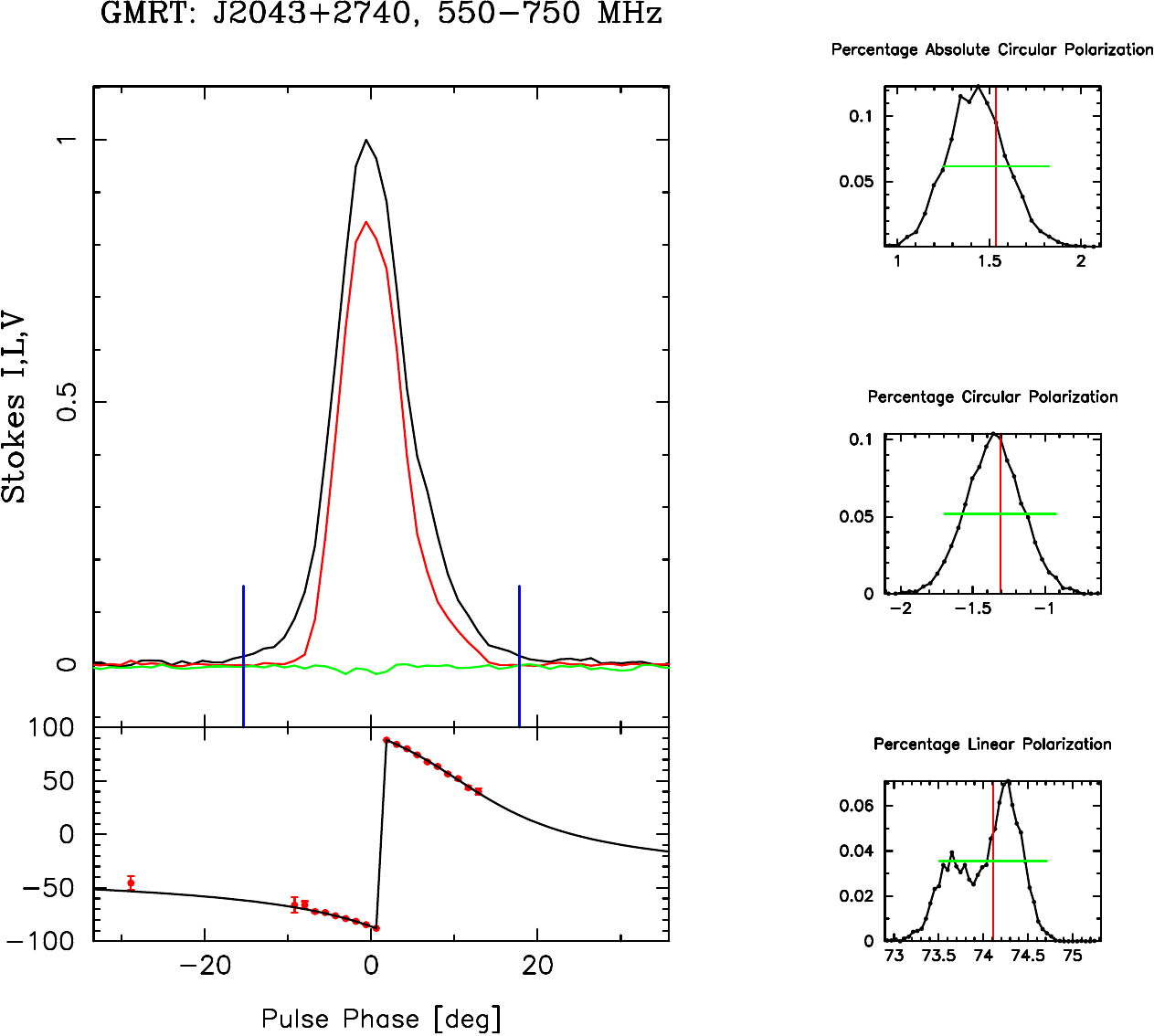}{0.3\textwidth}{(i) PSR J2043+2740 (550--750 MHz)}}
\gridline{\fig{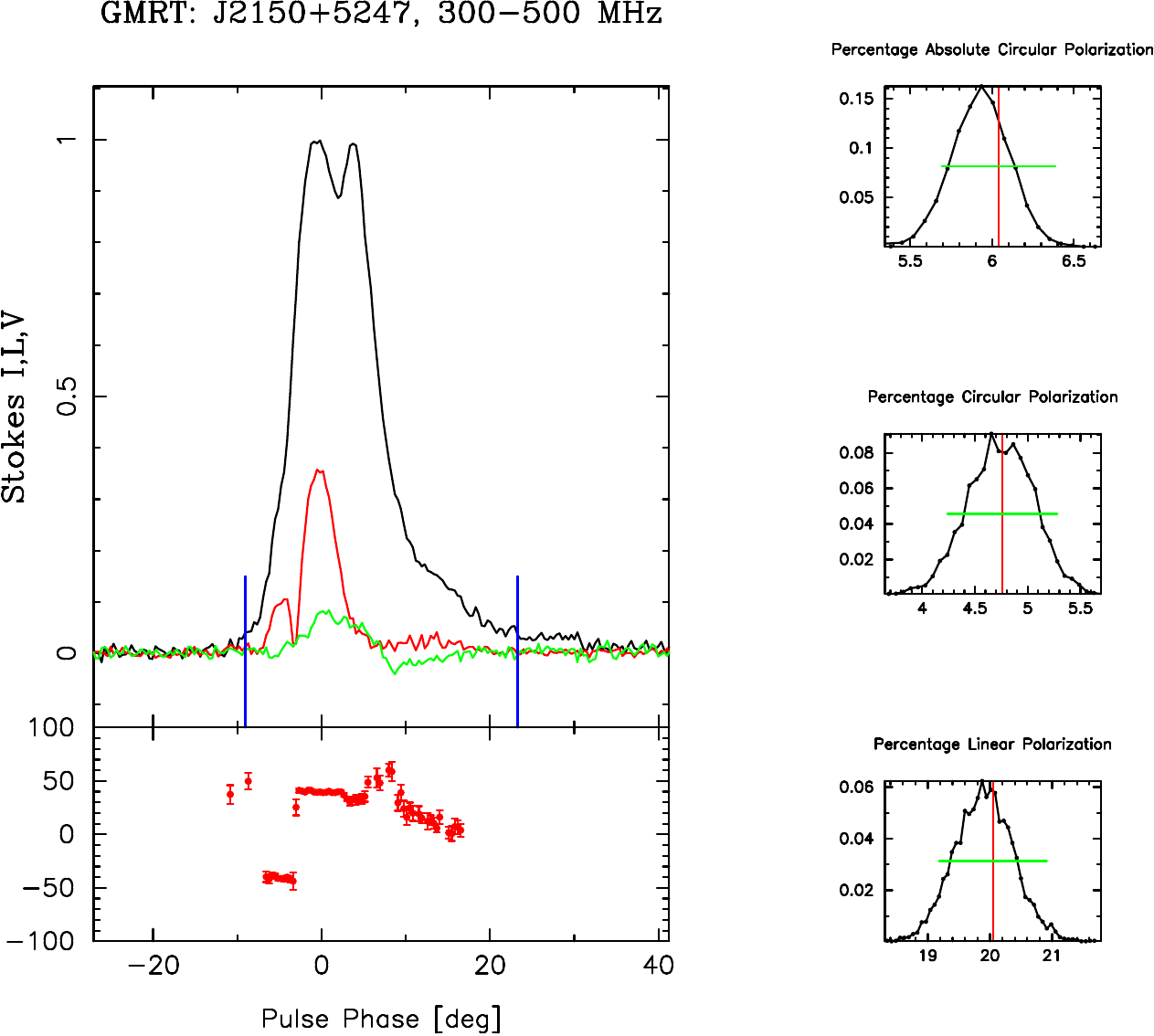}{0.3\textwidth}{(j) PSR J2150+5247 (300--500 MHz)}
          \fig{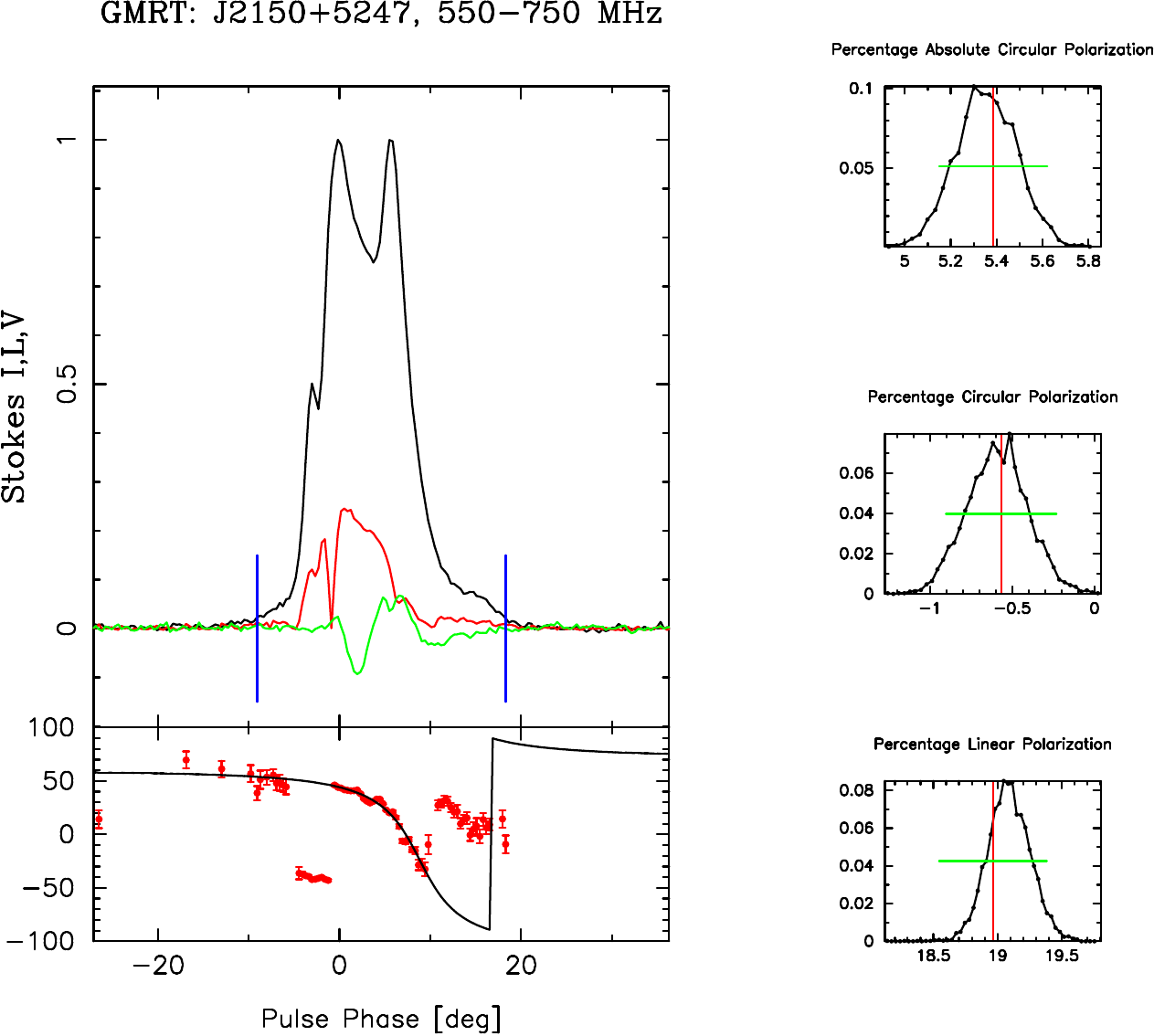}{0.3\textwidth}{(k) PSR J2150+5247 (550--750 MHz)}
          \fig{f1_48.pdf}{0.3\textwidth}{(l) PSR J2229+6114 (550--750 MHz)}}
\label{avgp4}
\caption{See caption in Fig.~\ref{fig:avgpol}.}
\end{figure}

\begin{figure}
\gridline{\fig{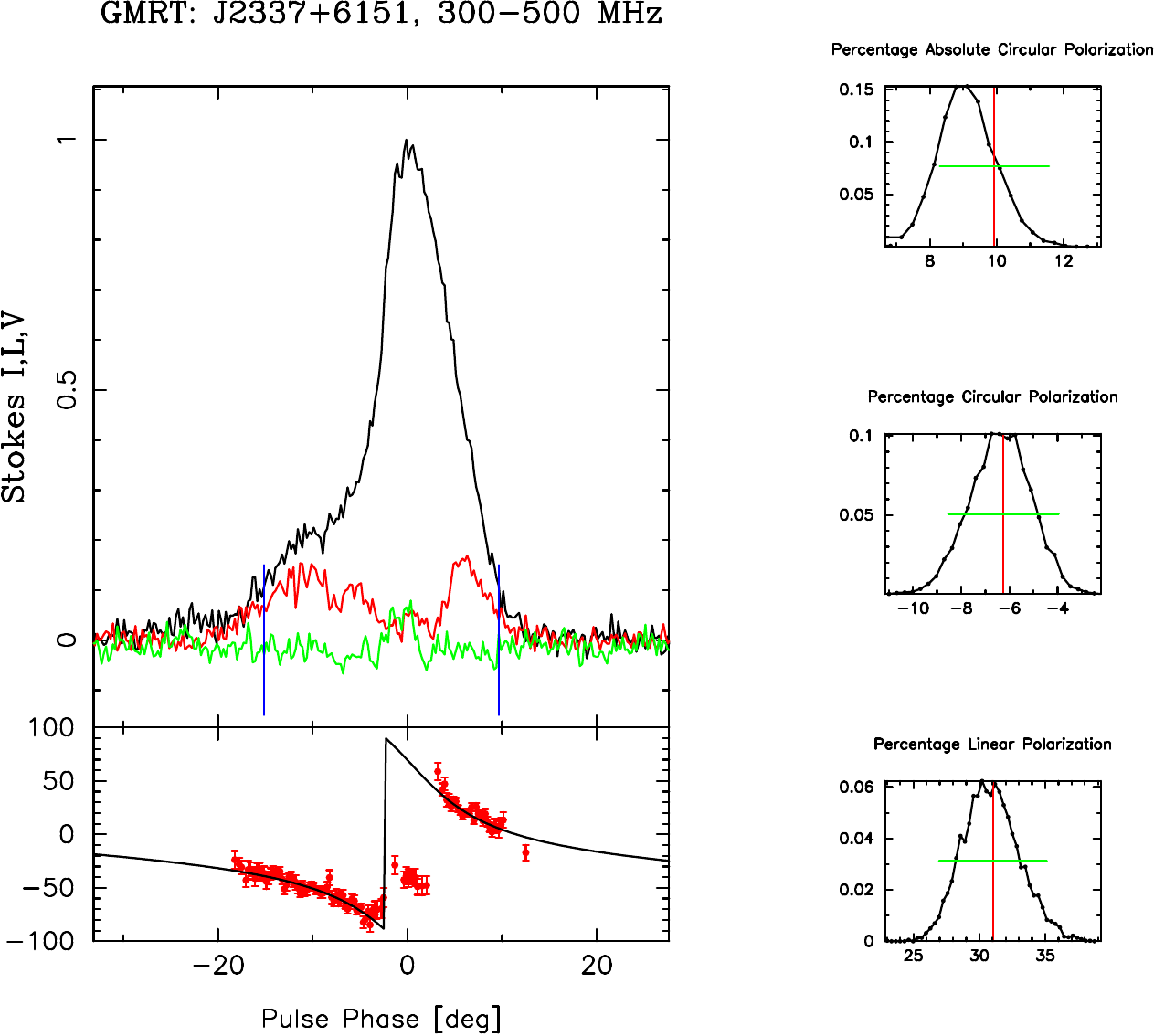}{0.3\textwidth}{(a) PSR J2337+6151 (300--500 MHz)}
          \fig{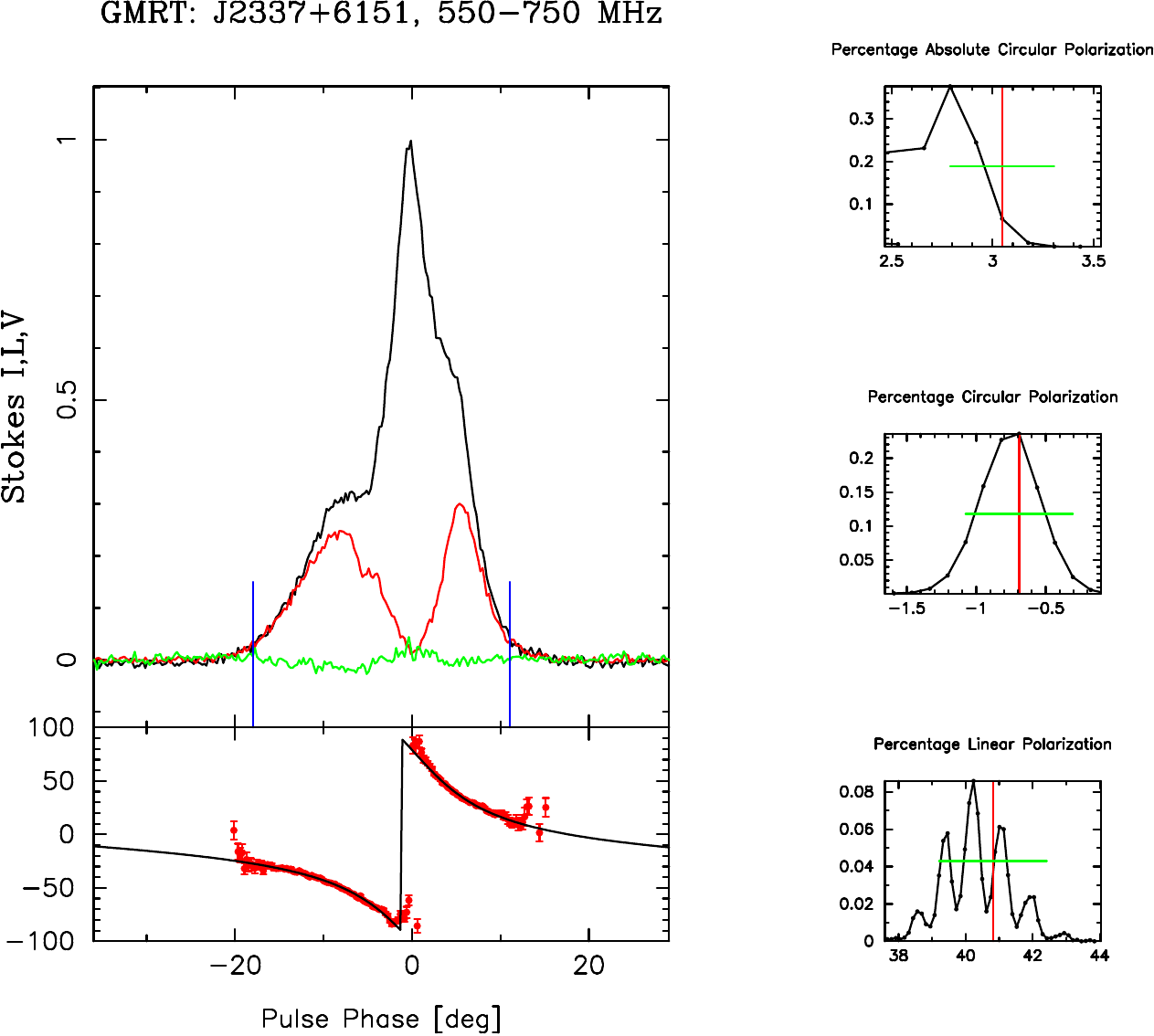}{0.3\textwidth}{(b) PSR J2337+6151 (550--750 MHz)}}
\label{avgp5}
\caption{See caption in Fig.~\ref{fig:avgpol}.}
\end{figure}

\end{document}